\documentclass[11pt,a4paper,english]{article}

\usepackage{float}
\usepackage{graphicx}
\usepackage{amsmath}

\usepackage{amsfonts}
\usepackage{latexsym}

\usepackage{babel}

\newcommand{\beq}{\begin{equation}}
\newcommand{\eeq}{\end{equation}}

\newcommand{\intd}{\mathrm{d}}

\newcommand{\mc}[1]{\mathcal{#1}}

\newcommand{\CC}{\mathbb{C}}

\newcommand{\RR}{\mathbb{R}}
\newcommand{\NN}{\mathbb{N}}

\newcommand{\hlf}{\frac{1}{2}}

\newcommand{\ii}{\mathrm{i}}

\newcommand{\ket}[1]{|#1 \rangle}
\newcommand{\braket}[2]{\langle #1 | #2 \rangle}

\newcommand{\brakettt}[3]{\langle #1 | #2 |#3 \rangle}
\newcommand{\braketv}[2]{\langle #1 , #2 \rangle}

\newcommand{\mass}{\mu}

\begin{document}

\title
{Projection operator approach to the quantization
of\\ higher spin fields}

\author{
G\'abor Zsolt T\'oth\thanks{email: tgzs@rmki.kfki.hu}
\\[4mm] 
\small Institute for Particle and Nuclear Physics \\
\small Wigner Research Centre for Physics, Hungarian Academy of Sciences \\
\small Konkoly Thege Mikl\'os \'ut 29-33\\
\small 1121 Budapest, Hungary\\
\date{}
}
\maketitle

\begin{abstract}
A general method to construct free quantum fields for massive particles of 
arbitrary definite spin in a canonical Hamiltonian framework is presented. 
The main idea of the method is as follows: a multicomponent Klein--Gordon field 
that satisfies canonical (anti)commutation relations and serves as an
auxiliary higher spin field is introduced, and the physical higher spin field 
is obtained by acting on the auxiliary field by a suitable differential operator.  
This allows the calculation of the (anti)commutation relations, the Green functions and the 
Feynman propagators of the higher spin fields. In addition, 
canonical equations of motions, 
which are expressed in terms of the auxiliary variables, can be obtained
also in the presence of interactions, if the interaction Hamiltonian operator is known.
The fields considered transform according to the $(n/2,m/2)\oplus (m/2,n/2)$
and $(n/2,m/2)$ representations of the Lorentz group. 
\end{abstract}

\thispagestyle{empty}

\newpage

\section{Introduction}

Higher spin particles and fields appear in various areas of high-energy physics. For example,  
supergravity models contain 
gravitinos, which have spin $3/2$,
some string theories have excitations corresponding to an infinite sequence 
of particle states of increasing spin, and several baryons and mesons with spins higher than $1$ 
have been observed experimentally.

Formalisms for particles of arbitrary spin were developed by several authors, including Dirac \cite{Dirac}, 
Fierz \cite{Fierz},  Fierz and Pauli \cite{FP}, Rarita and Schwinger \cite{RS}, Wigner and Bargmann \cite{BW}, 
Weinberg \cite{Weinberg,Weinberg2,WeinbergQFT}, Williams \cite{Williams}
(for further references see e.g.\ \cite{Lorce,DA,WR,Peng1,Peng2}). 
Fields transforming according to the $(n/2,0)\oplus (0,n/2)$ and $(n/2,m/2)$ representations  
of $SL(2,\CC)$ were considered in \cite{Weinberg,Weinberg2,WeinbergQFT,Williams}.   
The representations 
$(n/2,0)\oplus (0,n/2)$ in particular have the advantage that
their analogy with the Dirac representation (which corresponds to $n=1$) is very close, and they are also 
relatively simple.

In Weinberg's approach the main object is to obtain Feynman rules for perturbation calculations, 
and for this purpose it is
sufficient to construct free
higher spin quantum fields directly from particle creation and annihilation operators.
These fields can be used to build local interaction Hamiltonians, 
thus perturbation theory can be applied to compute S-matrix elements.  
The application of a canonical formalism is avoided in this way, which can be regarded as an advantage, 
because the canonical---in particular the Lagrangian---formalism is known to have serious difficulties in the case of 
higher spin fields (see e.g.\ \cite{VZ,Madore,DWP,SC,Frauendiener}). 
Nevertheless, a canonical formulation, and especially having field equations, 
may be useful for nonperturbative investigations.

In this paper we present a simple
construction of fields transforming according to the representations $(n/2,m/2)\oplus (m/2,n/2)$ and 
$(n/2,m/2)$ for massive particles of definite spin and mass
in a canonical Hamiltonian approach, 
which allows the derivation of canonical field equations also
in the presence of interactions.  
Our approach can also be regarded as a quantization method for higher spin fields,
and it provides a method somewhat different from that in \cite{Weinberg,WeinbergQFT} for obtaining 
the commutation or anticommutation relations, the Green functions, Feynman propagators, and the mode expansions. 
Otherwise, concerning the construction of interaction Hamiltonian operators and the perturbative computation of S-matrix elements,
our approach is compatible with that of \cite{Weinberg,WeinbergQFT}.

The main idea of our method can be stated briefly as follows:
we introduce an auxiliary multicomponent Klein--Gordon quantum 
field $\Psi$ that satisfies canonical commutation or anticommutation relations,
and obtain the sought higher spin field in the form
\beq
\label{eq.1}
\psi=\mc{D}\Psi\ ,
\eeq 
where $\mc{D}$ is a suitable differential operator, which projects out the physical degrees of freedom from 
$\Psi$. The properties of $\Psi$ and $\mc{D}$ determine then the properties
of $\psi$, and canonical Hamiltonian equations of motion can also be derived if the interaction Hamiltonian operator is
known. 
We note that projection operators were applied in the quantization of higher spin fields 
also in \cite{UV,AU}, and in the construction of wave functions for higher spin particles in \cite{Fronsdal}.

We use a formalism
in which reality properties are clear 
and it is easy to take the adjoints of the fields in a covariant way. 
An important part of the paper is an appendix where relevant 
parts of the representation theory of the $SL(2,\CC)$ group are discussed. 
In contrast with \cite{Weinberg,Weinberg2,WeinbergQFT,Williams}, we make use of the fact that 
the $(n/2,m/2)$ representations of $SL(2,\CC)$ can be obtained from 
the $(1/2,0)$ and $(0,1/2)$ representations
by taking tensor products. In particular, our treatment of the generalized Dirac gamma matrices is 
different from that in \cite{Weinberg,Weinberg2,WeinbergQFT,Williams}.

The paper is organized as follows. 
In Section \ref{sec.rkg} we start with a discussion of real and complex multicomponent bosonic and fermionic
(commuting and anticommuting)
Klein--Gordon fields in general, without fully specifying Lorentz transformation properties.
In Section \ref{sec.lt} we continue with specifying the Lorentz transformation properties of the auxiliary 
Klein--Gordon field $\Psi$
mentioned above.
In Section \ref{sec.cpt} the $C$, $P$, $T$ transformation properties of $\Psi$ are discussed.
In Section \ref{sec.hs} we introduce the proper 
higher spin fields $\psi$, which have only the desired physical degrees of freedom, and discuss their properties. 
We also discuss the description of interactions and the derivation of canonical Hamiltonian field equations.
In Section \ref{sec.qed} we discuss the case of quantum electrodynamics with a standard spin-$1/2$ (Dirac) field for illustration. 
Our conclusions are given in Section \ref{sec.concl}.
In Appendix \ref{app.a} we discuss relevant 
parts of the representation theory of the $SL(2,\CC)$ group, as mentioned above.
In Appendix \ref{app.b} we discuss, with the aim of further illustration, the calculation of
the equal-time anticommutator of the spin-$3/2$ field that transforms according 
to the representation $(3/2,0)\oplus (0,3/2)$.

Spinor indices are often written explicitly, but they are also suppressed at various places in the text, 
depending on which notation appears clearer. 
Index lowering and raising for spinors is not used. 
Upper indices are used for Minkowski vectors and lower indices for Minkowski covectors, 
whereas lower indices are used for spinors and upper indices for dual spinors. 
We often deal with vector spaces that arise as tensor products of other vector spaces. 
Denoting such a vector space with $V$, we append appropriate multiple indices to the elements $v$
of $V$ when the tensor product structure of $V$ is important, 
but otherwise we often append a single index to $v$,
which reflects only the vector space structure of $V$. The same 
notational method is applied also to linear mappings and tensors involving $V$. 
The dagger ${}^\dagger$ is used to denote the adjoint of operators in Hilbert spaces.  
The term Poincar\'e group refers in this paper to the simply connected covering group of the 
usual connected Poincar\'e group generated
by the translations and Lorentz transformations of the Minkowski space.

\section{Multicomponent Klein--Gordon fields}
\label{sec.rkg}

As the first step in the construction of higher spin fields,
the real bosonic and fermionic (i.e.\ commuting and anticommuting) 
multicomponent Klein--Gordon quantum fields $\Phi_\alpha$ are introduced.  
At this stage the dimension $D$ of space is kept arbitrary, 
and the Lorentz transformation properties of the fields are left unspecified.
The signature of the Minkowski metric $g_{\mu\nu}$ is taken to be $(+-\dots--)$.

The basic fields are 
the canonical fields $\Phi_\alpha$, $\Pi_\alpha$, which are Hermitian (real), i.e.\ 
$\Phi_\alpha^\dagger =\Phi_\alpha$, $\Pi_\alpha^\dagger =\Pi_\alpha$, 
and have $N$ components, indexed by $\alpha=1,\dots, N$. 
They satisfy the commutation or anticommutation relations
\begin{eqnarray}
\label{ac1}
[ \Phi_\alpha(x,t),\Phi_\beta(x',t) ]_\pm & = & 0\\
\label{ac2}
[ \Pi_\alpha(x,t), \Pi_\beta(x',t) ]_\pm & = & 0\\
\label{ac3}
[ \Phi_\alpha(x,t),\Pi_\beta(x',t) ]_\pm & = & \ii \epsilon_{\alpha\beta}\delta^D(x-x')\ .
\end{eqnarray}
The notation $[\ ,\ ]_\pm$ indicates that either 
a commutator (corresponding to the sign $-$) or an anticommutator (corresponding to the sign $+$) is meant;
commutators apply to the case of bosons
and anticommutators to the case of fermions.

In the fermionic case $N$ is even and $\epsilon_{\alpha\beta}$ 
is a nondegenerate antisymmetric and purely imaginary matrix: 
\begin{eqnarray}
\epsilon_{\alpha\beta} & = & -\epsilon_{\beta\alpha}\\
\epsilon_{\alpha\beta} & = & \epsilon_{\beta\alpha}^*\ ,
\end{eqnarray}
while in the bosonic case $\epsilon_{\alpha\beta}$ is a nondegenerate symmetric and real matrix: 
\begin{eqnarray}
\epsilon_{\alpha\beta} & = & \epsilon_{\beta\alpha}\\
\epsilon_{\alpha\beta} & = & \epsilon_{\alpha\beta}^*\ .
\end{eqnarray}
We denote the inverse of $\epsilon_{\alpha\beta}$ by $\epsilon^{\alpha\beta}$; thus we have 
\beq
\epsilon_{\alpha\beta}\epsilon^{\beta\gamma}={\delta_\alpha}^\gamma\ .
\eeq

$\Phi_\alpha$ and $\Pi_\alpha$ satisfy the free field equations 
\begin{eqnarray}
\label{y1}
\partial_t \Phi_\alpha &  = &   \Pi_\alpha \\
\label{y2}
\partial_t \Pi_\alpha & = & \partial_x\partial_x \Phi_\alpha -\mass^2\Phi_\alpha\ ,
\end{eqnarray}
where $\mass > 0$ is the mass of the fields $\Phi_\alpha$.
From (\ref{y1}) and (\ref{y2}) we obtain the Klein--Gordon equation for $\Phi_\alpha$:
\beq
\label{KG}
\partial_t\partial_t \Phi_\alpha-\partial_x\partial_x \Phi_\alpha +\mass^2\Phi_\alpha=0\ .
\eeq

It is a further defining property of $\Phi_\alpha$ that it admits the following 
representation in terms of plane wave solutions of the Klein--Gordon equation: 
\beq
\label{exp}
\Phi_\alpha (x,t)=\int \frac{\intd^D k}{\sqrt{2}(\sqrt{2\pi})^D \omega(k)} [e^{\ii kx}e^{\ii \omega(k)t}
      a_\alpha^\dagger (k)
+e^{-\ii kx}e^{-\ii \omega(k)t}
      a_\alpha (k)]\ ,
\eeq
where $a_\alpha^\dagger (k)$ and $a_\alpha (k)$ are mode creating and annihilating operators,  
\beq
\omega(k)=\sqrt{\mass^2+k^2}\ ,
\eeq
$k$ denotes a $D$-dimensional momentum covector, $kx=\sum_{i=1}^D k_i x^i$ 
denotes the contraction  
of the $D$-dimensional covector $k$ with the $D$-dimensional vector $x$, and 
$k^2=\sum_{i=1}^D k_i^2$.
The relativistic momentum covector corresponding to $k$ is $k_\mu = (\omega(k),k)$, and the 
position vector is, of course, $x^\mu=(t,x)$. 
The operators $a_\alpha(k)$ and $a_\alpha^\dagger (k)$ satisfy the (anti)commutation relations
\begin{eqnarray}
\label{anti1}
[ a_\alpha(k),a_\beta^\dagger (k') ]_\pm & = & \epsilon_{\alpha\beta}\delta^D(k-k')\omega(k)\\
\label{anti2}
[ a_\alpha(k),a_\beta(k') ]_\pm & = & 
[ a_\alpha^\dagger (k),a_\beta^\dagger (k') ]_\pm \ = \ 0\ .
\end{eqnarray}

$a_\alpha(k)$ and $a_\alpha(k)^\dagger$ can be expressed in terms of $\Phi_\alpha$ and $\Pi_\alpha$ as
\begin{eqnarray}
\label{f1}
\frac{\omega(k)}{\sqrt{2}(\sqrt{2\pi})^D}\int \intd x^D\ e^{-\ii kx}[\Phi_\alpha(x,0)+\frac{1}{\ii\omega(k)}\Pi_\alpha(x,0)] & = & a_\alpha^\dagger (k)\\
\label{f2}
\frac{\omega(k)}{\sqrt{2}(\sqrt{2\pi})^D}\int \intd x^D\ e^{\ii kx}[\Phi_\alpha(x,0)-\frac{1}{\ii \omega(k)}\Pi_\alpha(x,0)] & = & a_\alpha (k)\ .
\end{eqnarray}
The relations (\ref{anti1}) and (\ref{anti2}) follow from 
(\ref{ac1}), (\ref{ac2}),  (\ref{ac3}), (\ref{f1}) and (\ref{f2}). Conversely,
(\ref{exp}), the relation $\Pi_\alpha=\partial_t\Phi_\alpha$, and 
the anticommutation relations
(\ref{anti1}) and (\ref{anti2}) imply
(\ref{ac1}), (\ref{ac2}) and (\ref{ac3}).

The Hamiltonian operator  is given by
\beq
\label{H}
H  =  \frac{1}{2} \int \intd^D x\, :[ \Pi_\alpha \Pi_\beta \epsilon^{\alpha\beta}  + (\partial_x\Phi_\alpha) (\partial_x\Phi_\beta) \epsilon^{\alpha\beta}  + \mass^2\Phi_\alpha\Phi_\beta \epsilon^{\alpha\beta} ]:\, \\
\eeq
which is Hermitian, i.e.\  $H^\dagger =H$, as a consequence of the properties of $\epsilon^{\alpha\beta}$
and of the hermiticity of $\Phi_\alpha$ and $\Pi_\alpha$.
The notation $:{}:$ refers to the usual normal ordering for bosons or fermions, respectively.
The general Heisenberg equation of motion  
$\frac{d}{dt}O(t)=[\ii H,O(t)]$ becomes, using 
(\ref{H}), (\ref{ac1}), (\ref{ac2}) and (\ref{ac3}), equation (\ref{y1})
for $O=\Phi_\alpha$, and  (\ref{y2}) for $O=\Pi_\alpha$.
$H$ can be expressed in terms of 
$a_\alpha(k)$ and $a_\alpha(k)^\dagger$ as
\beq
\label{H2}
H =  \int \frac{\intd^D k}{\omega(k)}\ \omega(k)\, a_\alpha(k)^\dagger a_\beta(k) \epsilon^{\alpha\beta}\ .
\eeq

$\Phi_\alpha$ has the usual transformation property under space translations: 
\beq
[\ii P_i,\Phi_\alpha]  =  \partial_i\Phi_\alpha\ ,
\eeq
where $P_i$, $i=1,\dots, D$ are the Hermitian generators of space translations. 
$P_i$ can be expressed as
\beq
P_i  =   \int \intd^D x\ : \Pi_\alpha \partial_i\Phi_\beta \epsilon^{\alpha\beta}: \quad
 =   \int \frac{\intd^D k}{\omega(k)}\ k_i\,   a_\alpha(k)^\dagger a_\beta(k) \epsilon^{\alpha\beta}\ .
\eeq
These operators commute with each other and with $H$.

An energy-momentum tensor can be defined as 
\beq
\label{emt}
T_{\mu\nu}= : \partial_\mu\Phi_\alpha \partial_\nu\Phi_\beta \epsilon^{\alpha\beta}
-\frac{1}{2}g_{\mu\nu}(\partial_\lambda \Phi_\alpha \partial^\lambda \Phi_\beta-\mass^2\Phi_\alpha\Phi_\beta)\epsilon^{\alpha\beta} :\ .
\eeq
$T_{\mu\nu}$ is conserved, i.e.\ 
$\partial_\mu T^{\mu\nu}=0$, symmetric ($T^{\mu\nu}=T^{\nu\mu}$), and Hermitian ($T_{\mu\nu}^\dagger = T_{\mu\nu}$).
The energy and momentum density currents are $E_\mu=T_{\mu 0}$ and $P_{i\mu}=T_{\mu i}$.

A suitable Lagrangian function for the theory is
\beq
\label{LPhi}
L  =  \int \intd^D x\, \mc{L} = \frac{1}{2} \int \intd^D x\, [(\partial_t\Phi_\alpha) (\partial_t\Phi_\beta) \epsilon^{\alpha\beta}  - (\partial_x\Phi_\alpha) (\partial_x\Phi_\beta) \epsilon^{\alpha\beta}  - 
\mass^2\Phi_\alpha\Phi_\beta \epsilon^{\alpha\beta}]\ , 
\eeq
where $\Phi_\alpha$ is a real valued classical field in the bosonic case and a  
Grassmann-algebra valued field in the fermionic case.

The canonical momentum for $\Phi_\alpha$ is  
$\tilde{\Pi}^\alpha= \frac{\partial \mc{L}}{\partial (\partial_t \Phi_\alpha)}=\epsilon^{\alpha\beta} \partial_t \Phi_\beta $, and
the equal time commutation or anticommutation relations of $\Phi$ and $\tilde{\Pi}$ are then
$[ \Phi_\alpha(x,t),\tilde{\Pi}^\beta(x',t) ]_{-}  =  \ii {\delta_\alpha}^\beta \delta^D(x-x')$ and 
$[ \Phi_\alpha(x,t),\tilde{\Pi}^\beta(x',t) ]_{+}  =  -\ii {\delta_\alpha}^\beta \delta^D(x-x')$, respectively.
We could have taken the Lagrangian (\ref{LPhi}) and the latter relations as our starting point, 
however in the present paper we want to focus on 
the Hamiltonian formalism. It should  also be noted that 
although the standard canonical momentum 
for $\Phi_\alpha$ is $\tilde{\Pi}^\alpha$, we find it more convenient in this paper to use $\Pi_\alpha$  
instead. 
The relation between $\tilde{\Pi}^\alpha$ and $\Pi_\alpha$ is $\tilde{\Pi}^\alpha = \epsilon^{\alpha\beta}\Pi_\beta$.

The energy and momentum density currents
$E_\mu$ and $P_{i\mu}$
can be obtained as the Noether currents corresponding to
(\ref{LPhi}) and to the time and space translation symmetries.

\subsection{Hilbert space}
\label{sec.hilsp}

The Fock space is spanned by the vacuum state $\ket{0}$, which is annihilated by the operators 
$a_\alpha(k)$, 
and by the multi-particle states 
\beq
\ket{k_1,p_1;k_2,p_2;\dots;k_j,p_j}=\left(\prod_{i=1}^j  p_{i}^{\alpha}a_\alpha^\dagger(k_i)\right) \ket{0}\ ,
\eeq
where $k_1$, $k_2$, \dots , $k_j$ are the momenta of the particles and $p_1$, $p_2$, \dots , $p_j$ are their polarization vectors. The polarization vectors have $N$ components that can take arbitrary complex values, with the only 
restriction that none of the polarization vectors are allowed to be zero. 
We denote the ($N$ complex dimensional)
space of possible polarization vectors of a particle of momentum $k$ by $\hat{V}(k)$. 
On $\hat{V}(k)$ we introduce 
the scalar product 
\beq
\braketv{p_1}{p_2}=(p_1)^{\alpha *}(p_2)^\beta \epsilon_{\alpha\beta}\ .
\label{eq.sp00}
\eeq 
Here the ${}^*$ denotes componentwise complex conjugation.
Similarly, on the dual space of $\hat{V}(k)$, which we denote by $V(k)$, we introduce the scalar product
\beq
\braketv{p_1}{p_2}=(p_1)_\alpha^*(p_2)_\beta \epsilon^{\beta\alpha}\ .
\label{eq.sp01}
\eeq
With these definitions, if $\hat{u}_i$, $i=1,\dots, N$, is a basis in $\hat{V}(k)$ such that
$\braketv{\hat{u}_i}{\hat{u}_j}=\pm\delta_{ij}$, then the basis $u_i$ dual to $\hat{u}_i$ also satisfies 
$\braketv{u_i}{u_j}=\pm\delta_{ij}$, and vice versa.
(The dual basis is defined by the property $(\hat{u}_i)^\alpha (u_j)_\alpha = \delta_{ij}$.)

The (anti)commutation relation of the creation and annihilation operators for particles in general polarization states
is 
\beq
[(p_1)^{\alpha *} a_\alpha(k_1) , (p_2)^\beta a_\beta^\dagger(k_2) ]_\pm =\braketv{p_1}{p_2} \delta^D(k_1-k_2)\omega(k_1)\ .
\eeq 
This shows that the creation and annihilation operators of particles that have orthogonal polarizations 
(anti)commute.

For fermions the scalar products of the multi-particle states are given by
\begin{multline}
\braket{k_1,p_1;k_2,p_2; \dots ;k_j,p_j}{k_1',p_1';k_2',p_2'; \dots ;k_j',p_j'}=\\
\label{sc}
\sum_P \mathrm{sign}(P) \delta^D(k_1-k_{P_1}')\delta^D(k_2-k_{P_2}') \dots \delta^D(k_j-k_{P_j}')
\omega(k_1)\omega(k_2)\dots \omega(k_j)
\braketv{p_1}{p'_{P_1}}\dots \braketv{p_j}{p'_{P_j}},
\end{multline}
where $P$ is any permutation of the numbers $\{ 1,2, \dots ,j\}$. 
The same formula applies to bosons, but without the $\mathrm{sign}(P)$ factor.
Any two multi-particle
states that contain different number of particles are orthogonal.
The vacuum state is normalized as $\braket{0}{0}=1$.
The formula (\ref{sc}) follows from the anticommutation relations of the $a_\alpha(k)$ and $a_\alpha^\dagger(k)$ operators, from the normalization of $\ket{0}$ and from the property of $\ket{0}$ that it is annihilated by the $a_\alpha(k)$ operators.

The eigenvalue of $H$ on $\ket{k_1,p_1;k_2,p_2; \dots ;k_j,p_j}$ is
$\omega(k_1)+\omega(k_2)+ \dots +\omega(k_j)$, and
$H\ket{0}=0$;
the eigenvalue of $P_i$ on  $\ket{k_1,p_1;k_2,p_2; \dots ;k_j,p_j}$ is
$k_{1i}+k_{2i}+ \dots +k_{ji}$, and $P_i\ket{0}=0$.

The definiteness properties of the scalar product on the Hilbert space defined above are determined by the 
signature of the scalar product $\braketv{\ }{\ }$ given in (\ref{eq.sp00}).
In the case of fermions $\braketv{\ }{\ }$ is not positive definite; it has the signature
$(N/2,N/2)$.
In order to see this, let us assume that $\epsilon_{\alpha\beta}$ takes the canonical form, 
in which the nonzero matrix elements of $\epsilon_{\alpha\beta}$ are 
$\epsilon_{2n-1,2n}=-\epsilon_{2n,2n-1}=\ii$, $n=1,\dots, M$, where $M=N/2$. 
Any $\epsilon_{\alpha\beta}$ matrix can be brought to this form by considering the field $\Phi'_\alpha={S_\alpha}^\beta\Phi_\beta$ 
instead of $\Phi_\alpha$, where ${S_\alpha}^\beta$ is a suitable invertible real matrix. 
If the polarization vector $p$ is such that
its nonzero vector components are $p^{2n-1}=\ii/\sqrt{2}$ and $p^{2n}=1/\sqrt{2}$, 
where $n$ takes some fixed value between $1$ and $M$, then $\braketv{p}{p}=1$. 
We denote these vectors by $p_n^+$, $n=1,\dots, M$ ($n$ is not a vector component index here). 
However, if $p$ is such that
its nonzero vector components are $p^{2n-1}=-\ii/\sqrt{2}$ and $p^{2n}=1/\sqrt{2}$, then  $\braketv{p}{p}=-1$. 
We denote these polarization vectors by $p^-_n$, $n=1,\dots, M$.  
These vectors are orthogonal, i.e.\ $\braketv{p^+_n}{p^+_m}=\braketv{p^-_n}{p^-_m}=0$ if $n\ne m$, 
and $\braketv{p^+_n}{p^-_m}=0$ for all $n,m$. 
Denoting the space spanned by the $p^+_n$ vectors by $\hat{V}(k)_+$ and the space spanned by the $p^-_n$ vectors by $\hat{V}(k)_-$, we have 
$\hat{V}(k)=\hat{V}(k)_+\oplus \hat{V}(k)_-$. This means that $\hat{V}(k)$ can be decomposed into a direct sum of two orthogonal 
$N/2$-dimensional subspaces on which the scalar product $\braketv{\ }{\ }$ is positive and negative definite, respectively. 
We note that it is also 
easy to check that the one-particle states created by the 
$a_\alpha(k)^\dagger$, $\alpha=1,\dots, N$, operators have zero norm.

In the case of bosons, the definiteness of the scalar product on the Hilbert space depends 
on the signature of $\epsilon_{\alpha\beta}$.
In order to see this, let us assume that $\epsilon_{\alpha\beta}$ takes the canonical form, in which $\epsilon_{\alpha\beta}$ is diagonal.  
Any $\epsilon_{\alpha\beta}$ matrix can be brought to this form by considering the field $\Phi'_\alpha={S_\alpha}^\beta\Phi_\beta$ instead of $\Phi_\alpha$, where ${S_\alpha}^\beta$ is a suitable invertible 
real matrix. Now, if a polarization vector $p$ is such that
its only nonzero vector component is $p^{n}=1$, where $n$ takes some fixed value between $1$ and $N$, then $\braketv{p}{p}=\epsilon_{nn}$. 
We denote these vectors by $p_n$ ($n$ is not a vector component index here).   
These vectors are orthogonal, i.e.\ $\braketv{p_n}{p_m}=0$, if $n\ne m$. 
Denoting the space spanned by those $p_n$ vectors for which $\epsilon_{nn}=1$  by $\hat{V}(k)_+$,
and the space spanned by those $p_n$ vectors for which $\epsilon_{nn}=-1$ 
by $\hat{V}(k)_-$, we have 
$\hat{V}(k)=\hat{V}(k)_+\oplus \hat{V}(k)_-$. This means that $\hat{V}(k)$ can be decomposed into a direct sum of two orthogonal 
(with respect to $\braketv{\ }{\ }$)
subspaces, $\hat{V}(k)_+$ and $\hat{V}(k)_-$, on which the scalar product $\braketv{\ }{\ }$ is positive and negative definite, respectively, and the dimension of  $\hat{V}(k)_+$ and $\hat{V}(k)_-$ is given by the signature of 
$\epsilon_{\alpha\beta}$.
The signature of $\braketv{\ }{\ }$ is thus the same as that of $\epsilon_{\alpha\beta}$.

In summary, for any fixed $k$ if $N$ one-particle states with orthogonal polarization are taken, then 
in the fermionic case there are 
$N/2$ states that have positive scalar product with themselves
and $N/2$ states that have negative scalar product with themselves, 
whereas in the bosonic case the number of one-particle states that have positive or negative scalar product 
with themselves is given by the signature of $\epsilon_{\alpha\beta}$.

\subsection{Green function and propagator}

As can be seen from the mode expansion (\ref{exp}), the Green function of $\Phi_\alpha$ is  
\beq
\brakettt{0}{\Phi_\alpha(x,t_x)\Phi_\beta(y,t_y)}{0}=\epsilon_{\alpha\beta}
\int \frac{\intd^D k}{2(2\pi)^D \omega(k)} e^{-\ii k(x-y)}e^{-\ii \omega(k)(t_x-t_y)}  = 
\epsilon_{\alpha\beta} G(x-y,t_x-t_y)\ , 
\eeq
where the function $G$, defined by
\beq
G(x,t)=\int \frac{\intd^D k}{2(2\pi)^D \omega(k)} e^{-\ii kx}e^{-\ii \omega(k) t}\ ,
\eeq
is the Green function of the usual Klein--Gordon field.

The Feynman propagator for $\Phi_\alpha$ takes the form 
\begin{eqnarray}
\brakettt{0}{\mathrm{T}\Phi_\alpha(x,t_x)\Phi_\beta(y,t_y)}{0}
& = & \epsilon_{\alpha\beta}
\int \frac{\intd^D k\, \intd k_0}{(2\pi)^{D+1}}\, \frac{\ii }{k_0^2-k^2-\mass^2+\ii\epsilon}e^{-\ii k(x-y)}e^{-\ii k_0(t_x-t_y)}\\
& = & \epsilon_{\alpha\beta}D_F(x-y,t_x-t_y)\ ,
\end{eqnarray}
where the function $D_F$, defined by 
\beq
D_F(x,t)=
\int \frac{\intd^D k\, \intd k_0}{(2\pi)^{D+1}}\, \frac{\ii }{k_0^2-k^2-\mass^2+\ii\epsilon}e^{-\ii kx}e^{-\ii k_0t}\ ,
\eeq
is the Feynman propagator of the usual Klein--Gordon field.

The (anti)commutator $[\Phi_\alpha(x,t_x),\Phi_\beta(y,t_y)]_\pm$ is
\beq
[\Phi_\alpha(x,t_x),\Phi_\beta(y,t_y)]_\pm = \epsilon_{\alpha\beta}
[G(x-y,t_x-t_y)-G(y-x,t_y-t_x)]\ ,
\eeq
which is also just the commutator of the usual Klein--Gordon field multiplied by $\epsilon_{\alpha\beta}$.

\subsection{Complex fields}
\label{sec.ckg}

A complex bosonic or fermionic multicomponent Klein--Gordon field can be defined in terms of 
two real fields  $\Phi_{1\alpha}$ and $\Phi_{2\alpha}$ in the usual way as 
\beq
\Psi_\alpha=\frac{\Phi_{1\alpha}+\ii\Phi_{2\alpha}}{\sqrt{2}}\ .
\eeq
$\Phi_{1\alpha}$ and $\Phi_{2\alpha}$ are assumed to commute in the bosonic case and to anticommute
in the fermionic case. 
We also introduce the complex momentum fields 
\beq
\Pi_\alpha = \frac{\Pi_{1\alpha}+\ii\Pi_{2\alpha}}{\sqrt{2}}\ .
\eeq
The properties of real multicomponent Klein--Gordon fields can now be used to derive the properties of $\Psi$.

$\Psi_\alpha$ and $\Pi_\alpha$ satisfy the equal-time (anti)commutation relations
\begin{eqnarray}
\label{eq.ac5ps}
[ \Psi_\alpha^\dagger(x,t),\Pi_\beta(x',t) ]_\pm & = & \ii \epsilon_{\alpha\beta}\delta^D(x-x') \\
\label{eq.ac6ps}
[ \Psi_\alpha(x,t),\Pi_\beta^\dagger(x',t) ]_\pm & = & \ii \epsilon_{\alpha\beta}\delta^D(x-x')\ ;
\end{eqnarray}
all other (anti)commutators of $\Psi_\alpha$, $\Psi_\alpha^\dagger$, $\Pi_\alpha$, $\Pi_\alpha^\dagger$ are $0$.

$\Psi_\alpha$ has the mode expansion 
\beq
\Psi_\alpha (x,t)=\int \frac{\intd^D k}{\sqrt{2}(\sqrt{2\pi})^D \omega(k)} [e^{\ii kx}e^{\ii \omega(k)t}
      a_\alpha^\dagger(k)
+e^{-\ii kx}e^{-\ii\omega(k)t}
      b_\alpha (k)]\ ,
\eeq
where the creation and annihilation operators 
$a_\alpha^\dagger (k)$, $a_\alpha(k)$, $b_\alpha^\dagger (k)$ and $b_\alpha(k)$ are given by  
\beq
a_\alpha^\dagger(k)=\frac{a_{1\alpha}^\dagger(k)+\ii a_{2\alpha}^\dagger(k)}{\sqrt{2}}\ ,\quad
a_\alpha(k)=\frac{a_{1\alpha}(k)-\ii a_{2\alpha}(k)}{\sqrt{2}}\ ,
\eeq
\beq
b_\alpha^\dagger(k)=\frac{a_{1\alpha}^\dagger(k)-\ii a_{2\alpha}^\dagger(k)}{\sqrt{2}}\ ,\quad
b_\alpha(k)=\frac{a_{1\alpha}(k)+\ii a_{2\alpha}(k)}{\sqrt{2}}\ ,
\eeq
in terms of the creation and annihilation operators $a_{1\alpha}^\dagger (k)$, $a_{1\alpha}(k)$, $a_{2\alpha}^\dagger (k)$ and $a_{2\alpha}(k)$
appearing in the mode expansion of $\Phi_{1\alpha}$ and $\Phi_{2\alpha}$.
The nonzero (anti)commutators of $a_\alpha^\dagger (k)$, $a_\alpha(k)$, $b_\alpha^\dagger (k)$ and $b_\alpha(k)$
are
\beq
\label{eq.cc1}
[ a_\alpha(k),a_\beta^\dagger (k') ]_\pm =\epsilon_{\alpha\beta}\delta^D(k-k')\omega(k)
\eeq
and
\beq
\label{eq.cc2}
[ b_\alpha(k),b_\beta^\dagger (k') ]_\pm =\epsilon_{\alpha\beta}\delta^D(k-k')\omega(k)\ .
\eeq

The Hamiltonian operator, which is the sum of the Hamiltonian operators of the two real fields 
$\Phi_1$ and $\Phi_2$,
can be written as
\begin{eqnarray}
\label{HH}
H & = &  \int \intd^D x\, :[ \Pi_\alpha^\dagger \Pi_\beta \epsilon^{\alpha\beta}  + (\partial_x\Psi_\alpha^\dagger) (\partial_x\Psi_\beta) \epsilon^{\alpha\beta}  +\mass^2\Psi_\alpha^\dagger\Psi_\beta \epsilon^{\alpha\beta} ]:\\
\label{HH2}
& = & \int \frac{\intd^D k}{\omega(k)}\ \omega(k) [a_\alpha(k)^\dagger a_\beta(k) \epsilon^{\alpha\beta}
+b_\alpha(k)^\dagger b_\beta(k) \epsilon^{\alpha\beta}]\ .
\end{eqnarray}
The equations of motion generated for $\Psi_\alpha$, $\Psi_\alpha^\dagger$, $\Pi_\alpha$ and $\Pi_\alpha^\dagger$ by $H$
are 
\beq
\partial_t\Psi_\alpha=\Pi_\alpha\ ,\quad \partial_t\Psi_\alpha^\dagger=\Pi_\alpha^\dagger\ ,
\eeq
\beq
\partial_t\Pi_\alpha=\partial_x^2\Psi_\alpha-\mass^2\Psi_\alpha\ ,\quad 
\partial_t\Pi_\alpha^\dagger=\partial_x^2\Psi_\alpha^\dagger-\mass^2\Psi_\alpha^\dagger\ ,
\eeq 
and it follows from these equations that $\Psi_\alpha$ and $\Psi_\alpha^\dagger$ satisfy the Klein--Gordon equation.

The generators of spatial translations are
\begin{eqnarray}
P_i & = & \int \intd x^D\ : [\Pi_\alpha^\dagger \partial_i\Psi_\beta +
\Pi_\alpha \partial_i\Psi_\beta^\dagger]
\epsilon^{\alpha\beta}:\\
 & = & \int \frac{\intd k^D}{\omega(k)}\ k_i [a_\alpha^\dagger(k) a_\beta(k) 
+b_\alpha^\dagger(k) b_\beta(k)] \epsilon^{\alpha\beta}\ .
\end{eqnarray}

The energy-momentum tensor, which is the sum of the energy-momentum tensors of 
$\Phi_1$ and $\Phi_2$,
takes the form
\beq
\label{emt2}
T_{\mu\nu}= : (\partial_\mu\Psi_\alpha^\dagger \partial_\nu\Psi_\beta + \partial_\mu\Psi_\alpha \partial_\nu\Psi_\beta^\dagger )
 \epsilon^{\alpha\beta}-\frac{1}{2}g_{\mu\nu}
 (\partial_\lambda\Psi_\alpha^\dagger \partial^\lambda\Psi_\beta +
\partial_\lambda\Psi_\alpha \partial^\lambda\Psi^\dagger_\beta
-\mass^2\Psi_\alpha^\dagger\Psi_\beta-\mass^2\Psi_\alpha\Psi_\beta^\dagger) \epsilon^{\alpha\beta}  :\ .
\eeq

The Green function for $\Psi$ is
\beq
\brakettt{0}{\Psi_\alpha(x,t_x)\Psi_\beta^\dagger(y,t_y)}{0}=  \epsilon_{\alpha\beta} G(x-y,t_x-t_y)\ ,
\eeq
and 
\beq
\brakettt{0}{\Psi_\alpha^\dagger(x,t_x)\Psi_\beta(y,t_y)}{0}= \brakettt{0}{\Psi_\alpha(x,t_x)\Psi_\beta^\dagger(y,t_y)}{0}\ .
\eeq
$\brakettt{0}{\Psi_\alpha(x,t_x)\Psi_\beta(y,t_y)}{0}$ and 
$\brakettt{0}{\Psi_\alpha^\dagger(x,t_x)\Psi_\beta^\dagger(y,t_y)}{0}$ are zero.

The
Feynman propagator is
\beq
\brakettt{0}{\mathrm{T}\Psi_\alpha(x,t_x)\Psi_\beta^\dagger(y,t_y)}{0}=  \epsilon_{\alpha\beta} D_F(x-y,t_x-t_y)\ ,
\eeq
and
\beq
\brakettt{0}{\mathrm{T}\Psi_\alpha^\dagger(x,t_x)\Psi_\beta(y,t_y)}{0}=  
\brakettt{0}{\mathrm{T}\Psi_\alpha(x,t_x)\Psi_\beta^\dagger(y,t_y)}{0}\ .
\eeq
$\brakettt{0}{\mathrm{T}\Psi_\alpha(x,t_x)\Psi_\beta(y,t_y)}{0}$ and 
$\brakettt{0}{\mathrm{T}\Psi_\alpha^\dagger(x,t_x)\Psi_\beta^\dagger(y,t_y)}{0}$ are zero.

The (anti)commutator $[\Psi_\alpha(x,t_x),\Psi_\beta^\dagger(x,t_y)]_\pm$
is
\beq
[\Psi_\alpha(x,t_x),\Psi_\beta^\dagger(x,t_y)]_\pm = \epsilon_{\alpha\beta}
[G(x-y,t_x-t_y)-G(y-x,t_x-t_y)]\ ,
\eeq
\beq
[\Psi_\alpha^\dagger(x,t_x),\Psi_\beta(x,t_y)]_\pm = 
[\Psi_\alpha(x,t_x),\Psi_\beta^\dagger(x,t_y)]_\pm\ ,
\eeq
and 
$[\Psi_\alpha(x,t_x),\Psi_\beta(x,t_y)]_\pm =
[\Psi_\alpha^\dagger(x,t_x),\Psi_\beta^\dagger(x,t_y)]_\pm =0$.

\section{Lorentz transformations}
\label{sec.lt}

In this section we turn to the next step in the construction of higher spin fields, 
which is the definition of the $SL(2,\CC)$ transformation properties of the field $\Psi$ mentioned in the Introduction. 
In this and the following sections the discussion is restricted to the case of $D=3$ space dimensions. 
It is assumed that $\Psi$ is a complex multicomponent Klein--Gordon field in the sense of Section \ref{sec.ckg},
i.e.\ this is the starting point for the definition of the $SL(2,\CC)$ transformation properties.
Relevant parts of the representation theory of $SL(2,\CC)$ 
are collected in Appendix \ref{app.a}.

We consider the three possibilities that $\Psi$ transforms according to the representations $D^{(n)}$,
$D^{(n,m)}$ or $\tilde{D}^{(n)}$, where
\beq
D^{(n)}=(n/2,0)\oplus (0,n/2), \qquad n\in \NN^+ \ ,
\eeq
\beq
D^{(n,m)}=(n/2,m/2)\oplus (m/2,n/2),\qquad n,m\in \NN^+,\ \ n\ge m \ , 
\eeq
\beq 
\tilde{D}^{(n)}=(n/2,n/2),\qquad n\in \NN^+\ .
\eeq
The number of the 
components of $\Psi$ is $2(n+1)$, $2(n+1)(m+1)$ and $(n+1)^2$, respectively, in these cases, in accordance with the 
dimension of the representations.
In the cases of $D^{(n)}$ or $D^{(n,m)}$, $\Psi$ is bosonic if $n$ or $n+m$, respectively, is even, 
and fermionic if $n$ or $n+m$ is odd. $\Psi$ is bosonic for any value of $n$ in the case of $\tilde{D}^{(n)}$.

In the present and the following sections it is assumed that an arbitrary but fixed basis of $D^{(n)}$, $D^{(n,m)}$ or $\tilde{D}^{(n)}$
has been chosen, which is real with respect to the invariant complex conjugation defined in \ref{app.a5}.
The components of $\Psi$ are understood to be vector components
with respect to this basis. 
The reality of the basis is important because it implies that the
covariant complex conjugation coincides with the componentwise complex conjugation; in particular taking the 
adjoint of $\Psi$ (or other fields) componentwise is covariant. 
Otherwise, it is not necessary to explicitly specify the basis vectors.

The reason for taking the representation according to which $\Psi$ transforms to be
$D^{(n)}$, $D^{(n,m)}$ or $\tilde{D}^{(n)}$ is that these\footnote{with the exception of $D^{(n,n)}$} 
are the   
finite dimensional real irreducible representations of $SL(2,\CC)$. The representations 
$(n/2,m/2)$, $n\ne m$, are also irreducible, but not real, i.e.\ they do not admit an invariant complex conjugation,
therefore if $\Psi$ were defined to transform according to $(n/2,m/2)$ for some $n\ne m$, then the adjoint of 
$\Psi$ could not be defined in a covariant manner without getting the representation $(m/2,n/2)$ also involved.
Nevertheless, fields that transform according to the representations $(n/2,m/2)$ 
will also be considered later on.

We take $\epsilon^{\alpha\beta}$ 
to be the $SL(2,\CC)$-invariant bilinear form, 
described in \ref{app.a3},
on $D^{(n)}$, $D^{(n,m)}$ or $\tilde{D}^{(n)}$. 
$\epsilon^{\alpha\beta}$ is symmetric and real in the bosonic cases and  
antisymmetric and purely imaginary in the fermionic cases, as required by the definitions in Section
\ref{sec.rkg}.
The indices $\alpha$ and $\beta$ are understood to label the 
components of $\epsilon$ with respect to the real basis mentioned above.

The condition of the reality of the basis vectors could be dropped, but then
the $SL(2,\CC)$-covariant adjoint of $\Psi$ also includes a mixing of its components, and
a corresponding extension of the formalism of Section \ref{sec.rkg} is necessary.

In the case of $D^{(n)}$, the $SL(2,\CC)$ transformation law 
for $\Psi$ is 
\beq
\label{eq.lor}
U[\Lambda]^{-1}\Psi_\alpha(x) U[\Lambda]  = 
{(\Lambda_{D^{(n)}})_\alpha}^\beta \Psi_\beta (\Lambda_M^{-1}x)\ ,
\eeq
where $\Lambda$ is an element of $SL(2,\CC)$,
$U[\Lambda]$ is the unitary operator that represents $\Lambda$ in the Hilbert space, 
$\Lambda_{D^{(n)}}$ is the matrix that represents $\Lambda$ 
in $D^{(n)}$,
and $\Lambda_M$ represents $\Lambda$ in Minkowski space.
In the cases of $D^{(n,m)}$ and $\tilde{D}^{(n)}$, $\Lambda_{D^{(n)}}$ should be replaced by 
$\Lambda_{D^{(n,m)}}$ and $\Lambda_{\tilde{D}^{(n)}}$, respectively.
Due to the reality of $\Lambda_{D^{(n)}}$, $\Lambda_{D^{(n,m)}}$ and $\Lambda_{\tilde{D}^{(n)}}$,
the same transformation rule applies to $\Psi^\dagger$ and to $\Phi_1$ and $\Phi_2$.

It follows that $SL(2,\CC)$ transformations act on the creation operators $a_\alpha^\dagger(k)$ as
\beq
U[\Lambda]^{-1} a_\alpha^\dagger(k) U[\Lambda]={(\Lambda_{D^{(n)}})_\alpha}^\beta a_\beta^\dagger(\Lambda_M^T k)\ ,
\eeq 
and $a_\alpha(k)$, $b_\alpha^\dagger(k)$ and $b_\alpha(k)$ also have the same transformation property.
In this formula the $T$ in the superscript denotes transposition.  
By writing $\Lambda_M^T k$, we mean, of course, that 
$\Lambda_M^T$ acts on the dual four-vector $k_\mu=(\omega(k),k)$.
Again, $\Lambda_{D^{(n)}}$ should be replaced by 
$\Lambda_{D^{(n,m)}}$ or $\Lambda_{\tilde{D}^{(n)}}$, respectively, in the cases of $D^{(n,m)}$ and $\tilde{D}^{(n)}$.

For simplicity, we continue now with the discussion of the case of $D^{(n)}$, 
and we turn to the cases $D^{(n,m)}$ and $\tilde{D}^{(n)}$ afterwards.  
In order to see what kind of particles are described by $\Psi$, 
we introduce the creation operators 
\begin{eqnarray}
c_i^\dagger(k) & = & \hat{u}_i(k)^\alpha a_\alpha^\dagger(k),\qquad i=1,\dots, (n+1)\\
d_i^\dagger(k) & = & \hat{u}_i(k)^\alpha  b_\alpha^\dagger(k),\qquad i=1,\dots, (n+1)\\
f_i^\dagger(k) & = & \hat{v}_i(k)^\alpha a_\alpha^\dagger(k),\qquad i=1,\dots, (n+1)\\
h_i^\dagger(k) & = & \hat{v}_i(k)^\alpha  b_\alpha^\dagger(k),\qquad i=1,\dots, (n+1)\ .
\end{eqnarray}
The polarization vectors $\hat{u}_i(k)$ and $\hat{v}_i(k)$ appearing here are defined in \ref{app.a51}. 
They form a complete orthogonal set for any fixed $k$. 
$c_i^\dagger(k)$ is related to $c_i^\dagger(0)$ by the formula
\beq
\label{eq.57}
c_i^\dagger(k)  = U[\Lambda(k)]    c_i^\dagger(0)  U[\Lambda(k)]^{-1}\ ,
\eeq
where $\Lambda(k)$ is the $SL(2,\CC)$ element determined by the properties that 
$\Lambda(k)$ is a continuous function of $k$, $\Lambda(0)=I$, and
the Lorentz transformation corresponding to $\Lambda(k)$ is 
the Lorentz boost\footnote{This boost is understood to leave the plane spanned by 
$(\mass,0)$ and $(\omega(k),k)$ invariant and act as the identity in the orthogonal plane. 
The term `Lorentz boost' is used in the same sense throughout the paper.}
that takes the dual four-vector $(\mass,0)$ to $(\omega(k),k)$.  
The same formula (\ref{eq.57}) applies to
$d_i^\dagger(k)$, $f_i^\dagger(k)$ and $h_i^\dagger(k)$.

The operators $c_i(k)$, $d_i(k)$, $f_i(k)$, $h_i(k)$ satisfy the (anti)commutation relations
\begin{eqnarray}
&& [ c_i(k),c_j^\dagger (k') ]_\pm = [ d_i(k),d_j^\dagger (k') ]_\pm = \delta_{ij}\delta^3(k-k')\omega(k)\\
&& [ f_i(k),f_j^\dagger (k') ]_\pm = [ h_i(k),h_j^\dagger (k') ]_\pm = -\delta_{ij}\delta^3(k-k')\omega(k)\ ,
\end{eqnarray}
all other (anti)commutators of them are zero.
This shows that
$c_i^\dagger(k)$ and $d_i^\dagger(k)$ create states that have positive scalar product with themselves,
whereas 
$f_i^\dagger(k)$ and $h_i^\dagger(k)$ create states that have negative scalar product with themselves.

On the one-particle states $c_i^\dagger(k)\ket{0}$,  $d_i^\dagger(k)\ket{0}$, 
$f_i^\dagger(k)\ket{0}$,  $h_i^\dagger(k)\ket{0}$ the spin component operator 
$U[\Lambda(k)] \check{M}_3 U[\Lambda(k)]^{-1}$ has the eigenvalues
$\ii(n/2+1-i)$, $i=1,\dots, (n+1)$. Here $\check{M}_3$ denotes the (anti-Hermitian) operator representing 
the generator $M_3$ of $sl(2,\CC)$. 
Furthermore, the four one-particle subspaces spanned by $c_i^\dagger(k)\ket{0}$, $d_i^\dagger(k)\ket{0}$, 
$f_i^\dagger(k)\ket{0}$, $h_i^\dagger(k)\ket{0}$, $i=1,\dots, (n+1)$, respectively, are orthogonal and closed under 
$SL(2,\CC)$
transformations and space and time translations, i.e.\ they form irreducible unitary representations of 
the Poincar\'e group, characterized by mass $\mass$ and spin $n/2$.
Thus $\Psi$ describes four particles of mass $\mass$ and spin $n/2$, of which two are physical and
two are nonphysical (in the sense that the one-particle states have negative scalar product with themselves).

Using relations (\ref{eq.c1}) and (\ref{eq.c2}), $\Psi$ can be expressed in terms of $c_i$, $d_i$, $f_i$, $h_i$ as
\begin{eqnarray}
&& \Psi_\alpha (x,t)=\int \frac{\intd^3 k}{\sqrt{2}(\sqrt{2\pi})^3 \omega(k)} [e^{\ii kx}e^{\ii \omega(k)t}
      \sum_{i=1}^{n+1}  \{ u_{i\alpha}(k) c_i^\dagger(k) + v_{i\alpha}(k) f_i^\dagger(k) \} \nonumber \\
&& \hspace{5cm} +e^{-\ii kx}e^{-\ii\omega(k)t}
     \sum_{i=1}^{n+1} \{ u_{i\alpha}(k)^* d_i(k) +v_{i\alpha}(k)^* h_i(k) \} ]\ .
\label{eq.64}
\end{eqnarray}
For the real and imaginary parts $\Phi_1$ and $\Phi_2$ of $\Psi$ we have
\begin{eqnarray}
&& \Phi_{1\alpha} (x,t)=\int \frac{\intd^3 k}{\sqrt{2}(\sqrt{2\pi})^3 \omega(k)} [e^{\ii kx}e^{\ii \omega(k)t}
      \sum_{i=1}^{n+1}  \{ u_{i\alpha}(k) \frac{c_i^\dagger(k)+d_i^\dagger(k)}{\sqrt{2}} +
 v_{i\alpha}(k) \frac{f_i^\dagger(k) + h_i^\dagger(k) }{\sqrt{2}}   \} \nonumber \\
&& \hspace{3cm} +e^{-\ii kx}e^{-\ii\omega(k)t}
     \sum_{i=1}^{n+1} \{ u_{i\alpha}(k)^* \frac{c_i(k)+d_i(k)}{\sqrt{2}} +v_{i\alpha}(k)^* \frac{f_i(k)+h_i(k)}{\sqrt{2}} \} ]\ .
\end{eqnarray}
\begin{eqnarray}
&& \Phi_{2\alpha} (x,t)=\int \frac{\intd^3 k}{\sqrt{2}(\sqrt{2\pi})^3 \omega(k)} [e^{\ii kx}e^{\ii \omega(k)t}
      \sum_{i=1}^{n+1}  \{ u_{i\alpha}(k) \frac{c_i^\dagger(k)-d_i^\dagger(k)}{\ii\sqrt{2}} +
 v_{i\alpha}(k) \frac{f_i^\dagger(k) - h_i^\dagger(k) }{\ii\sqrt{2}}   \} \nonumber \\
&& \hspace{3cm} +e^{-\ii kx}e^{-\ii\omega(k)t}
     \sum_{i=1}^{n+1} \{ u_{i\alpha}(k)^* \frac{d_i(k)-c_i(k)}{\ii\sqrt{2}} +v_{i\alpha}(k)^* \frac{h_i(k)-f_i(k)}{\ii\sqrt{2}} \} ]\ .
\end{eqnarray}

The Hamiltonian operator and the generators of space translations can be expressed as
\beq
H=
\int \frac{\intd^3 k}{\omega(k)}\ \omega(k)  \sum_{i=1}^{n+1}[c_i(k)^\dagger c_i(k) + d_i(k)^\dagger d_i(k)
- f_i(k)^\dagger f_i(k) - h_i(k)^\dagger h_i(k)]
\eeq
and 
\beq
\label{eq.68}
P_j=
\int \frac{\intd^3 k}{\omega(k)}\ k_j  \sum_{i=1}^{n+1}[c_i(k)^\dagger c_i(k) + d_i(k)^\dagger d_i(k)
- f_i(k)^\dagger f_i(k) - h_i(k)^\dagger h_i(k)]\ , \qquad j=1,2,3.
\eeq

We turn now to the case of $D^{(n,m)}$, which is largely analogous to the case of $D^{(n)}$, 
therefore it is discussed in less detail.
The creation operators $c_{ij}^\dagger(k)$, $d_{ij}^\dagger(k)$, $f_{ij}^\dagger(k)$, $h_{ij}^\dagger(k)$,
$i=1,\dots, (n+1)$, $j=1,\dots, (m+1)$ can be defined as
\begin{eqnarray}
c_{ij}^\dagger(k) & = & \hat{u}_{ij}(k)^\alpha a_\alpha^\dagger(k)\\
d_{ij}^\dagger(k) & = & \hat{u}_{ij}(k)^\alpha  b_\alpha^\dagger(k)\\
f_{ij}^\dagger(k) & = & \hat{v}_{ij}(k)^\alpha a_\alpha^\dagger(k)\\
h_{ij}^\dagger(k) & = & \hat{v}_{ij}(k)^\alpha  b_\alpha^\dagger(k)\ ,
\end{eqnarray}
where $\hat{u}_{ij}(k)^\alpha$ are polarization vectors defined in \ref{app.a52}.
Formulas analogous to (\ref{eq.57}) hold also for these operators.

The (anti)commutation relations for $c_{ij}(k)$, $d_{ij}(k)$, $f_{ij}(k)$, $h_{ij}(k)$ are 
\begin{eqnarray}
\label{eq.cijkl1}
&& [ c_{ij}(k),c_{kl}^\dagger (k') ]_\pm = [ d_{ij}(k),d_{kl}^\dagger (k') ]_\pm = \delta_{ik}\delta_{jl}\delta^3(k-k')\omega(k)\\
\label{eq.cijkl2}
&& [ f_{ij}(k),f_{kl}^\dagger (k') ]_\pm = [ h_{ij}(k),h_{kl}^\dagger (k') ]_\pm = -\delta_{ik}\delta_{jl}\delta^3(k-k')\omega(k)\ .
\end{eqnarray}
This shows that
$c_{ij}^\dagger(k)$ and $d_{ij}^\dagger(k)$ create one-particle states that have positive scalar product with themselves,
whereas 
$f_{ij}^\dagger(k)$ and $h_{ij}^\dagger(k)$ create one-particle states that have negative scalar product with themselves.

Equations analogous to (\ref{eq.64}) - (\ref{eq.68}) can also be written down.
On the one-particle states $c_{ij}^\dagger(k)\ket{0}$,  $d_{ij}^\dagger(k)\ket{0}$, 
$f_{ij}^\dagger(k)\ket{0}$,  $h_{ij}^\dagger(k)\ket{0}$ the spin component operator 
$U[\Lambda(k)] \check{M}_3 U[\Lambda(k)]^{-1}$ has the eigenvalues
$\ii(\frac{n+m}{2}+2-i-j)$. 
The four one-particle subspaces spanned by $c_{ij}^\dagger(k)\ket{0}$, $d_{ij}^\dagger(k)\ket{0}$, 
$f_{ij}^\dagger(k)\ket{0}$, $h_{ij}^\dagger(k)\ket{0}$, $i=1,\dots, (n+1)$, $j=1,\dots, (m+1)$, respectively, 
are also closed under 
$SL(2,\CC)$
transformations and space and time translations, i.e.\ they form unitary representations of the Poincar\'e group. These representations are not irreducible; they are composed of irreducible representations of different
spins. These irreducible representations can be determined by looking at the action of the $SU(2)$ (rotation) subgroup that leaves
an arbitrary fixed momentum vector $k$ invariant on the finite dimensional spaces spanned by 
$c_{ij}^\dagger(k)\ket{0}$, $d_{ij}^\dagger(k)\ket{0}$, 
$f_{ij}^\dagger(k)\ket{0}$, $h_{ij}^\dagger(k)\ket{0}$, $i=1,\dots, (n+1)$, $j=1,\dots, (m+1)$. 
It is convenient to choose $k=0$, and then it can be seen that 
the decomposition of each of these four spaces into irreducible representations of
the $SU(2)$ subgroup leaving $k=0$ invariant is 
$((n+m)/2) \oplus ((n+m)/2-1) \oplus \dots \oplus ((n-m)/2)$.     
This means that $\Psi$ describes four particles of mass $\mass$ 
for each spin $(n+m)/2, (n+m)/2-1, \dots, (n-m)/2$,
of which two are physical and
two are nonphysical (in the sense that the one-particle states have negative scalar product with themselves).

In the case of $\tilde{D}^{(n)}$, creation operators $c_{ij}^\dagger(k)$, $d_{ij}^\dagger(k)$, 
$i=1,\dots, (n+1)$, $j=1,\dots, i$, 
$f_{ij}^\dagger(k)$, $h_{ij}^\dagger(k)$,
$i=1,\dots, (n+1)$, $j=1,\dots, (i-1)$, can be defined in the same way as above, and their commutation relations are also
given by (\ref{eq.cijkl1}), (\ref{eq.cijkl2}).
$c_{ij}^\dagger(k)$ and $d_{ij}^\dagger(k)$ create one-particle states that have positive scalar product with themselves,
whereas 
$f_{ij}^\dagger(k)$ and $h_{ij}^\dagger(k)$ create one-particle states that have negative scalar product with themselves.
On the one-particle states $c_{ij}^\dagger(k)\ket{0}$,  $d_{ij}^\dagger(k)\ket{0}$, 
$f_{ij}^\dagger(k)\ket{0}$,  $h_{ij}^\dagger(k)\ket{0}$ the spin component operator 
$U[\Lambda(k)] \check{M}_3 U[\Lambda(k)]^{-1}$ has the eigenvalues
$\ii(n+2-i-j)$. 
The four one-particle subspaces spanned by $c_{ij}^\dagger(k)\ket{0}$, $d_{ij}^\dagger(k)\ket{0}$, 
$f_{ij}^\dagger(k)\ket{0}$, $h_{ij}^\dagger(k)\ket{0}$, respectively, 
also form unitary representations of the Poincar\'e group.
The decomposition of these representations into irreducible ones is as follows:
the two spaces spanned by $c_{ij}^\dagger(k)\ket{0}$ and $d_{ij}^\dagger(k)\ket{0}$ have the decomposition
$(n)\oplus (n-2) \oplus (n-4)\oplus \dots \oplus (0)$ if $n$ is even and 
$(n)\oplus (n-2) \oplus (n-4)\oplus \dots \oplus (1)$ if $n$ is odd, and  
the two spaces spanned by $f_{ij}^\dagger(k)\ket{0}$ and $h_{ij}^\dagger(k)\ket{0}$
have the decomposition
$(n-1)\oplus (n-3) \oplus (n-5)\oplus \dots \oplus (1)$ if $n$ is even and 
$(n-1)\oplus (n-3) \oplus (n-5)\oplus \dots \oplus (0)$ if $n$ is odd. Here the numbers in the brackets denote the spin of the 
representations; the mass of these representations is $\mass$.   
Thus $\Psi$ describes two physical particles of mass $\mass$ 
for each spin $n, n-2, \dots$, and two nonphysical particles of mass $\mass$ 
for each spin $n-1, n-3, \dots$. 
This obviously means, in particular, that for a given value of $n$ the physical particles
that $\Psi$ and $\Psi^\dagger$
can create cannot have spins $n-1, n-3, \dots $. Nevertheless, 
$\Psi$ and $\Psi^\dagger$ fields which create physical particles of spin
$n-1, n-3, \dots $ can be obtained by changing the sign of $\epsilon^{\alpha\beta}$.

\section{$C$, $P$, $T$ transformations}
\label{sec.cpt}

In this section the transformation properties of $\Psi$ under 
charge conjugation and space and time reflections are defined, although
the exploration of all possibilities for defining these transformation properties is not attempted.

\noindent
{\it Charge conjugation}\\
Charge conjugation is represented on the Hilbert space by a unitary operator $C$ that has the properties 
\beq
C^2=I\ ,\qquad C\ket{0}=\ket{0}\ ,
\eeq
\beq 
CHC^{-1}=H\ ,\quad CP_iC^{-1}=P_i\ ,\quad                 
C\check{M}_iC^{-1}=\check{M}_i\ ,\quad C\check{N}_iC^{-1}=\check{N}_i\ ,\quad i=1,2,3\ ,
\eeq
where $H$, $P_i$ are the Hermitian generators of time and space translations, and  
$\check{M}_i$ and $\check{N}_i$ 
are the anti-Hermitian operators representing the generators of $sl(2,\CC)$ on the Hilbert space. 
The behaviour of the fields $\Phi_{1\alpha}$, $\Phi_{2\alpha}$ and $\Psi_\alpha$ 
under the action of $C$ is given by the equations
\begin{eqnarray}
C^{-1}\Phi_{1\alpha} C & = &  \Phi_{1\alpha}\\
C^{-1}\Phi_{2\alpha} C & = & -\Phi_{2\alpha}\\
C^{-1}\Psi_\alpha C & = &  \Psi_\alpha^\dagger \ .
\end{eqnarray}
$\Phi_1$ is thus a self-conjugate field, whereas $\Phi_2$ is anti-self-conjugate.
$C$ acts 
on the creation operators $a_{1\alpha}^\dagger (k)$, $a_{2\alpha}^\dagger (k)$,
$a_{\alpha}^\dagger (k)$,  $b_{\alpha}^\dagger (k)$   as
\begin{eqnarray}
&& C^{-1}a_{1\alpha}^\dagger (k)C=a_{1\alpha}^\dagger(k)\ ,
 \hspace{1cm} C^{-1}a_{2\alpha}^\dagger (k)C=-a_{2\alpha}^\dagger(k)\ ,\\
&& C^{-1}a_{\alpha}^\dagger (k)C= b_{\alpha}^\dagger(k)\ ,
 \hspace{1.34cm} C^{-1}b_{\alpha}^\dagger (k)C= a_{\alpha}^\dagger(k)\ .
\end{eqnarray}
In the case of the representations $D^{(n)}$,
\begin{eqnarray}
C^{-1}c_i^\dagger (k) C & = &  d_i^\dagger (k)\\
C^{-1}d_i^\dagger (k) C & = &  c_i^\dagger (k)\\
C^{-1}f_i^\dagger (k) C & = &  h_i^\dagger (k)\\
C^{-1}h_i^\dagger (k) C & = &  f_i^\dagger (k)\ ,
\end{eqnarray}
i.e.\ charge conjugation interchanges $c_i^\dagger(k) $ with $d_i^\dagger(k)$ and $f_i^\dagger(k)$ with $h_i^\dagger(k)$, respectively. 
Completely analogous formulas apply also in the cases of $D^{(n,m)}$ and $\tilde{D}^{(n)}$.

\noindent
{\it Parity}\\
The space reflection transformation $(t,x)\to (t,-x)$ is represented on the Hilbert space by a unitary operator $P$, which has the properties 
\beq
P^2=I\ ,\qquad 
P\ket{0}=\ket{0}\ .
\eeq 
It is related to the generators $M_i$, $N_i$, $i=1,2,3$ of $sl(2,\CC)$ as
\beq
P\check{M}_iP^{-1}=\check{M}_i\ ,\qquad
P\check{N}_iP^{-1}=-\check{N}_i\ ,
\eeq                                   
and $H$ and $P_i$ transform as
\beq
PHP^{-1}=H\ ,\qquad PP_iP^{-1}=-P_i\ .
\eeq

If $\Psi_\alpha$ is fermionic, then the transformation of the fields 
$\Phi_{1\alpha}$, $\Phi_{2\alpha}$ and $\Psi_\alpha$ under $P$ is given by the equations
\begin{eqnarray}
\label{eq.p1}
P^{-1}\Phi_{1\alpha}(x,t)P & = & - {\mc{P}_\alpha}^\beta\Phi_{2\beta}(-x,t)\\
\label{eq.p2}
P^{-1}\Phi_{2\alpha}(x,t)P & = & {\mc{P}_\alpha}^\beta\Phi_{1\beta}(-x,t)\\ 
\label{eq.p3}
P^{-1}\Psi_\alpha(x,t)P & = & \ii {\mc{P}_\alpha}^\beta\Psi_\beta(-x,t)\ ,
\end{eqnarray}
where $\mc{P}=-\ii  \gamma^{00\dots 0}$ (the $\gamma^{00\dots 0}$ appearing here is defined in \ref{app.a4}).
$\mc{P}$ has the property $\mc{P}^2= -I$.
$a_{\alpha}^\dagger(k)$ and  $b_{\alpha}^\dagger(k)$ 
transform under $P$ as 
\beq
P^{-1}a_{\alpha}^\dagger (k)P=\ii {\mc{P}_\alpha}^\beta a_{\beta}^\dagger(-k)\ ,\qquad 
P^{-1}b_{\alpha}^\dagger (k)P=-\ii {\mc{P}_\alpha}^\beta b_{\beta}^\dagger (-k)\ ,
\eeq
and in the case of $D^{(n)}$ we have
\begin{eqnarray}
P^{-1}c_i^\dagger (k) P & = &  - c_i^\dagger (-k)\\
P^{-1}d_i^\dagger (k) P & = &   d_i^\dagger (-k)\\
P^{-1}f_i^\dagger (k) P & = &   f_i^\dagger (-k)\\
P^{-1}h_i^\dagger (k) P & = &  - h_i^\dagger (-k)\ .
\end{eqnarray}
Completely analogous equations apply also in the case of $D^{(n,m)}$.
Equations (\ref{eq.p1}) and (\ref{eq.p2}) show that space reflections link 
the real fields  
$\Phi_1$ and $\Phi_2$ together. 
However, if  $P^2=-I$ is also allowed on states containing an odd number of particles, 
then one can define space reflections 
on $\Phi_1$ and $\Phi_2$ separately:
\begin{eqnarray}
\label{eq.p4}
P^{-1}\Phi_{1\alpha}(x,t)P & = &  {\mc{P}_\alpha}^\beta\Phi_{1\beta}(-x,t)\\
\label{eq.p5}
P^{-1}\Phi_{2\alpha}(x,t)P & = & {\mc{P}_\alpha}^\beta\Phi_{2\beta}(-x,t)\\ 
\label{eq.p6}
P^{-1}\Psi_\alpha(x,t)P & = &  {\mc{P}_\alpha}^\beta\Psi_\beta(-x,t)\ ,
\end{eqnarray}
where $\mc{P}=-\ii \gamma^{00\dots 0}$, as above. 
$a_{\alpha}^\dagger(k)$ and  $b_{\alpha}^\dagger(k)$ 
transform under $P$ in this case as 
\beq
P^{-1}a_{\alpha}^\dagger (k)P= {\mc{P}_\alpha}^\beta a_{\beta}^\dagger(-k)\ ,\qquad 
P^{-1}b_{\alpha}^\dagger (k)P= {\mc{P}_\alpha}^\beta b_{\beta}^\dagger (-k)\ .
\eeq
In the case of $D^{(n)}$ we have
\begin{eqnarray}
P^{-1}c_i^\dagger (k) P & = & \ii  c_i^\dagger (-k)\\
P^{-1}d_i^\dagger (k) P & = & \ii  d_i^\dagger (-k)\\
P^{-1}f_i^\dagger (k) P & = & -\ii f_i^\dagger (-k)\\
P^{-1}h_i^\dagger (k) P & = & -\ii h_i^\dagger (-k)\ ,
\end{eqnarray}
and completely analogous equations apply also in the case of $D^{(n,m)}$.
On the space of those states that contain an odd number of particles, $P^2=-I$ holds instead of $P^2=I$.
Real fermion fields with this kind of transformation law under space reflections are called Majorana fermion fields 
(see \cite{WeinbergQFT}).

If $\Psi_\alpha$ is bosonic, then the transformation of the fields 
$\Phi_{1\alpha}$, $\Phi_{2\alpha}$ and $\Psi_\alpha$ under $P$ is given by the equations
\begin{eqnarray}
P^{-1}\Phi_{1\alpha}(x,t)P & = &  {\mc{P}_\alpha}^\beta\Phi_{1\beta}(-x,t)\\
P^{-1}\Phi_{2\alpha}(x,t)P & = & {\mc{P}_\alpha}^\beta\Phi_{2\beta}(-x,t)\\ 
\label{eq.p3even}
P^{-1}\Psi_\alpha(x,t)P & = &  {\mc{P}_\alpha}^\beta\Psi_\beta(-x,t)\ ,
\end{eqnarray}
where $\mc{P}= \gamma^{00\dots 0}$ (in the case of $\tilde{D}^{(n)}$, $\mc{P}=  \tau^{00\dots 0}$;  
see Appendix \ref{app.a4} for the definition of 
the $\tau$ tensors).
$\mc{P}$ has the property $\mc{P}^2=I$.

$a_{\alpha}^\dagger(k)$ and  $b_{\alpha}^\dagger(k)$ 
transform under $P$ in this case as 
\beq
P^{-1}a_{\alpha}^\dagger (k)P={\mc{P}_\alpha}^\beta a_{\beta}^\dagger(-k)\ ,\qquad 
P^{-1}b_{\alpha}^\dagger (k)P= {\mc{P}_\alpha}^\beta b_{\beta}^\dagger (-k)\ .
\eeq
In the case of $D^{(n)}$ we have
\begin{eqnarray}
P^{-1}c_i^\dagger (k) P & = & c_i^\dagger (-k)\\
P^{-1}d_i^\dagger (k) P & = & d_i^\dagger (-k)\\
P^{-1}f_i^\dagger (k) P & = & -f_i^\dagger (-k)\\
P^{-1}h_i^\dagger (k) P & = & -h_i^\dagger (-k)\ ,
\end{eqnarray}
and completely analogous equations apply also in the cases of $D^{(n,m)}$ and $\tilde{D}^{(n)}$.

In all cases,
$\epsilon^{\alpha\beta}$ has the invariance property  
\beq
\epsilon^{\alpha\beta}{\mc{P}_\alpha}^\rho {\mc{P}_\beta}^\delta =\epsilon^{\rho\delta}
\eeq
and $\mc{P}$ is real.
$\mc{P}$ has the following commutation relations with the generators of $sl(2,\CC)$ in $D^{(n)}$: 
\begin{eqnarray}
\mc{P}(M_i)_{D^{(n)}}\mc{P}^{-1} & = & (M_i)_{D^{(n)}}\\
\mc{P}(N_i)_{D^{(n)}}\mc{P}^{-1} & = & -(N_i)_{D^{(n)}}\ .
\end{eqnarray}
The same relations apply also to the cases of $D^{(n,m)}$ and $\tilde{D}^{(n)}$.

\noindent
{\it Time reversal}\\
The time reversal transformation $(t,x)\to (-t,x)$ 
is represented on the Hilbert space by an antiunitary operator $T$ that has the property
\beq
T\ket{0}=\ket{0}
\eeq 
and
\beq
THT^{-1}=H\ ,\qquad TP_iT^{-1}=-P_i\ ,
\eeq
\beq
T\check{M}_iT^{-1}=\check{M}_i\ ,\qquad
T\check{N}_iT^{-1}=-\check{N}_i\ ,\qquad i=1,2,3\ .
\eeq
The transformation of the field $\Psi$ under $T$ is given by the equations
\begin{eqnarray}
T^{-1}\Phi_{1\alpha}(x,t) T & = & {\mc{T}_\alpha}^\beta \Phi_{1\beta}(x,-t)\\
T^{-1}\Phi_{2\alpha}(x,t) T & = & -{\mc{T}_\alpha}^\beta \Phi_{2\beta}(x,-t)\\
T^{-1}\Psi_\alpha(x,t) T & = & {\mc{T}_\alpha}^\beta \Psi_\beta(x,-t)\ ,
\end{eqnarray}
where $\mc{T}=\ii \gamma^{00\dots 0}\gamma^5$ if $\Psi$ is fermionic, and 
$\mc{T}=\gamma^{00\dots 0}$ if $\Psi$ is bosonic and transforms according to a $D^{(n)}$ or $D^{(n,m)}$ type representation. 
In the case of $\tilde{D}^{(n)}$, $\mc{T}=\tau^{00\dots 0}$. 
In the cases of $D^{(n)}$, $D^{(n,m)}$ and $\tilde{D}^{(n)}$, $\mc{T}$ has the property  
$\mc{T}^2 = (-1)^n I$, $\mc{T}^2 = (-1)^{n+m} I$ or $\mc{T}^2 = I$, respectively, and it is also real.

The action of $T$ on the creation operators is given by the equations
\beq
T^{-1}a_\alpha^\dagger(k)T={\mc{T}_\alpha}^\beta a_\beta^\dagger(-k)\ ,\qquad
T^{-1}b_\alpha^\dagger(k)T={\mc{T}_\alpha}^{\beta } b_\beta^\dagger(-k)\ . 
\eeq
In the case of $D^{(n)}$,
\begin{eqnarray}
T^{-1}c_i^\dagger (k) T & = & (-1)^{i+1}c_{n+2-i}^\dagger (-k)\\
T^{-1}d_i^\dagger (k) T & = & (-1)^{i+1}d_{n+2-i}^\dagger (-k)\\
T^{-1}f_i^\dagger (k) T & = & (-1)^{i+1}f_{n+2-i}^\dagger (-k)\\
T^{-1}h_i^\dagger (k) T & = & (-1)^{i+1}h_{n+2-i}^\dagger (-k)\ .
\end{eqnarray}
In the case of $D^{(n,m)}$,
\begin{eqnarray}
T^{-1}c_{ij}^\dagger (k) T & = & (-1)^{m+i+j}c_{n+2-i,m+2-j}^\dagger (-k)\\
T^{-1}d_{ij}^\dagger (k) T & = & (-1)^{m+i+j}d_{n+2-i,m+2-j}^\dagger (-k)\\
T^{-1}f_{ij}^\dagger (k) T & = & (-1)^{m+i+j}f_{n+2-i,m+2-j}^\dagger (-k)\\
T^{-1}h_{ij}^\dagger (k) T & = & (-1)^{m+i+j}h_{n+2-i,m+2-j}^\dagger (-k)\ .
\end{eqnarray}
In the case of $\tilde{D}^{(n)}$,
\begin{eqnarray}
T^{-1}c_{ij}^\dagger (k) T & = & (-1)^{n+i+j}c_{n+2-j,n+2-i}^\dagger (-k)\\
T^{-1}d_{ij}^\dagger (k) T & = & (-1)^{n+i+j}d_{n+2-j,n+2-i}^\dagger (-k)\\
T^{-1}f_{ij}^\dagger (k) T & = & (-1)^{n+i+j+1}f_{n+2-j,n+2-i}^\dagger (-k)\\
T^{-1}h_{ij}^\dagger (k) T & = & (-1)^{n+i+j+1}h_{n+2-j,n+2-i}^\dagger (-k)\ .
\end{eqnarray}

$\epsilon^{\alpha\beta}$ has the invariance property  
$
\epsilon^{\alpha\beta}{\mc{T}_\alpha}^\rho {\mc{T}_\beta}^\delta =- \epsilon^{\rho\delta}\ 
$
if $\Psi$ is fermionic and 
$
\epsilon^{\alpha\beta}{\mc{T}_\alpha}^\rho {\mc{T}_\beta}^\delta = \epsilon^{\rho\delta}\ 
$
if $\Psi$ is bosonic.

$\mc{T}$ has the same commutation relations with the generators of $sl(2,\CC)$ in $D^{(n)}$ as $\mc{P}$: 
\begin{eqnarray}
\mc{T}(M_i)_{D^{(n)}}\mc{T}^{-1} & = & (M_i)_{D^{(n)}}\\
\mc{T}(N_i)_{D^{(n)}}\mc{T}^{-1} & = & -(N_i)_{D^{(n)}}\ .
\end{eqnarray}
The same relations apply also to the cases of $D^{(n,m)}$ and $\tilde{D}^{(n)}$. 

\noindent
{\it CPT transformation}\\
Under $CPT$, the fields $\Phi_1$, $\Phi_2$, $\Psi$ have the following transformation properties:
if $\Psi$ is fermionic and the parity transformations are defined by (\ref{eq.p1}) - (\ref{eq.p3}), then   
\begin{eqnarray}
(CPT)^{-1}\Phi_1(x,t) (CPT) & = & \gamma^5 \Phi_2(-x,-t)\\
(CPT)^{-1}\Phi_2(x,t) (CPT) & = & -\gamma^5 \Phi_1(-x,-t)\\
(CPT)^{-1}\Psi(x,t) (CPT) & = & \ii\gamma^5 \Psi^\dagger(-x,-t)\ .
\end{eqnarray}
If $\Psi$ is fermionic, and the space reflections are Majorana type, 
then
\begin{eqnarray}
(CPT)^{-1}\Phi_1(x,t) (CPT) & = & \gamma^5 \Phi_1(-x,-t)\\
(CPT)^{-1}\Phi_2(x,t) (CPT) & = & \gamma^5 \Phi_2(-x,-t)\\
(CPT)^{-1}\Psi(x,t) (CPT) & = & \gamma^5 \Psi^\dagger(-x,-t)\ .
\end{eqnarray}
If $\Psi$ is bosonic, then 
\begin{eqnarray}
(CPT)^{-1}\Phi_1(x,t) (CPT) & = &  \Phi_1(-x,-t)\\
(CPT)^{-1}\Phi_2(x,t) (CPT) & = &  \Phi_2(-x,-t)\\
(CPT)^{-1}\Psi(x,t) (CPT) & = &  \Psi^\dagger(-x,-t)\ .
\end{eqnarray}

\section{Fields of arbitrary spin}
\label{sec.hs}

In Section \ref{sec.lt} it was seen that the field $\Psi$ can create 
nonphysical states that have negative scalar product with themselves.
Moreover, in the cases of the representations $D^{(n,m)}$ and $\tilde{D}^{(n)}$
$\Psi$ can create particles for several spin values. 
The final step in the construction of higher spin fields 
is the elimination of the nonphysical degrees of freedom, and also the 
elimination of those degrees of freedom that have a spin 
different from the desired value.
This is achieved by applying a suitable differential operator to $\Psi$, so that the physical
higher spin field will be $\psi=\mc{D}\Psi$.

\subsection{Fields transforming according to $D^{(n)}$}
\label{sec.dn}

We begin with the case when $\Psi$ transforms according to the representations $D^{(n)}$. In this case the 
differential operator is
$\mc{D}^+ = \frac{1}{2\mass^n} (\mass^n+\ii^n \gamma^{{\mu_1}{\mu_2}\dots{\mu_n}}\partial_{{\mu_1}{\mu_2}\dots{\mu_n}})$,
i.e.\ 
we define the field $\psi$ as
\beq
\label{eq.psidef}
\psi_\alpha={(\mc{D}^+)_\alpha}^\beta \Psi_\beta = \frac{1}{2\mass^n}{(\mass^n+\ii^n \gamma^{{\mu_1}{\mu_2}\dots{\mu_n}}\partial_{{\mu_1}{\mu_2}\dots{\mu_n}})_\alpha}^\beta\Psi_\beta\ .
\eeq
The generalized gamma tensor ${(\gamma^{{\mu_1}{\mu_2}\dots{\mu_n}})_\alpha}^\beta$ 
appearing here is defined in \ref{app.a4}. 
In the following we discuss the properties of $\psi$, which will justify the choice of $\mc{D}$.

Since $\Psi$ satisfies the Klein--Gordon equation, $\psi$ also satisfies it:
\beq
\label{eq.psikg}
(\mass^2+\partial_\mu\partial^\mu)\psi_\alpha=0\ .
\eeq
The differential operator 
$(\mass^n+\ii^n \gamma^{{\mu_1}{\mu_2}\dots{\mu_n}}\partial_{{\mu_1}{\mu_2}\dots{\mu_n}})$ satisfies the relation
\beq
\label{eq.prop}
{(\mass^n-\ii^n\gamma^{{\mu_1}{\mu_2}\dots{\mu_n}}\partial_{{\mu_1}{\mu_2}\dots{\mu_n}})_\alpha}^\beta
{(\mass^n+\ii^n\gamma^{{\rho_1}{\rho_2}\dots{\rho_n}}\partial_{{\rho_1}{\rho_2}\dots{\rho_n}})_\beta}^\xi
=[\mass^{2n}+(-1)^{n+1}(\partial_\mu\partial^\mu)^n] {\delta_\alpha}^\xi\ , 
\eeq
therefore equation (\ref{eq.psidef}) together with the Klein--Gordon equation for $\Psi$ imply
\beq
\label{eq.psidirac}
{(\mass^n-\ii^n \gamma^{{\mu_1}{\mu_2}\dots{\mu_n}}\partial_{{\mu_1}{\mu_2}\dots{\mu_n}})_\alpha}^\beta\psi_\beta=0\ ,
\eeq
which was found also in \cite{Weinberg} and is known as the Weinberg (or Joos--Weinberg) equation. 
For $n=1$ (\ref{eq.psidirac}) is the usual Dirac equation.

It is worth noting that (\ref{eq.psidirac}) implies, by virtue of (\ref{eq.prop}),   
\beq
\label{eq.133}
[\mass^{2n} +(-1)^{n+1} (\partial_\mu\partial^\mu)^n]\psi_\alpha=0\ .
\eeq
In the case of $n=1$
this is identical with the Klein--Gordon equation. 
For $n>1$, (\ref{eq.psikg}) implies (\ref{eq.133}), 
but, obviously, the reverse is not true. 
It is also known that (\ref{eq.psidirac}) in itself does not imply 
(\ref{eq.psikg}) and has nonphysical solutions if $n>1$ \cite{SSS,AE}.

It is also important to note that on the space of the solutions of the Klein--Gordon equation
the differential operators\\ 
$\mc{D}^+ = \frac{1}{2\mass^n} (\mass^n+\ii^n \gamma^{{\mu_1}{\mu_2}\dots{\mu_n}}\partial_{{\mu_1}{\mu_2}\dots{\mu_n}})$
and 
$\mc{D}^- = \frac{1}{2\mass^n} (\mass^n-\ii^n \gamma^{{\mu_1}{\mu_2}\dots{\mu_n}}\partial_{{\mu_1}{\mu_2}\dots{\mu_n}})$\\ 
are projection operators, i.e.\ $\mc{D}^+\mc{D}^+ = \mc{D}^+$, $\mc{D}^-\mc{D}^- = \mc{D}^-$, and 
the relations $\mc{D}^+\mc{D}^- = \mc{D}^-\mc{D}^+ = 0$ and $\mc{D}^+ +  \mc{D}^- = I$ also hold.

Due to the reality properties of $\gamma^{{\mu_1}{\mu_2}\dots{\mu_n}}$, $\psi_\alpha^\dagger$ is given by 
\beq
\psi_\alpha^\dagger=\frac{1}{2\mass^n}{(\mass^n+\ii^n \gamma^{{\mu_1}{\mu_2}\dots{\mu_n}}\partial_{{\mu_1}{\mu_2}\dots{\mu_n}})_\alpha}^\beta\Psi_\beta^\dagger\ ,
\eeq
and it satisfies the same generalized Dirac equation 
\beq
{(\mass^n-\ii^n \gamma^{{\mu_1}{\mu_2}\dots{\mu_n}}\partial_{{\mu_1}{\mu_2}\dots{\mu_n}})_\alpha}^\beta\psi_\beta^\dagger=0
\eeq
as $\psi_\alpha$.

The real and imaginary parts $\phi_1$ and $\phi_2$  of $\psi$ can also be introduced as 
\beq
\phi_1=\frac{\psi+\psi^\dagger}{\sqrt{2}}\ , \qquad \phi_2=\frac{\psi-\psi^\dagger}{\ii\sqrt{2}}\ .
\eeq
$\phi_1$ and $\phi_2$ are related to 
$\Phi_1$ and $\Phi_2$, respectively, by equations of the same form as (\ref{eq.psidef}). 
Obviously, they satisfy the Klein--Gordon equation, and they also satisfy the generalized Dirac equation 
(\ref{eq.psidirac}).

We define the field $\psi_\alpha^-$ as 
\beq
\label{eq.psimindef}
\psi_\alpha^-=\frac{1}{2\mass^n}{(\mass^n-\ii^n \gamma^{{\mu_1}{\mu_2}\dots{\mu_n}}\partial_{{\mu_1}{\mu_2}\dots{\mu_n}})_\alpha}^\beta\Psi_\beta\ .
\eeq
It will be seen below that $\psi_\alpha^-$ can be called the nonphysical part of $\Psi$. 
$\psi^-$ obviously satisfies the Klein--Gordon equation and the generalized Dirac equation with 
reverse sign,
\beq
{(\mass^n+\ii^n \gamma^{{\mu_1}{\mu_2}\dots{\mu_n}}\partial_{{\mu_1}{\mu_2}\dots{\mu_n}})_\alpha}^\beta\psi_\beta^-=0\ .
\eeq
We also have $\Psi=\psi+\psi^-$.

The property of ${(\gamma^{{\mu_1}{\mu_2}\dots{\mu_n}})_\alpha}^\beta$ that it is an invariant tensor under $SL(2,\CC)$
transformations implies that the $SL(2,\CC)$ transformation law 
for $\psi$, $\phi_1$, $\phi_2$, $\psi^-$ has the same form as the $SL(2,\CC)$ transformation law 
(\ref{eq.lor}) for
$\Psi$. It is not difficult to verify that $\psi$ and $\psi^-$ also have the same $C$, $P$, $T$ transformation properties as $\Psi$, and $\phi_1$ and $\phi_2$ have the same $C$, $P$, $T$ transformation properties
as $\Phi_1$ and $\Phi_2$, respectively.

In terms of mode creation and annihilation operators 
$\psi$ can be expressed as 
\beq
\psi_\alpha (x,t)=\int \frac{\intd^3 k}{\sqrt{2}(\sqrt{2\pi})^3 \omega(k)} [e^{\ii kx}e^{\ii\omega(k)t}
      \tilde{a}_\alpha^\dagger(k)
+e^{-\ii kx}e^{-\ii\omega(k)t}
      \tilde{b}_\alpha (k)]\ ,
\eeq
where
\begin{eqnarray}
\tilde{a}_\alpha^\dagger(k) & = & \frac{1}{2\mass^n}{(\mass^n+(-1)^n k_{\mu_1} k_{\mu_2} \dots k_{\mu_n}  \gamma^{{\mu_1}{\mu_2}\dots{\mu_n}} )_\alpha}^\beta a_\beta^\dagger(k)\\
\tilde{b}_\alpha(k) & = & \frac{1}{2\mass^n}{(\mass^n+ k_{\mu_1} k_{\mu_2} \dots k_{\mu_n} \gamma^{{\mu_1}{\mu_2}\dots{\mu_n}} )_\alpha}^\beta b_\beta (k)\ ,
\end{eqnarray}
and 
$k_\mu$ is the vector $(\omega(k),k)$.
Taking into consideration the identities (\ref{eq.pp1})-(\ref{eq.pp4}), 
the mode expansion of $\psi$ can also be written as
\begin{eqnarray}
&& \psi_\alpha (x,t)=\int \frac{\intd^3 k}{\sqrt{2}(\sqrt{2\pi})^3 \omega(k)}  \sum_{i=1}^{n+1}
[e^{\ii kx}e^{\ii \omega(k)t}
        u_{i\alpha}(k) c_i^\dagger(k) \nonumber \\
&& \hspace{6cm} +e^{-\ii kx}e^{-\ii\omega(k)t}
      u_{i\alpha}(k)^* d_i(k)  ]\ .
\label{eq.psimodex}
\end{eqnarray}
This mode expansion shows that $\psi$ does not have nonphysical degrees of freedom, and that 
$\psi$ and $\psi^\dagger$ create two kinds of particle of mass $\mass$ and spin $n/2$, 
which form a particle-antiparticle pair. 
The mode creation operators for these particles are $c_i^\dagger(k)$ and $d_i^\dagger(k)$, respectively.

For $\phi_1$ and $\phi_2$ we have 
\begin{eqnarray}
&& \phi_{1\alpha} (x,t)=\int \frac{\intd^3 k}{\sqrt{2}(\sqrt{2\pi})^3 \omega(k)}
\sum_{i=1}^{n+1}
 [e^{\ii kx}e^{\ii \omega(k)t}
        u_{i\alpha}(k) \frac{c_i^\dagger(k)+d_i^\dagger(k)}{\sqrt{2}} 
     \nonumber \\
&& \hspace{6cm} +e^{-\ii kx}e^{-\ii\omega(k)t}
      u_{i\alpha}(k)^* \frac{c_i(k)+d_i(k)}{\sqrt{2}} ] 
\end{eqnarray}
and
\begin{eqnarray}
&& \phi_{2\alpha} (x,t)=\int \frac{\intd^3 k}{\sqrt{2}(\sqrt{2\pi})^3 \omega(k)} \sum_{i=1}^{n+1}
 [e^{\ii kx}e^{\ii \omega(k)t}
        u_{i\alpha}(k) \frac{c_i^\dagger(k)-d_i^\dagger(k)}{\ii\sqrt{2}} 
    \nonumber \\
&& \hspace{6cm} +e^{-\ii kx}e^{-\ii\omega(k)t}
    u_{i\alpha}(k)^* \frac{d_i(k)-c_i(k)}{\ii\sqrt{2}}  ]\ .
\end{eqnarray}
This shows that the real fields $\phi_1$ and $\phi_2$ create 
a single type of self-conjugate or anti-self-conjugate particle of mass $\mass$ and spin $n/2$. 
The mode creation operators are 
$\frac{c_i^\dagger(k)+d_i^\dagger(k)}{\sqrt{2}}$ for the self-conjugate particle
and 
$\frac{c_i^\dagger(k)-d_i^\dagger(k)}{\ii\sqrt{2}}$ for the anti-self-conjugate particle.
In the fermionic case,
if the action of parity transformations is defined by (\ref{eq.p1}) - (\ref{eq.p3}), 
then parity transformations interchange the particles created by  $\phi_1$ and $\phi_2$, whereas with the Majorana type
action of  parity transformations (\ref{eq.p4}) - (\ref{eq.p6}) the particles created by 
$\phi_1$ and $\phi_2$ are not mixed.

The mode expansion of $\psi^-$ is 
\begin{eqnarray}
&& \psi^-_\alpha (x,t)=\int \frac{\intd^3 k}{\sqrt{2}(\sqrt{2\pi})^3 \omega(k)}  \sum_{i=1}^{n+1}
[e^{\ii kx}e^{\ii \omega(k)t}
        v_{i\alpha}(k) f_i^\dagger(k) \nonumber \\
&& \hspace{6cm} +e^{-\ii kx}e^{-\ii\omega(k)t}
      v_{i\alpha}(k)^* h_i(k)  ]\ .
\end{eqnarray}
This shows that $\psi^-$ is the nonphysical part of $\Psi$, since $f_i^\dagger(k)$ and 
$h_i^\dagger(k)$ create nonphysical particles.

$(I-\ii\gamma^5)/2$ and $(I+\ii\gamma^5)/2$ are projectors on left and right handed spinors, respectively.
(The matrix $\gamma^5$ appearing here is defined in \ref{app.a4}.) 
Thus, chiral fields transforming according to the representations $(n/2,0)$ or $(0,n/2)$ can be constructed from $\Psi$ as 
\beq
\Psi_L=\frac{I-\ii\gamma^5}{2}\Psi\ ,\qquad  \Psi_R=\frac{I+\ii\gamma^5}{2}\Psi\ ,
\eeq
and in the same way also from the other fields $\psi$, $\phi_1$, $\phi_2$, $\psi^-$.

From the (anti)commutation properties of $\Psi$, from the definition (\ref{eq.psidef}) and (\ref{eq.psimindef})
of $\psi$ and $\psi^-$, and from the 
properties of $\gamma^{{\mu_1}{\mu_2}\dots{\mu_n}}$ it follows that
\beq
[\psi_\alpha(x,t_x),\psi^{-}_\beta(y,t_y)]_{\pm}=
[\psi_\alpha^\dagger(x,t_x),\psi^{-\dagger}_\beta(y,t_y)]_{\pm}=0\ ,
\eeq
\beq
[\psi_\alpha^\dagger(x,t_x),\psi^{-}_\beta(y,t_y)]_{\pm}=
[\psi_\alpha(x,t_x),\psi^{-\dagger}_\beta(y,t_y)]_{\pm}=0\ ,
\eeq
\beq
[\psi_\alpha(x,t_x),\psi_\beta(y,t_y)]_\pm=[\Psi_\alpha(x,t_x),\psi_\beta(y,t_y)]_\pm
=[\psi_\alpha(x,t_x),\Psi_\beta(y,t_y)]_\pm=0\ ,
\eeq
\beq
[\psi_\alpha^\dagger(x,t_x),\psi_\beta^\dagger(y,t_y)]_\pm=[\Psi_\alpha^\dagger(x,t_x),\psi_\beta^\dagger(y,t_y)]_\pm
=[\psi_\alpha^\dagger(x,t_x),\Psi_\beta^\dagger(y,t_y)]_\pm=0\ ,
\eeq
\beq
[\psi_\alpha(x,t_x),\psi_\beta^\dagger(y,t_y)]_\pm=[\Psi_\alpha(x,t_x),\psi_\beta^\dagger(y,t_y)]_\pm
=[\psi_\alpha(x,t_x),\Psi_\beta^\dagger(y,t_y)]_\pm\ ,
\eeq
\beq
[\psi_\alpha^\dagger(x,t_x),\psi_\beta(y,t_y)]_\pm=[\Psi_\alpha^\dagger(x,t_x),\psi_\beta(y,t_y)]_\pm
=[\psi_\alpha^\dagger(x,t_x),\Psi_\beta(y,t_y)]_\pm\ ,
\label{eq.174}
\eeq
\beq
[\psi_\alpha(x,t_x),\psi_\beta^\dagger(y,t_y)]_\pm=
[\psi_\alpha^\dagger(x,t_x),\psi_\beta(y,t_y)]_\pm\ ,
\eeq
and
\begin{eqnarray}
[\psi_\alpha(x,t_x),\psi_\beta^\dagger(y,t_y)]_\pm & = &
[{(\mc{D}^+)_\alpha}^\delta \Psi_\delta (x,t_x),\Psi_\beta^\dagger(y,t_y)]_{\pm} \nonumber  \\
&&\hspace{-4.2cm} 
=\ \frac{1}{2\mass^n} {(\mass^n +\ii^n \gamma^{{\mu_1}{\mu_2}\dots{\mu_n}}\partial_{\mu_1}\partial_{\mu_2}\dots\partial_{\mu_n})_\alpha}^\delta
\epsilon_{\delta\beta} [G(x-y,t_x-t_y) - G(y-x,t_y-t_x) ]\ . \nonumber \\
\label{eq.153}
\end{eqnarray}

The equal-time (anti)commutator 
$[\psi_\alpha(x,t),\psi_\beta^\dagger(y,t)]_\pm$ can be calculated in the following way: 
in the defining expression (\ref{eq.psidef}) for $\psi$ one eliminates the time derivatives by applying 
\beq
\label{eq.elim}
\partial_t\Psi=\Pi\ ,\qquad
\partial_{t}\Pi=\partial_x\partial_x\Psi-\mass^2\Psi\ ,
\eeq
which gives an expression for $\psi$ in terms of $\Psi$, $\Pi$, and their spatial derivatives. 
Using this expression and the (anti)commutation relations (\ref{eq.ac5ps}) and (\ref{eq.ac6ps}), 
the equal-time (anti)commutator 
$[\psi_\alpha(x,t),\psi_\beta^\dagger(y,t)]_\pm$ can be evaluated.
Alternatively, the formula (\ref{eq.153}) can be used directly.
(Anti)commutators involving the derivatives of $\psi$ and $\psi^\dagger$ can be calculated in the same way.
The case of spin $3/2$ is discussed in some detail in Appendix \ref{app.b} as example.

In a similar way as for the (anti)commutators, for the Green function we obtain 
\beq
\brakettt{0}{\psi_\alpha(x,t_x)\psi_\beta(y,t_y)}{0}
=
\brakettt{0}{\Psi_\alpha(x,t_x)\psi_\beta(y,t_y)}{0}
=
\brakettt{0}{\psi_\alpha(x,t_x)\Psi_\beta(y,t_y)}{0}
=0
\eeq
\beq
\brakettt{0}{\psi_\alpha^\dagger(x,t_x)\psi_\beta^\dagger(y,t_y)}{0}
=
\brakettt{0}{\Psi_\alpha^\dagger(x,t_x)\psi_\beta^\dagger(y,t_y)}{0}
=
\brakettt{0}{\psi_\alpha^\dagger(x,t_x)\Psi_\beta^\dagger(y,t_y)}{0}
=0
\eeq
\beq
\brakettt{0}{\psi_\alpha^\dagger(x,t_x)\psi_\beta(y,t_y)}{0}
=
\brakettt{0}{\Psi_\alpha^\dagger(x,t_x)\psi_\beta(y,t_y)}{0}
=
\brakettt{0}{\psi_\alpha^\dagger(x,t_x)\Psi_\beta(y,t_y)}{0}
\eeq
\beq
\brakettt{0}{\psi_\alpha(x,t_x)\psi_\beta^\dagger(y,t_y)}{0}
=
\brakettt{0}{\Psi_\alpha(x,t_x)\psi_\beta^\dagger(y,t_y)}{0}
=
\brakettt{0}{\psi_\alpha(x,t_x)\Psi_\beta^\dagger(y,t_y)}{0}
\eeq
\beq
\brakettt{0}{\psi_\alpha(x,t_x)\psi_\beta^\dagger(y,t_y)}{0}
=\brakettt{0}{\psi_\alpha^\dagger(x,t_x)\psi_\beta(y,t_y)}{0}
\eeq
and 
\begin{eqnarray}
\brakettt{0}{\psi_\alpha(x,t_x)\psi_\beta^\dagger(y,t_y)}{0}
& = & \brakettt{0}{{(\mc{D}^+)_\alpha}^\delta\Psi_\delta(x,t_x)\Psi_\beta^\dagger(y,t_y)}{0} \nonumber \\
& = & \frac{1}{2\mass^n} {(\mass^n +\ii^n \gamma^{\mu_1\mu_2\dots\mu_n}\partial_{\mu_1}\partial_{\mu_2}\dots\partial_{\mu_n})_\alpha}^\delta
\epsilon_{\delta\beta} G(x-y,t_x-t_y)\ . \nonumber \\
\end{eqnarray}

The covariant part $D_F(x-y,t_x-t_y)_{\alpha\beta}^C$ of the Feynman propagator 
\begin{eqnarray}
&& D_F(x-y,t_x-t_y)_{\alpha\beta}=\brakettt{0}{\mathrm{T}\psi_\alpha(x,t_x)\psi_\beta^\dagger(y,t_y)}{0} \nonumber \\
&&\hspace{2cm} = \brakettt{0}{\mathrm{T}\psi_\alpha^\dagger(x,t_x)\psi_\beta(y,t_y)}{0}
= \brakettt{0}{\mathrm{T}\psi_\alpha^\dagger(x,t_x)\Psi_\beta(y,t_y)}{0}\nonumber \\
&&\hspace{2cm} = \brakettt{0}{\mathrm{T}{(\mc{D}^+)_\alpha}^\delta\Psi_\delta^\dagger(x,t_x)\Psi_\beta(y,t_y)}{0}  
\end{eqnarray}
for $\psi_\alpha$ is 
\beq
\frac{1}{2\mass^n} {(\mass^n +\ii^n \gamma^{\mu_1\mu_2\dots\mu_n}\partial_{\mu_1}\partial_{\mu_2}\dots\partial_{\mu_n})_\alpha}^\delta
\epsilon_{\delta\beta} D_F(x-y,t_x-t_y)\ ,
\eeq 
which can also be written as 
\beq
\label{eq.cfeyn}
D_F(x-y,t_x-t_y)_{\alpha\beta}^C=
\frac{1}{2\mass^n}
\int \frac{\intd^3 k\, \intd k_0}{(2\pi)^4}\, \frac{\ii
{(\mass^n+\gamma^{{\mu_1}{\mu_2}\dots{\mu_n}}k_{\mu_1} k_{\mu_2}\dots k_{\mu_n})_\alpha}^\delta \epsilon_{\delta\beta}
 }{k_0^2-k^2-\mass^2+\ii\epsilon}e^{-\ii k(x-y)}e^{-\ii k_0(t_x-t_y)}\ .
\eeq
This is obtained from the complete (raw) Feynman propagator 
$\brakettt{0}{\mathrm{T}{(\mc{D}^+)_\alpha}^\delta\Psi_\delta^\dagger(x,t_x)\Psi_\beta(y,t_y)}{0}$ by  
pulling the differential operator $\mc{D}^+$ out of the time ordered product and neglecting the contact terms that arise.   
In principle there is no difficulty in calculating these contact terms, but we do not discuss this
in detail here. 
(\ref{eq.cfeyn}) shows that for large $|k|$ the Fourier transform of the propagator behaves as $|k|^{n-2}$.  
Interactions involving the fields $\psi$ with high spins can therefore be expected to 
have strong ultraviolet divergences.

The (anti)commutators, Green functions and Feynman propagators of the chiral fields 
$\frac{I\mp\ii\gamma^5}{2}\psi$, $\frac{I\mp\ii\gamma^5}{2}\phi_1$, $\frac{I\mp\ii\gamma^5}{2}\phi_2$ can be obtained in a straightforward way from those of $\psi$, $\phi_1$, $\phi_2$.

\subsubsection{Electromagnetic current}

We have seen that $\psi$ satisfies the field equation (\ref{eq.psidirac}), 
which is a generalization of the Dirac equation. 
This equation can be derived from a Lagrangian, which is invariant under the 
global $U(1)$  
transformation $\psi\to e^{\ii\alpha} \psi$, $\alpha\in\RR$, thus
Noether's theorem can be applied to find a conserved current corresponding to this symmetry.

The Lagrangian is
\beq
L= \int \intd^3 x\  \psi_\alpha^\dagger {(\mass^n-\ii^n\gamma^{{\mu_1}{\mu_2}\dots{\mu_n}}\partial_{{\mu_1}{\mu_2}\dots{\mu_n}})_\beta}^\rho \psi_\rho \epsilon^{\alpha\beta}\ , 
\eeq
and the corresponding conserved current that Noether's theorem gives  is (up to normalization)
\begin{eqnarray} 
j^\mu = \frac{-2(-\ii)^{n-1}}{n \mass^{n-2} } (\psi_\alpha^\dagger \partial_{{\nu_1}{\nu_2}\dots{\nu_{n-1}}} \psi_\gamma
-\partial_{\nu_1}\psi_\alpha^\dagger  \partial_{{\nu_2}\dots{\nu_{n-1}}} \psi_\gamma
+\partial_{{\nu_1}{\nu_2}}\psi_\alpha^\dagger  \partial_{{\nu_3}\dots{\nu_{n-1}}} \psi_\gamma \nonumber \\
-\dots 
+(-1)^{n+1} \partial_{{\nu_1}{\nu_2}\dots{\nu_{n-1}}} \psi_\alpha^\dagger\, \psi_\gamma){(\gamma^{\mu{\nu_1}{\nu_2}\dots{\nu_{n-1}}})_\beta}^\alpha \epsilon^{\beta\gamma}\ .
\end{eqnarray}
This current was found also in \cite{Williams}.

$j^\mu$ is real:  $j^{\mu\dagger}=j^\mu$.
It transforms under $P$ and $T$ as 
\begin{eqnarray} 
&& P^{-1}j^0(x,t)P=j^0(-x,t)\ ,\quad P^{-1}j^i(x,t)P=-j^i(-x,t)\ ,\nonumber\\ 
&& T^{-1}j^0(x,t)T=j^0(x,-t)\ ,\quad T^{-1}j^i(x,t)T=-j^i(x,-t)\ ,
\end{eqnarray}
where
$i=1,2,3$, and the transformation property of $\psi$ under $P$ is assumed to be given by 
(\ref{eq.p3}) or (\ref{eq.p3even}).
$:j^\mu :$ changes sign under $C$: 
\beq
C:j^\mu :C \ = \ -:j^\mu :\ .
\eeq
For $n=1$, $j^\mu$ is the usual electromagnetic current of the Dirac field.

The charge corresponding to $j^\mu$ is 
\beq
Q=\int \intd^3 x\ :j^0:\ \ =\, \sum_{i=1}^{n+1} \int \frac{d^3 k}{\omega(k)} [-c_i^\dagger(k)c_i(k) + d_i^\dagger(k)d_i(k)]\ .
\eeq
$Q$ has the eigenvalue $-1$ on one-particle states created by $c_i^\dagger(k)$ and 
$+1$ on one-particle states created by $d_i^\dagger(k)$.

\subsection{Fields transforming according to $D^{(n,m)}$}

We turn now to the case when $\Psi$ transforms according to one of the representations $D^{(n,m)}$.
In this case the field
\beq
\hat{\psi}_\alpha={(\mc{D}^+)_\alpha}^\beta \Psi_\beta=
\frac{1}{2\mass^{n+m}}{(\mass^{n+m}+\ii^{n+m} \gamma^{\mu_1\nu_2\dots\mu_{n+m}}\partial_{\mu_1\mu_2\dots\mu_{n+m}})_\alpha}^\beta\Psi_\beta\ , 
\eeq
where
\beq
{(\mc{D}^+)_\alpha}^\beta = \frac{1}{2\mass^{n+m}}{(\mass^{n+m}+\ii^{n+m} \gamma^{{\mu_1}{\mu_2}\dots{\mu_{n+m}}}\partial_{{\mu_1}{\mu_2}\dots{\mu_{n+m}}})_\alpha}^\beta\ , 
\eeq
is very similar to the field $\psi$ defined in the case of the representations $D^{(n)}$; in particular
it describes only physical particles, it has similar mode expansion, (anti)commutation relations, 
Green function and propagator, and it satisfies the generalized Dirac equation 
\beq
{(\mc{D}^-)_\alpha}^\beta \hat{\psi}_\beta    =
\frac{1}{2\mass^{n+m}}{(\mass^{n+m}-\ii^{n+m} \gamma^{{\mu_1}{\mu_2}\dots{\mu_{n+m}}}\partial_{{\mu_1}{\mu_2}\dots{\mu_{n+m}}})_\alpha}^\beta \hat{\psi}_\beta= 0
\eeq
in addition to the Klein--Gordon equation.
However, it describes several pairs of particles with different spins; 
specifically one pair of particles for each spin 
$(n+m)/2$, $(n+m)/2-1$, $\dots$, $(n-m)/2$. 
In order to obtain fields that describe particles with a specific spin it is necessary 
to apply further projection operators to $\hat{\psi}$.
For a specific spin $s$ a suitable projection operator $\mc{D}_s$ can be obtained as a real linear combination of the 
differential
operators 
\begin{eqnarray}
&& \ii^{n+m}\gamma_{(1)}^{\mu_1\mu_2\dots\mu_{n+m-2}}\partial_{\mu_1\mu_2\dots\mu_{n+m-2}}\ , \nonumber \\
&& \ii^{n+m}\gamma_{(2)}^{\mu_1\mu_2\dots\mu_{n+m-4}}\partial_{\mu_1\mu_2\dots\mu_{n+m-4}}\ , \nonumber \\
&& \dots\ , \nonumber \\
&& \ii^{n+m}\gamma_{(m)}^{\mu_1\mu_2\dots\mu_{n-m}}\partial_{\mu_1\mu_2\dots\mu_{n-m}}\ , \nonumber
\end{eqnarray} 
where 
$\gamma_{(1)}^{\mu_1\mu_2\dots\mu_{n+m-2}}$,  
$\gamma_{(2)}^{\mu_1\mu_2\dots\mu_{n+m-4}}$,
$\dots$, 
$\gamma_{(m)}^{\mu_1\mu_2\dots\mu_{n-m}}$
are invariant tensors defined in \ref{app.a4}.
The coefficients needed in the linear combination can be determined by direct calculations for 
not very large values of $m$, but
we leave it as an open problem to find general formulas for these coefficients.

The field that describes a pair of particles of spin $s$ can be defined as  
\beq
\label{eq.psis}
\psi_{(s)}=\mc{D}_s \mc{D}^+ \Psi\ .
\eeq
Both $\mc{D}^+$ and $\mc{D}_s$ are real, therefore 
\beq
\psi_{(s)}^\dagger=\mc{D}_s \mc{D}^+ \Psi^\dagger \ .
\eeq
We also have $\Psi=\hat{\psi}^- + \psi_{((n+m)/2)} + \psi_{((n+m)/2-1)} + \dots +\psi_{(n-m)}$, where 
$\hat{\psi}^- = \mc{D}^- \Psi$.
In addition to the Klein--Gordon equation, $\psi_{(s)}$ satisfies the differential equations
$\mc{D}^- \psi_{(s)}=0$ and 
$\mc{D}_{s'}\psi_{(s)}=0$ for any $s'\ne s$.
We have $[ \psi_{(s)},\psi_{(s')} ]_\pm=0$ and $[ \psi_{(s)},\psi_{(s')}^\dagger ]_\pm=0$ if $s\ne s'$, 
and $[ \hat{\psi}^-,\psi_{(s)} ]_\pm=0$, $[ \hat{\psi}^-,\psi_{(s)}^\dagger ]_\pm=0$.
$\hat{\psi}$, $\hat{\psi}^-$ and $\psi_{(s)}$ have the same Lorentz and $C$, $P$, $T$ transformation properties as 
$\Psi$. 
Real and imaginary parts   
$\phi_{1(s)}$ and $\phi_{2(s)}$ of $\psi_{(s)}$ can be introduced in the same way as in the case of the representations
$D^{(n)}$. 
The (anti)commutation relations, 
the Green function and the propagator of $\psi_{(s)}$ can also be calculated similarly as in the case of $D^{(n)}$. 
Specifically,
\begin{eqnarray}
[\psi_{(s)\alpha}(x,t_x),\psi_{(s)\beta}^\dagger(y,t_y)]_\pm & = &
[\psi_{(s)\alpha}^\dagger(x,t_x),\psi_{(s)\beta}(y,t_y)]_\pm \nonumber \\
& = & [{(\mc{D}_{s}\mc{D}^+)_\alpha}^\delta \Psi_\delta(x,t_x), \Psi_\beta^\dagger(y,t_y)]_\pm \nonumber \\
& = & {(\mc{D}_{s}\mc{D}^+)_\alpha}^\delta
\epsilon_{\delta\beta} [G(x-y,t_x-t_y) - G(y-x,t_y-t_x) ]\ ,
\label{eq.aa1}
\end{eqnarray}
\begin{eqnarray}
\brakettt{0}{\psi_{(s)\alpha}(x,t_x)\psi_{(s)\beta}^\dagger(y,t_y)}{0} & = &
\brakettt{0}{\psi_{(s)\alpha}^\dagger(x,t_x)\psi_{(s)\beta}(y,t_y)}{0} \nonumber \\
& = & \brakettt{0}{{(\mc{D}_{s}\mc{D}^+)_\alpha}^\delta\Psi_\delta(x,t_x)\Psi_\beta^\dagger(y,t_y)}{0} \nonumber \\
& = & {(\mc{D}_{s}\mc{D}^+)_\alpha}^\delta
\epsilon_{\delta\beta} G(x-y,t_x-t_y)\ ,
\label{eq.aa2}
\end{eqnarray}
and the covariant part of the Feynman propagator 
\begin{eqnarray}
\brakettt{0}{\mathrm{T}\psi_{(s)\alpha}(x,t_x)\psi_{(s)\beta}^\dagger(y,t_y)}{0} = 
\brakettt{0}{\mathrm{T}\psi_{(s)\alpha}^\dagger(x,t_x)\psi_{(s)\beta}(y,t_y)}{0} \nonumber \\
= \brakettt{0}{\mathrm{T}{(\mc{D}_{s}\mc{D}^+)_\alpha}^\delta\Psi_\delta(x,t_x)\Psi_\beta^\dagger(y,t_y)}{0}
\end{eqnarray}
is 
\beq
{(\mc{D}_{s}\mc{D}^+)_\alpha}^\delta
\epsilon_{\delta\beta} D_F(x-y,t_x-t_y)  \ .
\label{eq.aa3}
\eeq
The operators $\mc{D}_{s}$ and $\mc{D}^+$ differentiate with respect to $t_x$ and $x$ in the formulas above.

Mode expansions analogous to (\ref{eq.psimodex}) can also be written down for $\psi_{(s)}$, 
using polarization vectors that transform according to the $(s)$ representation of the relevant $SU(2)$ little group. 
These polarization vectors are described in more detail in \ref{app.a52}.  The mode expansion for $\psi_{(s)}$
reads 
\begin{eqnarray}
&& \psi_{(s)\alpha} (x,t)=\int \frac{\intd^3 k}{\sqrt{2}(\sqrt{2\pi})^3 \omega(k)}  \sum_{i=1}^{2s+1}
[e^{\ii kx}e^{\ii \omega(k)t}
        u_{(s),i\alpha}(k) c_{(s),i}^\dagger(k) \nonumber \\
&& \hspace{6cm} +e^{-\ii kx}e^{-\ii\omega(k)t}
      u_{(s),i\alpha}(k)^* d_{(s),i}(k)  ]\ ,
\label{eq.psi(s)modex}
\end{eqnarray}
where 
\beq
c_{(s),i}^\dagger(k) =  \hat{u}_{(s),i}(k)^\alpha a_\alpha^\dagger(k)\ ,\qquad
d_{(s),i}^\dagger(k) =  \hat{u}_{(s),i}(k)^\alpha b_\alpha^\dagger(k)\ .
\eeq

The fields $\frac{I\mp \ii\gamma^5}{2}\psi_{(s)}$, $\frac{I\mp \ii\gamma^5}{2}\phi_{1(s)}$,
$\frac{I\mp \ii\gamma^5}{2}\phi_{2(s)}$,
which describe the same kind of particles as 
$\psi_{(s)}$, $\phi_{1(s)}$, $\phi_{2(s)}$, transform according to the $(n/2,m/2)$ and $(m/2,n/2)$ representations, 
respectively. The (anti)commutation relations, Green functions and Feynman propagators of these fields can be obtained from those of $\psi_{(s)}$, $\phi_{1(s)}$, $\phi_{2(s)}$.

\subsection{Fields transforming according to $\tilde{D}^{(n)}$}

Finally we discuss the case when $\Psi$ transforms according to one of the representations $\tilde{D}^{(n)}$. 
In this case one can define for any specific spin $s= 0, 1, \dots, n$ a projection operator $\mc{D}_s$ by 
taking a suitable real linear combination of  
\begin{eqnarray}
&& \ii^{2n} \tau^{\mu_1\mu_2\dots \mu_{2n}}\partial_{\mu_1\mu_2\dots \mu_{2n}}\ , \nonumber \\ 
&& \ii^{2n} \tau_{(1)}^{\mu_1\mu_2\dots \mu_{2n-2}}\partial_{\mu_1\mu_2\dots \mu_{2n-2}}\ , \nonumber \\ 
&& \ii^{2n} \tau_{(2)}^{\mu_1\mu_2\dots \mu_{2n-4}}\partial_{\mu_1\mu_2\dots \mu_{2n-4}}\ , \nonumber \\
&& \dots\ , \nonumber \\
&& \ii^{2n} \tau_{(n-1)}^{\mu_1\mu_2}\partial_{\mu_1\mu_2} \nonumber
\end{eqnarray}
and the identity matrix, so that the  
field
\beq
\label{eq.psis2}
\psi_{(s)}=\mc{D}_s \Psi
\eeq
describes a pair of particles of spin $s$ (see Appendix \ref{app.a4} for the definition of the $\tau$ tensors). 
Again, we leave it as an open problem to find general explicit formulas
for the coefficients needed to obtain $\mc{D}_s$. 
$\psi_{(s)}$ and $\psi_{(s)}^\dagger$ 
create physical particles for $s=n,n-2,\dots$ and 
nonphysical particles (i.e.\ one-particle states that have negative scalar product with themselves) for $s=n-1,n-2,\dots$, 
but this can be reversed by changing the sign of $\epsilon$.

The operators $\mc{D}_s$ are real, therefore
\beq
\psi_{(s)}^\dagger=\mc{D}_s \Psi^\dagger\ .
\eeq
$\Psi$ has the decomposition $\Psi=\psi_{(n)}+\psi_{(n-1)}+\psi_{(n-2)}+ \dots + \psi_{(0)}$. 
In addition to the Klein--Gordon equation,
$\psi_{(s)}$ satisfies the differential equations 
$\mc{D}_{s'} \psi_{(s)} = 0$ for all $s'\ne s$. 
We also have $[ \psi_{(s)},\psi_{(s')} ]=0$ and $[ \psi_{(s)},\psi_{(s')}^\dagger ]=0$ if $s\ne s'$.
$\psi_{(s)}$ has the same Lorentz and $C$, $P$, $T$ transformation properties as $\Psi$. Real and imaginary parts of 
$\psi_{(s)}$ can be introduced in the same way as in the case of the representations $D^{(n)}$.
The commutation relations, the Green function and the propagator of $\psi_{(s)}$ can 
be calculated similarly as in the case of $D^{(n)}$. 
Equations (\ref{eq.aa1}), (\ref{eq.aa2}) and (\ref{eq.aa3}) apply with the replacement $\mc{D}_{s}\mc{D}^+ \to \mc{D}_{s}$.
Mode expansions analogous to (\ref{eq.psimodex}) and (\ref{eq.psi(s)modex}) can also be written down for $\psi_{(s)}$, 
using polarization vectors that transform according to the $(s)$ representation of the relevant $SU(2)$ little group. 
These polarization vectors are described in \ref{app.a53}.

\subsection{Interactions}
\label{sec.int}

In this section it is discussed briefly how the fields defined in the previous sections can be applied
to describe in a Hamiltonian framework
interactions of massive higher spin particles with themselves and with possible other particles 
described by different types of fields. We consider the case when 
the interactions are local.
The same notation is used for the Heisenberg fields of interacting systems as 
for the free fields, but this should not cause confusion.
   
If the physical system contains 
$N$ pairs of particles that can be described by the $\psi$ or $\psi_{(s)}$ type fields, then 
one should take $N$ corresponding fields $\Psi_i$, $i=1,\dots, N$, 
which are the type of fields defined in Section \ref{sec.lt}, 
and a full Hamiltonian operator that has the form
\beq
H= \sum_{i=1}^N H_{\Psi_i} + H_\Phi + H_{\mathrm{int}} \ , 
\eeq
where $H_{\Psi_i}$ are the free particle Hamiltonian operators having the form given in (\ref{HH}), 
$H_\Phi$ is the collective free Hamiltonian operator for the other types of fields, if there are any, 
and $H_{\mathrm{int}} = \int \intd^3 x\ \mc{H}_{\mathrm{int}} $.
$\mc{H}_{\mathrm{int}}$ is a self-adjoint interaction term that may depend on 
$\Psi_i$ only through the corresponding $\psi_i$ (or $\psi_{({s_i})i}$) field and its derivatives.
The latter condition is imposed because the interactions should involve only physical degrees of freedom. 
The gauge theory -like situation in which $\mc{H}_{\mathrm{int}}$ depends on 
$\Psi_i$ and its derivatives directly, i.e.\ not only through $\psi_i$, is also conceivable, 
but we do not discuss this possibility in any detail here. 
We also assume that $\mc{H}_{\mathrm{int}}$ is even with respect to the collective change of sign of the fermionic fields. 
In accordance with \cite{WeinbergQFT} (chapter 7), 
the Hamiltonian operator is constructed from free fields, which are taken at $t=0$.
The Heisenberg fields of the interacting system are defined as the fields whose time evolution is determined by $H$ 
and which are equal to the corresponding 
free fields at $t=0$.

Since the use of the fields $\Psi_i$ implies that the Hilbert space contains nonphysical and superfluous 
polarization states, 
it should be kept in mind that 
in the calculation of physical S-matrix elements only those states in the Hilbert space
should be taken as initial and final states which have suitable polarization.

Canonical Hamiltonian equations of motion for the Heisenberg fields $\Psi_i$, $\Pi_i$ 
of the interacting system
can be obtained in the following way:
in the expression for $H$ one eliminates the time derivatives of the (free) $\Psi_i$ fields by applying the rules 
\beq
\label{eq.elim2}
\partial_t\Psi_i \ \to\  \Pi_i\ ,\qquad
\partial_{t}\Pi_i \ \to\  \partial_x^2\Psi_i-\mass^2\Psi_i\ ,
\eeq
and than one takes the standard Heisenberg equations
\beq
\label{eq.cnl}
\partial_t \Psi_{i\alpha} = [\ii H, \Psi_{i\alpha}]\ , \qquad
\partial_t \Pi_{i\alpha} = [\ii H, \Pi_{i\alpha}]
\eeq
of quantum mechanics,
where the commutators on the right hand side can be calculated using
(\ref{eq.ac5ps}) and (\ref{eq.ac6ps}). 
These equations do not contain higher than first order time derivatives of $\Psi_i$ and $\Pi_i$, 
and there are no time derivatives of $\Psi_i$ and $\Pi_i$
on the right hand side.

The relation between the Heisenberg fields $\Psi_i$, $\Pi_i$ of the interacting system 
and $\psi_i$ or $\psi_{({s_i})i}$ (also of the interacting system) is given by that version of 
(\ref{eq.psidef}) or (\ref{eq.psis}) or (\ref{eq.psis2})
which 
is obtained from
(\ref{eq.psidef}) or (\ref{eq.psis}) or (\ref{eq.psis2}), respectively,   
by eliminating all time derivatives on the right hand side by the application of (\ref{eq.elim2}). 
The fields  
$\psi_i^-$ or $\hat{\psi}_i^-$ or $\psi_{(s')i}$, $s'\ne s_i$,
can be obtained in a similar way.
These are free fields, of course, 
since the interaction terms in the Hamiltonian operator commute with them. 
We note that $\psi_i^-=\Psi_i-\psi_i$ 
in the case when $\psi_i$ transforms according to a representation $D^{(n)}$.

In principle, a Lagrangian can also be obtained in the usual way by Legendre transformation. This, however, 
involves the elimination of $\Pi_i$ for $\Psi_i$ and their derivatives, 
which can be very difficult in practice, because the interaction Hamiltonian, 
when expressed in terms of $\Psi_i$ and $\Pi_i$, generally contains high derivatives 
(depending on the spin of the interacting particles) of these fields, 
even if in terms of $\psi_i$ or $\psi_{({s_i})i}$ it does not contain derivatives at all. 
The reason for this is that by virtue of the relations 
(\ref{eq.psidef}), (\ref{eq.psis}) and (\ref{eq.psis2}), 
the expression for $\psi$ or $\psi_{(s)}$  in terms of $\Psi$ and $\Pi$ contains high derivatives
(depending on the spin of $\psi$ or $\psi_{(s)}$)
of $\Psi$ and $\Pi$.

If the equations of motion are studied at the classical level, i.e.\ if they are considered as classical equations for classical fields, then the nonphysical and superfluous fields  
$\psi_i^-$ or $\hat{\psi}_i^-$ or $\psi_{(s')i}$, $s'\ne s_i$,  can be prescribed
arbitrarily at the initial time; in particular they can be set to zero. Such a prescription 
for the nonphysical  and superfluous fields implies nontrivial constraints for $\Psi_i$ and $\Pi_i$. 
We also note, without going into details, that 
setting initial values is straightforward as long as only $\Psi_i$ and $\Pi_i$ are considered, 
but the situation is generally more complicated if one wants to set the initial values directly 
in terms of the physical fields
$\psi_i$ or $\psi_{({s_i})i}$, 
due to the complicated relation between these fields and $\Psi_i$, $\Pi_i$.

\section{Illustration: Dirac field and quantum electrodynamics}
\label{sec.qed}

In order to illustrate the derivation of canonical Hamiltonian equations of motion, 
we consider in this section the simple and familiar case of standard quantum electrodynamics
with a single Dirac field, which transforms according to $D^{(1)}$. 
As in Section \ref{sec.int}, the same notation is used for the Heisenberg fields of the interacting system as 
for the free fields.

We replace the definition (\ref{eq.psidef}) of the free $\psi$ field by 
\beq
\label{eq.psidefdirac}
\psi=\frac{1}{\sqrt{2\mass}}(\mass+\ii\gamma^\mu\partial_\mu)\Psi\ ,
\eeq
i.e.\ we introduce the factor $\sqrt{2\mass}$, 
in order to have the usual normalization.
We also use the notation  
$\bar{\psi}^\beta=\psi^\dagger_\alpha \epsilon^{\alpha\beta}$ in this section.

$\psi$ satisfies the Dirac equation 
\beq
(\mass-\ii\gamma^\mu\partial_\mu)\psi=0\ ,
\eeq
and has
anticommutation relations  
\begin{eqnarray} 
\label{eq.dcom1}
\{\psi_\alpha(x,t),\psi_\beta(x',t)\} & = & 0\\
\label{eq.dcom2}
\{\psi_\alpha(x,t),\psi_\beta^\dagger(x',t)\} & = &
 {(\gamma^0)_\alpha}^\rho \epsilon_{\rho\beta}\delta^3(x-x')\ .
\end{eqnarray}
In the basis defined in (\ref{eq.realb}), 
${(\gamma^0)_\alpha}^\rho \epsilon_{\rho\beta} = \delta_{\alpha\beta}$,
thus (\ref{eq.dcom1}) and (\ref{eq.dcom2}) are the standard anticommutation relations of the Dirac field.

We also introduce $\psi^-$ as
\beq
\psi^-=\frac{1}{\sqrt{2\mass}}(\mass-\ii\gamma^\mu\partial_\mu)\Psi\ .
\eeq
$\psi^-$ satisfies the equation 
\beq
(\mass+\ii\gamma^\mu\partial_\mu)\psi^-=0\ .
\eeq

The Hamiltonian operator for quantum electrodynamics can be written as 
\beq
H= H_{EM} + H_{\Psi} + H_{int}\ ,
\eeq
where
$H_\Psi$ is given by (\ref{HH}),  
\beq
H_{EM}=-\frac{1}{2}\int \intd^3 x\ :[\Pi^A_\mu \Pi^{A\mu}
+(\partial_x A_\mu)(\partial_x A^\mu)]:
\eeq
is the Hamiltonian operator of the free electromagnetic field, and 
\beq
H_{int}=\int \intd^3 x\ eA_\mu j^\mu\ 
\eeq
is the interaction term. $e=-|e|$ is the 
electron charge, $j^\mu=\bar{\psi}\gamma^\mu \psi$ is the electromagnetic current.

The electromagnetic field is understood to be quantized in the Gupta--Bleuler formalism. $\Pi^A_\mu$ is 
the canonical momentum field conjugate to $A_\mu$; the equal-time commutation relation of 
these fields is
\beq
[A_\mu(x,t), \Pi^A_\nu(x',t)]  =  -\ii g_{\mu\nu}\delta^3(x-x')\ .
\eeq

The equations of motion for $A_\mu$ and $\Pi^A_\mu$ in the presence of interaction are 
\begin{eqnarray}
[\ii H,A_\mu] & = & \partial_t A_\mu \ = \ \Pi^A_\mu \\
\ [\ii H, \Pi^A_\mu] & = & \partial_t \Pi^A_\mu\ =\ 
\partial_x^2 A_\mu + ej_\mu\ ,
\end{eqnarray}
thus $A_\mu$ satisfies the usual equation  
\beq
(\partial_t^2-\partial_x^2)A_\mu=ej_\mu\ .
\eeq

From $\partial_t \Psi_\alpha = [\ii H, \Psi_\alpha]$ and
$\partial_t \Pi_\alpha = [\ii H, \Pi_\alpha]$ 
the following equations of motion are obtained 
for the $\Psi_\alpha$ and $\Pi_\alpha$ fields of the interacting system:
\begin{eqnarray}
\label{eq.me1}
\partial_t\Psi & = & \Pi-e\frac{\ii}{\sqrt{2\mass}} A_\mu \gamma^0\gamma^\mu  \psi\\
\label{eq.me2}
\partial_t\Pi & = & \partial_x^2\Psi-\mass^2\Psi
-e\frac{\mass}{\sqrt{2\mass}}A_\mu\gamma^\mu\psi
-e\frac{\ii}{\sqrt{2\mass}}
\partial_x(A_\mu\gamma^x\gamma^\mu\psi)\ .
\end{eqnarray}
Of course, $\psi$ means in these equations and in $H$ the expression
\beq
\label{eq.psi}
\psi=\frac{1}{\sqrt{2\mass}}(\mass+\ii\gamma^x\partial_x)\Psi+\frac{1}{\sqrt{2\mass}}\ii\gamma^0\Pi\ ,
\eeq
which is obtained from (\ref{eq.psidefdirac}) by the substitution $\partial_t \Psi \to \Pi$.
The analogous expression for $\psi^-$ is 
\beq
\label{eq.psi-}
\psi^- = \frac{1}{\sqrt{2\mass}}(\mass-\ii\gamma^x\partial_x)\Psi-\frac{1}{\sqrt{2\mass}}\ii\gamma^0\Pi\  .
\eeq
(\ref{eq.me1}) and (\ref{eq.me2}) constitute the canonical Hamiltonian equations of motion 
in the example under consideration.

From (\ref{eq.psi}) and (\ref{eq.psi-}) one can express $\Psi$ and $\Pi$ as
\begin{eqnarray}
\Psi & = & \frac{1}{\sqrt{2\mass}}(\psi+\psi^-)\\
\Pi & = & -\ii \sqrt{\frac{\mass}{2}}\gamma^0(\psi-\psi^-)-
\frac{1}{\sqrt{2\mass}}\gamma^0\gamma^x\partial_x(\psi+\psi^-)\ .
\end{eqnarray}
Using this relation between $\Psi$, $\Pi$ and $\psi$, $\psi^-$, 
the equations of motion
(\ref{eq.me1}) and (\ref{eq.me2}) can be rewritten as
\begin{eqnarray}
(\mass+\ii\gamma^\mu\partial_\mu)\psi^- + (-\mass+\ii\gamma^\mu\partial_\mu -eA_\mu\gamma^\mu)\psi & = & 0\\
(\mass-\ii\gamma^x\partial_x)(\mass+\ii\gamma^\mu\partial_\mu)\psi^- + 
(-\mass-\ii\gamma^x\partial_x)(-\mass+\ii\gamma^\mu\partial_\mu -eA_\mu\gamma^\mu)\psi & = & 0\ ,
\end{eqnarray}
which are equivalent with 
\begin{eqnarray}
\label{eq.184}
(\mass-\ii\gamma^\mu\partial_\mu +eA_\mu\gamma^\mu )\psi & = & 0\\ 
\label{eq.185}
(\mass+\ii\gamma^\mu\partial_\mu)\psi^- & = & 0\ ,
\end{eqnarray}
as expected. (\ref{eq.184}) is the usual equation of motion for the Dirac field in quantum electrodynamics, and 
(\ref{eq.185}) is the free equation of motion for $\psi^-$.

We turn now to the determination of the Lagrangian function. From equations 
(\ref{eq.me1}) and (\ref{eq.psi}) one can express $\Pi$ in terms of $A$ and $\Psi$:
\beq
\Pi=\frac{1-\frac{e}{2\mass}A_\nu\gamma^0\gamma^\nu\gamma^0}{1-\frac{e^2}{4\mass^2}A_\mu A^\mu}
[\partial_t\Psi +\frac{\ii e}{2\mass}A_\lambda\gamma^0\gamma^\lambda(\mass+\ii\gamma^x\partial_x)\Psi]\ .
\eeq
(\ref{eq.psi}) gives then the following expression for $\psi$ in terms of $\Psi$ and $A$: 
\beq
\label{eq.psix}
\psi=\frac{1}{\sqrt{2\mass}}\frac{1-\frac{e}{2\mass}A_\mu\gamma^\mu}{1-\frac{e^2}{4\mass^2}A_\nu A^\nu}
(\mass+\ii\gamma^\lambda \partial_\lambda)\Psi\ .
\eeq

Finally, for the Lagrangian 
$L=\int \intd^3 x\, [-\partial_t A_\mu \Pi^{A\mu}   
+\partial_t \Psi_\alpha^\dagger \Pi_\beta \epsilon^{\alpha\beta}
+\partial_t \Psi_\alpha \Pi_\beta^\dagger \epsilon^{\alpha\beta}]
-H$, which is given by the Legendre transformation, 
one finds the result
\beq
\label{L}
L  = L_A+L_\Psi - \int \intd^3 x\, \left[
\frac{e^2}{2\mass}A_\mu A^\mu \bar{\psi}\psi
+eA_\mu \bar{\psi} \gamma^\mu \psi\right]
, 
\eeq
where 
\begin{eqnarray}
L_A & = & -\frac{1}{2} \int\intd^3 x\, \partial_\mu A_\nu \partial^\mu A^\nu\\
L_\Psi & = & \int \intd^3 x\, [\partial_\mu\Psi_\alpha^\dagger \partial^\mu\Psi_\beta 
-\mass^2\Psi_\alpha^\dagger\Psi_\beta ]\epsilon^{\alpha\beta}\ .
\end{eqnarray}
The interaction term is the third term on the right hand side of (\ref{L}), 
and $\psi$ is understood to be an abbreviation for 
(\ref{eq.psix}).

\section{Summary and conclusion}
\label{sec.concl}

In this paper we gave a construction of free quantum fields $\psi$ for massive particles of arbitrary spin 
in a canonical Hamiltonian framework. 
The fields that we used transform according to the $(n/2,m/2)\oplus (m/2,n/2)$  and  $(n/2,n/2)$ representations of 
$SL(2,\CC)$. In order to construct the desired fields, we introduced auxiliary fields 
$\Psi$, which transform according to the same representation as $\psi$. These auxiliary fields are constrained only by the 
Klein--Gordon equation, and are quantized canonically. 
The fields $\psi$ are obtained then by applying a suitable
differential operator $\mc{D}$ on $\Psi$. The application of $\mc{D}$ on $\Psi$ is necessary because 
$\Psi$ has more degrees of freedom then are needed. 
The (anti)commutator, the Green function and the Feynman propagator of $\psi$ turn out to be closely related to $\mc{D}$. 
Fields transforming according to  $(n/2,m/2)$ can also be obtained in a straightforward way by applying a projector
to the fields transforming according to $(n/2,m/2)\oplus (m/2,n/2)$.

The simplest case is that of the fields which transform according to the representations $(n/2,0)\oplus (0,n/2)$;
this is the case when the most explicit formula for $\mc{D}$ could be given. 
For $(n/2,m/2)\oplus (m/2,n/2)$  and  $(n/2,n/2)$, $n,m\ne 0$, it remains an open mathematical problem to find 
general formulas for certain coefficients appearing in $\mc{D}$; nevertheless for not very large values of 
$m$ these coefficients can be determined by direct calculation if necessary.

Although we restricted ourselves to fields transforming according to the\\ 
$(n/2,m/2)\oplus (m/2,n/2)$ and $(n/2,m/2)$ 
representations, 
it is important to note that 
the main ideas of the construction are not specific to these representations; one could take more complicated reducible representations of $SL(2,\CC)$, such as those that are used, for example, in the Rarita--Schwinger 
or in the Wigner--Bargmann formalisms 
(for recent developments on higher spin fields in these formalisms, see \cite{Peng1,Peng2}). 

The presented method allows the derivation of canonical Hamiltonian equations of motion 
in a straightforward way also in the presence of interactions,  
if the interaction Hamiltonian operator is known as an expression in terms of the fields $\psi$. 
These equations are expressed in terms of $\Psi$ and the corresponding canonical momentum fields $\Pi$, 
but $\psi$ can be expressed in terms of these fields in a straightforward way.

The use of the fields $\Psi$ has the consequence that Hilbert spaces with indefinite metric have to be introduced, 
nevertheless the physical states span a subspace on which the scalar product is positive definite.  
Although the equations of motion describe some nonphysical and superfluous degrees of freedom in addition to the proper physical ones, there is a clear separation between these degrees of freedom, and 
the nonphysical and superfluous degrees of freedom remain decoupled from the
physical ones.

Finally, it should be noted that the Hamiltonian formalism has the well known disadvantage that it 
is not manifestly covariant (in contrast with the Lagrangian formalism), and in general it is not easy
to construct interaction Hamiltonians that ensure the Lorentz covariance of the corresponding physical systems. 
This disadvantage of the Hamiltonian formalism limits the practical applicability
of our method for the derivation of field equations for interacting systems.

\appendix

\section{Representations of  $SL(2,\CC)$}
\label{app.a}

\renewcommand{\theequation}{A.\arabic{equation}} 
\setcounter{equation}{0}

In this appendix we discuss those parts of the representation theory 
of $SL(2,\CC)$, the covering group of the Lorentz group $SO(3,1)$, that are relevant for this paper.
The main topics are the classification of finite dimensional representations, 
invariant bilinear forms,  generalized gamma matrices, invariant complex conjugation,
useful special basis vectors and their orthogonality properties, completeness relations and 
projectors.

\subsection{Classification of the finite dimensional representations of $SL(2,\CC)$}
\label{app.a1}

The 
Lie algebra of $SL(2,\CC)$, denoted by $sl(2,\CC)$, 
is generated by $M_i$, $N_i$, $i=1,2,3$, which satisfy the commutation relations
\begin{eqnarray}
[M_i,M_j] & = & \epsilon_{ijk}M_k\\
\ [N_i,N_j] & = & -\epsilon_{ijk}M_k\\
\ [M_i,N_j] & = & \epsilon_{ijk}N_k\ ,
\end{eqnarray}
where $\epsilon_{ijk}$ is completely antisymmetric and $\epsilon_{123}=1$. We consider representations 
of this algebra in complex vectors spaces.

In Hilbert spaces $M_i$ and $N_i$ are represented by anti-Hermitian operators, which we denote by $\check{M}_i$, 
$\check{N}_i$. The usual Hermitian generators are $\ii\check{M}_i$, $\ii\check{N}_i$.
The commutators of $\check{M}_i$, $\check{N}_i$ and the Hermitian generators of the translations
$H$, $P_i$ are 
\begin{eqnarray}
[\check{M}_i, H] & = & 0\\
\ [\check{M}_i, P_j] & = & \epsilon_{ijk} P_k  \\
\ [\check{N}_i, H] & = & -P_i  \\
\ [\check{N}_i, P_j] & = & -\delta_{ij} H\  .  
\end{eqnarray}
In the case of finite dimensional representations we generally do not use distinct notation for 
$M_i$, $N_i$ and their
representations.

The finite dimensional irreducible representations of $sl(2,\CC)$ in complex vector spaces
can be labeled by two half-integers:
$(n/2,m/2)$, $n,m=0,1,2,\dots$. The (complex) dimension of the representation $(n/2,m/2)$ is $(n+1)(m+1)$. 
The tensor product of two irreducible representations can be decomposed into a sum of irreducible 
representations according to the rule
\beq
\left(\frac{n_1}{2},\frac{m_1}{2}\right)\otimes \left(\frac{n_2}{2},\frac{m_2}{2}\right) = 
\bigoplus_{n_3=|n_1-n_2|}^{n_1+n_2}\ \bigoplus_{m_3=|m_1-m_2|}^{m_1+m_2} \left(\frac{n_3}{2},\frac{m_3}{2}\right).
\eeq
The dual of the representation $(n/2,m/2)$ is equivalent to the representation $(n/2,m/2)$, the complex conjugate of 
$(n/2,m/2)$ is equivalent to $(m/2,n/2)$, and thus the adjoint of $(n/2,m/2)$ is also equivalent to 
$(m/2,n/2)$.

As is also well known, the finite dimensional irreducible representations of $su(2)$ in complex vector spaces
can be labeled by a single half-integer:
$(n/2)$, $n=0,1,2,\dots$. The (complex) dimension of the representation $(n/2)$ is $n+1$. 
The tensor product of two irreducible representations can be decomposed into a sum of irreducible 
representations according to the rule
\beq
\left(\frac{n_1}{2}\right)\otimes \left(\frac{n_2}{2}\right) = 
\bigoplus_{n_3=|n_1-n_2|}^{n_1+n_2}\  \left(\frac{n_3}{2}\right).
\eeq
The dual, the complex conjugate and the adjoint of the representation $(n/2)$ 
is equivalent to the same representation $(n/2)$. 

A representation $(n/2,m/2)$ of $sl(2,\CC)$ is also a representation of the $su(2)$ subalgebra generated by 
$M_1$, $M_2$, $M_3$, and as such it is equivalent to the representation $(n/2)\otimes (m/2)$.

The left and right handed Weyl spinors are vectors (in the general linear algebraic sense) 
that transform according to the two-dimensional representations
$(1/2,0)$ and $(0,1/2)$, respectively. Since these are complex representations, i.e.\ their complex conjugates
are not equivalent to themselves,
it is not possible in these representations to find  a basis with respect to which the generators $M_i$, $N_i$ are real matrices.
The representations  $(n/2,m/2)$, $n\ne m$ are also complex in this sense, whereas  
$(n/2,n/2)$ and $(n/2,m/2)\oplus (m/2,n/2)$ are real representations.  $(1/2,1/2)$ is the (complexified) 
Minkowski representation.
To be specific, we take the following matrices in the Minkowski representation for
the generators $M_i$, $N_i$:  
\beq
M_{1}=\left( \begin{array}{cccc}
0 & 0 & 0 & 0\\
0 & 0 & 0 & 0\\
0 & 0 & 0 & -1\\
0 & 0 & 1 & 0\\
\end{array}\right) ,\quad 
M_{2}=\left( \begin{array}{cccc}
0 & 0 & 0 & 0\\
0 & 0 & 0 & 1\\
0 & 0 & 0 & 0\\
0 & -1 & 0 & 0\\
\end{array}\right) ,\quad 
M_{3}=\left( \begin{array}{cccc}
0 & 0 & 0 & 0\\
0 & 0 & -1 & 0\\
0 & 1 & 0 & 0\\
0 & 0 & 0 & 0\\
\end{array}\right) 
\eeq

\beq
N_{1}=\left( \begin{array}{cccc}
0 & -1 & 0 & 0\\
-1 & 0 & 0 & 0\\
0 & 0 & 0 & 0\\
0 & 0 & 0 & 0\\
\end{array}\right) ,\quad 
N_{2}=\left( \begin{array}{cccc}
0 & 0 & -1 & 0\\
0 & 0 & 0 & 0\\
-1 & 0 & 0 & 0\\
0 & 0 & 0 & 0\\
\end{array}\right) ,\quad 
N_{3}=\left( \begin{array}{cccc}
0 & 0 & 0 & -1\\
0 & 0 & 0 & 0\\
0 & 0 & 0 & 0\\
-1 & 0 & 0 & 0\\
\end{array}\right) .
\eeq

We call the vectors transforming under the representations $(n/2,0)$ and $(0,n/2)$ 
left and right 
handed Weyl multispinors, respectively. 
$(n/2,0)$ and $(0,n/2)$ can be obtained as symmetric
tensor products of  $(1/2,0)$ or $(0,1/2)$, respectively:
\begin{eqnarray}
(n/2,0) & = & \vee^n (1/2,0)\\
(0,n/2) & = & \vee^n (0,1/2)\ .
\end{eqnarray}
Furthermore,
\beq
(n/2,m/2)= (n/2,0)\otimes (0,m/2)\ .
\eeq
We introduce the notations $D^{(n)}$, $D^{(n,m)}$, $\tilde{D}^{(n)}$ as 
\begin{eqnarray}
D^{(n)} & = & (n/2,0) \oplus (0,n/2) \\
D^{(n,m)} & = & (n/2,m/2) \oplus (m/2,n/2) \\
\tilde{D}^{(n)} & = & (n/2,n/2)\ ,
\end{eqnarray}
where $n\ge m$.
In the case $n=m$ it should be kept in mind that $D^{(n,n)}$ is composed of two distinct 
copies of $(n/2,n/2)$, although this will not be mentioned explicitly in the following.
$D^{(1)} = (1/2,0) \oplus (0,1/2)$ is known as the Dirac representation.
The representations $D^{(n)}$, $D^{(n,m)}$, $\tilde{D}^{(n)}$ are real, 
i.e.\ it is possible to find basis vectors 
such that $M_i$, $N_i$ are real matrices.

\subsection{Basis vectors for the representations $(n/2,m/2)$} 
\label{app.a2}

A specific matrix representation of $M_i$, $N_i$ on left and right handed Weyl spinors can be given as follows.
For $(1/2,0)$, 
\beq
M_{i}=\frac{-\ii}{2}\sigma_{i}\ ,\qquad N_{i}=\hlf \sigma_{i}\ ,\qquad i=1,2,3\ ,
\eeq
for $(0,1/2)$,
\beq
M_{i}=\frac{-\ii}{2}\sigma_{i}\ ,\qquad N_{i}=-\hlf \sigma_{i}\ ,\qquad i=1,2,3\ ,
\eeq
where $\sigma_i$ are the Pauli sigma matrices  
\beq
\sigma_1= \left( \begin{array}{cc}
0 & 1\\
1 & 0\\
\end{array}\right) \qquad \sigma_2= \left( \begin{array}{cc}
0 & -\ii\\
\ii & 0\\
\end{array}\right) \qquad \sigma_3= \left( \begin{array}{cc}
1 & 0\\
0 & -1\\
\end{array}\right),
\eeq
which have the well known properties 
\beq
[\sigma _{i},\sigma _{j}]=2\ii\epsilon _{ijk}\sigma _{k}\ ,\qquad
\sigma _{i}\sigma _{j}=\ii\epsilon _{ijk}\sigma _{k}\ ,\qquad
\mathrm{Tr}\, \, \sigma _{i}=0\ .
\eeq
In the following $e_1$, $e_2$ and $e_3$, $e_4$ denote fixed basis vectors of 
$(1/2,0)$ and $(0,1/2)$ with respect to which $M_i$, $N_i$ take the form above 
($e_1$, $e_2$ are the basis vectors of $(1/2,0)$ and $e_3$, $e_4$ are the basis vectors of $(0,1/2)$). 
We call $e_1$, $e_2$, $e_3$, $e_4$ Weyl basis vectors.
The corresponding dual basis vectors are denoted by $\hat{e}_i$, $i=1,\dots, 4$.
These vectors are defined by the property $(\hat{e}_i)^\alpha (e_j)_\alpha = \delta_{ij}$, $i,j=1,2$,
$(\hat{e}_i)^{\dot{\alpha}} (e_j)_{\dot{\alpha}} = \delta_{ij}$, $i,j=3,4$.
Here and throughout the paper dotted indices are used for right handed spinors.

Basis vectors for $(n/2,0)$ and $(0,n/2)$ can be obtained by taking symmetric tensor products of $e_1$, $e_2$ and $e_3$, $e_4$.
Specifically, we define the basis vectors
\begin{eqnarray}
&& E_l=\frac{1}{n!} {n\choose {l-1}}^{1/2}\, \mathrm{Sym}_{i_1,i_2,\dots, i_n} e_{i_1}\otimes e_{i_2}\otimes \dots \otimes e_{i_n}\ , 
\qquad l=1,\dots, (n+1) \nonumber \\
&& i_1=1,\ i_2=1,\ \dots,\ i_{n-l+1}=1,\quad
i_{n-l+2}=2,\ i_{n-l+3}=2,\ \dots,\  i_n=2
\end{eqnarray}
\begin{eqnarray}
&& F_l=\frac{1}{n!} {n\choose {l-1}}^{1/2}\, \mathrm{Sym}_{i_1,i_2,\dots, i_n} e_{i_1}\otimes e_{i_2}\otimes \dots \otimes e_{i_n}\ , 
\qquad l=1,\dots, (n+1) \nonumber \\
&& i_1=3,\ i_2=3,\ \dots,\ i_{n-l+1}=3,\quad
i_{n-l+2}=4,\ i_{n-l+3}=4,\ \dots,\ i_n=4\ .
\end{eqnarray}
Here we have introduced the notation $\mathrm{Sym}$ for symmetrization, which means summation over all 
permutations of the indices in the subscript. In the formulas above this means that there are $n!$ 
terms on the right hand side, of which many are actually identical. 
The dual basis vectors $\hat{E}_l$, $\hat{F}_l$ are obtained by replacing the vectors $e_i$ by $\hat{e}_i$ in these formulas.

The vectors $E_l\otimes F_k$, $l=1,\dots, (n+1)$, $k=1,\dots, (m+1)$ constitute a basis for $(n/2,m/2)$, and 
the vectors $\hat{E}_l\otimes \hat{F}_k$ are dual basis vectors.

The action of $M_i$, $N_i$ on the elements of $(n/2,0)$ and $(0,n/2)$ is defined in the usual way, i.e.\ if 
$v=\sum_{i_1,i_2,\dots, i_n} c_{i_1,i_2,\dots, i_n} e_{i_1}\otimes e_{i_2}\otimes \dots \otimes e_{i_n}$, where 
$c_{i_1,i_2,\dots, i_n}$ are complex coefficients, then 
\begin{eqnarray}
M_i v = \sum_{i_1,i_2,\dots, i_n} c_{i_1,i_2,\dots, i_n} 
[(M_ie_{i_1})\otimes e_{i_2}\otimes \dots \otimes e_{i_n} + 
e_{i_1}\otimes (M_ie_{i_2})\otimes \dots \otimes e_{i_n} +
\dots \nonumber \\
+ e_{i_1}\otimes e_{i_2}\otimes \dots \otimes (M_ie_{i_n})
]\ ,  
\end{eqnarray}
and the same formula applies also to $N_i$. The definition is similar for $(n/2,m/2)$; if 
$v=\sum_{l,k} c_{l,k} E_l\otimes F_k$, then 
$M_i v=\sum_{l,k} c_{l,k} [(M_i E_l) \otimes F_k + E_l \otimes (M_i F_k)]$.

\subsection{Invariant bilinear forms}
\label{app.a3}

In this subsection invariant bilinear forms on the representations $(n/2,m/2)$,
$D^{(n)}$, $D^{(n,m)}$ and $\tilde{D}^{(n)}$ are introduced.

$(1/2,0)$ and $(0,1/2)$ admit unique (up to multiplication by constant) invariant
bilinear forms, which are nondegenerate and antisymmetric. Specifically, we use the invariant bilinear forms given by the following matrices
with respect to the basis vectors $e_1$, $e_2$ and $e_3$, $e_4$:
\beq
\epsilon^L =\left( \begin{array}{cc}
0 & -1\\
1 & 0\\
\end{array}\right),
\qquad \epsilon^R= \left( \begin{array}{cc}
0 & 1\\
-1 & 0\\
\end{array}\right)  .
\eeq
(This means $\epsilon^L(e_1,e_2)=-\epsilon^R(e_3,e_4)=-1$.)
The matrices of the inverses of $\epsilon^L$ and $\epsilon^R$ (with respect to the dual basis vectors) are the negative of the 
matrices of $\epsilon^L$ and $\epsilon^R$.

$(n/2,m/2)$ also admit unique (up to multiplication by constant) invariant
bilinear forms, which are nondegenerate. Specifically, on $(n/2,0)$ and on $(0,n/2)$ 
we take the bilinear forms $\epsilon^L$ and $\epsilon^R$ defined as
\begin{eqnarray}
\label{eq.A16}
\epsilon^{L\alpha_1\alpha_2\dots\alpha_n;\delta_1\delta_2\dots\delta_n} & = &  \frac{1}{n!}\,
\mathrm{Sym}_{\delta_1\delta_2\dots\delta_n}\, \epsilon^{L\alpha_1\delta_1} \epsilon^{L\alpha_2\delta_2}\dots
\epsilon^{L\alpha_n\delta_n}\\
\label{eq.A17}
\epsilon^{R\dot{\alpha}_1\dot{\alpha}_2\dots\dot{\alpha}_n;\dot{\delta}_1\dot{\delta}_2\dots\dot{\delta}_n} & = &  \frac{1}{n!}\, 
\mathrm{Sym}_{\dot{\delta}_1\dot{\delta}_2\dots\dot{\delta}_n}\, \epsilon^{R\dot{\alpha}_1\dot{\delta}_1} \epsilon^{R\dot{\alpha}_2\dot{\delta}_2}\dots
\epsilon^{R\dot{\alpha}_n\dot{\delta}_n}\ .
\end{eqnarray}
$\epsilon^{L\alpha_1\alpha_2\dots\alpha_n;\delta_1\delta_2\dots\delta_n}$ and 
$\epsilon^{R\dot{\alpha}_1\dot{\alpha}_2\dots\dot{\alpha}_n;\dot{\delta}_1\dot{\delta}_2\dots\dot{\delta}_n}$ 
are completely symmetric in their 
indices $\alpha_1\alpha_2\dots\alpha_n$, $\delta_1\delta_2\dots\delta_n$, $\dot{\alpha}_1\dot{\alpha}_2\dots\dot{\alpha}_n$
and $\dot{\delta}_1\dot{\delta}_2\dots\dot{\delta}_n$. 
$\epsilon^{L\alpha_1\alpha_2\dots\alpha_n;\delta_1\delta_2\dots\delta_n}$ and 
$\epsilon^{R\dot{\alpha}_1\dot{\alpha}_2\dots\dot{\alpha}_n;\dot{\delta}_1\dot{\delta}_2\dots\dot{\delta}_n}$, 
as defined in (\ref{eq.A16}) and (\ref{eq.A17}),
have several $(1/2,0)$- or $(0,1/2)$-indices, corresponding to the fact that 
$(n/2,0)$ and $(0,n/2)$ were defined as spaces of symmetric tensors having $n$ 
$(1/2,0)$- or $(0,1/2)$-indices.
Nevertheless, 
$\epsilon^L$ and $\epsilon^R$ can also be regarded as tensors having two upper $(n/2,0)$- or 
$(0,n/2)$-indices. In the following and in other sections of the paper we often use a notation 
that corresponds to this view of $\epsilon^L$ and $\epsilon^R$, as we also mentioned at the end of the Introduction. 
 
On $(n/2,m/2)$ we take the invariant bilinear form 
\beq
\label{eq.bilnm}
\epsilon^{\alpha_1\alpha_2\dots\alpha_n \dot{\alpha}_1\dot{\alpha}_2\dots\dot{\alpha}_m;
\delta_1\delta_2\dots\delta_n \dot{\delta}_1\dot{\delta}_2\dots\dot{\delta}_m}=
\epsilon^{L\alpha_1\alpha_2\dots\alpha_n;\delta_1\delta_2\dots\delta_n} 
\epsilon^{R\dot{\alpha}_1\dot{\alpha}_2\dots\dot{\alpha}_m;\dot{\delta}_1\dot{\delta}_2\dots\dot{\delta}_m}\ .
\eeq
These bilinear forms are antisymmetric for $n+m$ odd and symmetric for $n+m$ even.  

The inverses of  $\epsilon^L$ and $\epsilon^R$ can also be obtained as
\begin{eqnarray}
\epsilon^{L}_{\alpha_1\alpha_2\dots\alpha_n;\delta_1\delta_2\dots\delta_n} & = &  \frac{1}{n!}\,
\mathrm{Sym}_{\delta_1\delta_2\dots\delta_n}\, \epsilon^{L}_{\alpha_1\delta_1} \epsilon^{L}_{\alpha_2\delta_2}\dots
\epsilon^{L}_{\alpha_n\delta_n}\\
\epsilon^{R}_{\dot{\alpha}_1\dot{\alpha}_2\dots\dot{\alpha}_n;\dot{\delta}_1\dot{\delta}_2\dots\dot{\delta}_n} & = &  \frac{1}{n!}\,
\mathrm{Sym}_{\dot{\delta}_1\dot{\delta}_2\dots\dot{\delta}_n}\, \epsilon^{R}_{\dot{\alpha}_1\dot{\delta}_1} \epsilon^{R}_{\dot{\alpha}_2\dot{\delta}_2}\dots
\epsilon^{R}_{\dot{\alpha}_n\dot{\delta}_n}\ .
\end{eqnarray}
The inverse of $\epsilon$ (defined in (\ref{eq.bilnm})) is 
\beq
\epsilon_{\alpha_1\alpha_2\dots\alpha_n \dot{\alpha}_1\dot{\alpha}_2\dots\dot{\alpha}_m;
\delta_1\delta_2\dots\delta_n \dot{\delta}_1\dot{\delta}_2\dots\dot{\delta}_m}=
\epsilon^L_{\alpha_1\alpha_2\dots\alpha_n;\delta_1\delta_2\dots\delta_n} 
\epsilon^R_{\dot{\alpha}_1\dot{\alpha}_2\dots\dot{\alpha}_m;\dot{\delta}_1\dot{\delta}_2\dots\dot{\delta}_m}\ .
\eeq

The nonzero matrix elements of $\epsilon^{L}$ and $\epsilon^{R}$ with respect to the basis vectors $E_l$, $F_l$ defined 
above
are 
\begin{eqnarray}
\epsilon^L(E_l,E_{n-l+2}) & = & (-1)^{n+l+1},\qquad l=1,\dots, (n+1)\\
\epsilon^R(F_l,F_{n-l+2}) & = & (-1)^{l+1},\qquad l=1,\dots, (n+1)\ .
\end{eqnarray} 
The inverses of $\epsilon^L$ and $\epsilon^R$ have the nonzero matrix elements
\begin{eqnarray}
\epsilon^L(\hat{E}_l,\hat{E}_{n-l+2}) & = & (-1)^{l+1},\qquad l=1,\dots, (n+1)\\
\epsilon^R(\hat{F}_l,\hat{F}_{n-l+2}) & = & (-1)^{n+l+1},\qquad l=1,\dots, (n+1) 
\end{eqnarray} 
with respect to the dual basis. We do not introduce distinct
notation for the inverses of $\epsilon^L$, and $\epsilon^R$, but this should
not cause confusion.

On $D^{(1)}$, we take the antisymmetric invariant bilinear form $\epsilon$ 
that coincides with $\epsilon^L$ and $\epsilon^R$ on the subspaces 
$(1/2,0)$ and $(0,1/2)$, respectively, and is diagonal with respect to the decomposition 
$D^{(1)}=(1/2,0)\oplus (0,1/2)$, i.e.\ that has the matrix
\beq
\epsilon=\left( \begin{array}{cc}
\epsilon^L & 0 \\
0 & \epsilon^R \\
\end{array}\right)
\eeq
with respect to the basis  $e_1$, $e_2$, $e_3$, $e_4$. We also take invariant bilinear forms $\epsilon$
on $D^{(n)}$ and $D^{(n,m)}$ defined in the same way.

\subsection{Generalized gamma matrices}
\label{app.a4}

With respect to the Weyl basis, the Dirac gamma matrices are given by 
\beq
\gamma^0=\left( \begin{array}{cc}
0 & I \\
I & 0 \\
\end{array}\right),\qquad
\gamma^i=\left( \begin{array}{cc}
0 & \sigma_i \\
-\sigma_i & 0 \\
\end{array}\right),\qquad i=1,2,3\ ,
\eeq
where $I$ denotes the $2\times 2$ identity matrix.
These matrices satisfy the well known identity
\beq
\label{eq.gammaid}
\{\gamma^\mu,\gamma^\nu \}=2g^{\mu\nu}\ .
\eeq
In addition, they are related to the representation $M_{Di}$, $N_{Di}$ of the generators $M_i$, $N_i$ on $D^{(1)}$ by the equations
\beq
L_D^{\mu\nu}=\frac{1}{4}[\gamma^\mu,\gamma^\nu]\ ,
\eeq
\beq
L_D^{0i}=-N_{Di}\ ,\qquad  
L_D^{12}=M_{D3}\ ,\quad 
L_D^{23}=M_{D1}\ ,\quad 
L_D^{13}=-M_{D2}\ ,\quad
L_D^{\mu\nu}=-L_D^{\nu\mu}\ .
\eeq
One can also define the matrix $\gamma^5$ as
\beq
\gamma^5=\gamma^0\gamma^1\gamma^2\gamma^3\ .
\eeq
It satisfies the identity
\beq
\{ \gamma^5,\gamma^\mu \}=0\ ,
\eeq
and its matrix form is
\beq
\gamma^5=\left( \begin{array}{cc}
\ii I & 0 \\
0 & -\ii I \\
\end{array}\right).
\eeq

$\gamma^\mu$, regarded as a tensor with an upper Minkowski vector index and an upper and a lower Dirac spinor index 
(the latter two are suppressed), is an $SL(2,\CC)$-invariant tensor. As usual, this means 
\beq
{(\Lambda_M)^\nu}_\mu {(\Lambda_D)_\delta}^\alpha {(\Lambda_D^{-1})_\beta}^\rho {(\gamma^\mu)_\alpha}^\beta =
{(\gamma^\nu)_\delta}^\rho \ ,
\eeq
where $\Lambda_M$ and $\Lambda_D$ denote the representation of an element $\Lambda$ of $SL(2,\CC)$ in the Minkowski and Dirac spinor spaces, respectively.
$\gamma^5$ is also an invariant tensor 
having one upper and one lower Dirac spinor index.

The Dirac gamma matrices take the form
\beq
\gamma^\mu=\left( \begin{array}{cc}
0 & \eta^\mu \\
\bar{\eta}^\mu & 0 \\
\end{array}\right)
\eeq
with respect to the decomposition $D^{(1)}=(1/2,0)\oplus (0,1/2)$,
where $\eta^\mu$ is an invariant tensor with one Minkowski vector index, one upper $(0,1/2)$-index and one lower $(1/2,0)$-index, and
$\bar{\eta}^\mu$ is an invariant tensor with one Minkowski vector index, one lower $(0,1/2)$-index and one upper $(1/2,0)$-index (however, the spinor indices are suppressed here).
It follows from (\ref{eq.gammaid}) that
$\eta^\mu$ and $\bar{\eta}^\mu$ satisfy the identities
\begin{eqnarray}
\label{eq.a46-1}
\eta^\mu\bar{\eta}^\nu + \eta^\nu\bar{\eta}^\mu & = & 2g^{\mu\nu}\\
\label{eq.a46-2}
\bar{\eta}^\mu \eta^\nu + \bar{\eta}^\nu\eta^\mu & = & 2g^{\mu\nu}\ .
\end{eqnarray}

As generalizations of $\eta^\mu$ and $\bar{\eta}^\mu$, the invariant tensors 
$\tau^{\mu_1\mu_2\dots\mu_n}$ and
$\bar{\tau}^{\mu_1\mu_2\dots\mu_n}$ can be defined as
\begin{eqnarray}
{(\tau^{\mu_1\mu_2\dots\mu_n})_{\alpha_1\alpha_2\dots\alpha_n}}^{\dot{\delta}_1\dot{\delta}_2\dots\dot{\delta}_n}
& = & \nonumber \\
&& \hspace{-2cm} \frac{1}{(n!)^2}\, \mathrm{Sym_{\alpha_1\alpha_2\dots\alpha_n}}\, \mathrm{Sym_{\dot{\delta}_1\dot{\delta}_2\dots\dot{\delta}_n}}\,
{(\eta^{\mu_1})_{\alpha_1}}^{\dot{\delta}_1} {(\eta^{\mu_2})_{\alpha_2}}^{\dot{\delta}_2} \dots {(\eta^{\mu_n})_{\alpha_n}}^{\dot{\delta}_n}\\[4mm]
{(\bar{\tau}^{\mu_1\mu_2\dots\mu_n})_{\dot{\alpha}_1\dot{\alpha}_2\dots\dot{\alpha}_n}}^{\delta_1\delta_2\dots\delta_n}
& = & \nonumber \\
&& \hspace{-2cm} \frac{1}{(n!)^2}\, \mathrm{Sym_{\dot{\alpha}_1\dot{\alpha}_2\dots\dot{\alpha}_n}}\, \mathrm{Sym_{\delta_1\delta_2\dots\delta_n}}\,
{(\bar{\eta}^{\mu_1})_{\dot{\alpha}_1}}^{\delta_1} {(\bar{\eta}^{\mu_2})_{\dot{\alpha}_2}}^{\delta_2} \dots {(\bar{\eta}^{\mu_n})_{\dot{\alpha}_n}}^{\delta_n}\ .
\end{eqnarray} 
${(\tau^{\mu_1\mu_2\dots\mu_n})_{\alpha_1\alpha_2\dots\alpha_n}}^{\dot{\delta}_1\dot{\delta}_2\dots\dot{\delta}_n}$
and
${(\bar{\tau}^{\mu_1\mu_2\dots\mu_n})_{\dot{\alpha}_1\dot{\alpha}_2\dots\dot{\alpha}_n}}^{\delta_1\delta_2\dots\delta_n}$
are completely 
symmetric in the indices $\alpha_1\alpha_2\dots\alpha_n$, $\delta_1\delta_2\dots\delta_n$, 
$\dot{\alpha}_1\dot{\alpha}_2\dots\dot{\alpha}_n$ and $\dot{\delta}_1\dot{\delta}_2\dots\dot{\delta}_n$, therefore 
$\tau$ can be regarded as a tensor having $n$ upper Minkowski indices, one upper $(0,n/2)$-index and one lower $(n/2,0)$-index, and
$\bar{\tau}$ can be regarded as a tensor having $n$ upper Minkowski indices, 
one lower $(0,n/2)$-index and one upper $(n/2,0)$-index.
In the following and in other sections of the paper we often use a notation
which corresponds to this interpretation of $\tau$ and
$\bar{\tau}$. Both
$\tau$ and $\bar{\tau}$
are also completely symmetric and traceless in their Minkowski indices;  
\beq
g_{\mu\nu}\tau^{\mu\nu\dots\lambda}  = g_{\mu\nu}\bar{\tau}^{\mu\nu\dots\lambda} = 0\ .
\eeq

The Fierz identities
\begin{eqnarray}
{(\eta^{\mu})_\alpha}^{\dot{\beta}} {(\eta_{\mu})_\gamma}^{\dot{\delta}} & = &  2 \epsilon^L_{\alpha\gamma}\epsilon^{R\dot{\beta}\dot{\delta}}\\
{(\bar{\eta}^{\mu})_{\dot{\alpha}}}^\beta {(\bar{\eta}_{\mu})_{\dot{\gamma}}}^\delta & = & 2 \epsilon^R_{\dot{\alpha}\dot{\gamma}}\epsilon^{L\beta\delta}
\end{eqnarray}
can be generalized as
\begin{eqnarray}
{(\tau^{\mu_1\mu_2\dots\mu_n})_\alpha}^{\dot{\beta}} {(\tau_{\mu_1\mu_2\dots\mu_n})_\gamma}^{\dot{\delta}} & = &  2^n\epsilon^L_{\alpha\gamma}\epsilon^{R\dot{\beta}\dot{\delta}}\\
{(\bar{\tau}^{\mu_1\mu_2\dots\mu_n})_{\dot{\alpha}}}^\beta {(\bar{\tau}_{\mu_1\mu_2\dots\mu_n})_{\dot{\gamma}}}^\delta & = & 2^n\epsilon^R_{\dot{\alpha}\dot{\gamma}}\epsilon^{L\beta\delta}\ .
\end{eqnarray}
We also have
\beq
{(\tau^{\mu_1\mu_2\dots\mu_n})_\alpha}^{\dot{\beta}} {(\bar{\tau}_{\mu_1\mu_2\dots\mu_n})_{\dot{\gamma}}}^\rho  =   2^n{\delta_\alpha}^\rho {\delta_{\dot{\gamma}}}^{\dot{\beta}} \ .
\eeq

For $(n/2,m/2)$, $n\ne 0$, $m\ne 0$, $n\ge m$, one can define the invariant tensor
\begin{eqnarray}
{(\tau^{\mu_1\mu_2\dots\mu_n\mu'_1\mu'_2\dots\mu'_m})_{\alpha_1\alpha_2\dots\alpha_n\dot{\alpha}_1\dot{\alpha}_2\dots\dot{\alpha}_m}}^{\delta_1\delta_2\dots\delta_m\dot{\delta}_1\dot{\delta}_2\dots\dot{\delta}_n}
& = & \nonumber \\ 
&& \hspace{-5cm} {(\tau^{\mu_1\mu_2\dots\mu_n})_{\alpha_1\alpha_2\dots\alpha_n}}^{\dot{\delta}_1\dot{\delta}_2\dots\dot{\delta}_n}
{(\bar{\tau}^{\mu'_1\mu'_2\dots\mu'_m})_{\dot{\alpha}_1\dot{\alpha}_2\dots\dot{\alpha}_m}}^{\delta_1\delta_2\dots\delta_m}\ ,
\end{eqnarray}
which can be regarded as a tensor with  $n+m$ upper Minkowski indices, 
one upper $(m/2,n/2)$-index and one lower $(n/2,m/2)$-index. 
$\tau$ is obviously completely symmetric and traceless
in the first $n$ and last $m$ Minkowski indices.
We introduce the notation  $\bar{\tau}$ for the tensor defined in the same way as $\tau$, but with
$n\ne 0$, $m\ne 0$, $n\le m$.

Generalized gamma tensors for $D^{(n,m)}$ can be defined as
\beq
\gamma^{\mu_1\mu_2\dots\mu_n\mu'_1\mu'_2\dots\mu'_m}=\left( \begin{array}{cc}
0 & \tau^{\mu_1\mu_2\dots\mu_n\mu'_1\mu'_2\dots\mu'_m} \\
\bar{\tau}^{\mu'_1\mu'_2\dots\mu'_m \mu_1\mu_2\dots\mu_n} & 0 \\
\end{array}\right)\ ,
\eeq
where the matrix form corresponds to the decomposition 
$D^{(n,m)}=(n/2,m/2)\oplus (m/2,n/2)$.
$\gamma^{\mu_1\mu_2\dots\mu_n\mu'_1\mu'_2\dots\mu'_m}$ is an invariant tensor having 
$n+m$ upper Minkowski indices, one upper $D^{(n,m)}$-index and one lower $D^{(n,m)}$-index.
Generalized gamma tensors for $D^{(n)}$ can be defined in the same way (with $m=0$).

We also introduce the tensors 
$$\gamma_{(1)}^{\mu_1\mu_2\dots\mu_{n-1}\mu'_1\mu'_2\dots\mu'_{m-1}},\quad  
\gamma_{(2)}^{\mu_1\mu_2\dots\mu_{n-2}\mu'_1\mu'_2\dots\mu'_{m-2}},\quad
\dots\quad , \quad 
\gamma_{(m)}^{\mu_1\mu_2\dots\mu_{n-m}}$$ 
(assuming $m\ge 1$). 
$\gamma_{(1)}^{\mu_1\mu_2\dots\mu_{n-1}\mu'_1\mu'_2\dots\mu'_{m-1}}$ 
is obtained by contracting the $\mu_1$ and $\mu'_1$ Minkowski indices of 
$\gamma^{\mu_1\mu_2\dots\mu_n\mu'_1\mu'_2\dots\mu'_m}$ with $g_{\mu_1\mu_1'}$;
$\gamma_{(2)}^{\mu_1\mu_2\dots\mu_{n-2}\mu'_1\mu'_2\dots\mu'_{m-2}}$
is obtained by contracting the $\mu_1$ and $\mu'_1$ Minkowski indices of 
$\gamma_{(1)}^{\mu_1\mu_2\dots\mu_{n-1}\mu'_1\mu'_2\dots\mu'_{m-1}}$ 
with $g_{\mu_1\mu_1'}$, and so on.
The tensors 
$$\tau_{(1)}^{\mu_1\mu_2\dots\mu_{n-1}\mu'_1\mu'_2\dots\mu'_{m-1}},\quad  
\tau_{(2)}^{\mu_1\mu_2\dots\mu_{n-2}\mu'_1\mu'_2\dots\mu'_{m-2}},\quad
\dots\quad ,\quad 
\tau_{(m)}^{\mu_1\mu_2\dots\mu_{n-m}}$$
and 
$$\bar{\tau}_{(1)}^{\mu_1\mu_2\dots\mu_{n-1}\mu'_1\mu'_2\dots\mu'_{m-1}},\quad 
\bar{\tau}_{(2)}^{\mu_1\mu_2\dots\mu_{n-2}\mu'_1\mu'_2\dots\mu'_{m-2}},\quad
\dots\quad,\quad 
\bar{\tau}_{(m)}^{\mu_1\mu_2\dots\mu_{n-m}}$$ 
can be defined similarly. 
Any further contractions of $\tau_{(m)}^{\mu_1\mu_2\dots\mu_{n-m}}$  and $\bar{\tau}_{(m)}^{\mu_1\mu_2\dots\mu_{n-m}}$ 
(or of $\tau^{\mu_1\mu_2\dots\mu_{n}}$  and $\bar{\tau}^{\mu_1\mu_2\dots\mu_{n}}$, if $m=0$) with $g_{\mu\nu}$
give zero.

The $\gamma^{\mu_1\mu_2\dots\mu_n\mu'_1\mu'_2\dots\mu'_m}$ tensors have the following important properties:
\begin{eqnarray}
\label{eq.s1}
{(\gamma^{\mu_1\mu_2\dots\mu_n\mu'_1\mu'_2\dots\mu'_m})_\alpha}^\rho \epsilon_{\rho\beta} & = & {(\gamma^{\mu_1\mu_2\dots\mu_n\mu'_1\mu'_2\dots\mu'_m})_\beta}^\rho \epsilon_{\rho\alpha}\\
\label{eq.s2}
{(\gamma^{\mu_1\mu_2\dots\mu_n\mu'_1\mu'_2\dots\mu'_m})_\rho}^\alpha \epsilon^{\rho\beta} & = & {(\gamma^{\mu_1\mu_2\dots\mu_n\mu'_1\mu'_2\dots\mu'_m})_\rho}^\beta \epsilon^{\rho\alpha}\ ,
\end{eqnarray}
i.e.\ 
${(\gamma^{\mu_1\mu_2\dots\mu_n\mu'_1\mu'_2\dots\mu'_m})_\alpha}^\rho \epsilon_{\rho\beta}$ and 
${(\gamma^{\mu_1\mu_2\dots\mu_n\mu'_1\mu'_2\dots\mu'_m})_\rho}^\alpha \epsilon^{\rho\beta}$ are 
symmetric in $\alpha$ and $\beta$. \\ 
$\gamma_{(1)}^{\mu_1\mu_2\dots\mu_{n-1}\mu'_1\mu'_2\dots\mu'_{m-1}}$,  
$\gamma_{(2)}^{\mu_1\mu_2\dots\mu_{n-2}\mu'_1\mu'_2\dots\mu'_{m-2}}$,
$\dots$, 
$\gamma_{(m)}^{\mu_1\mu_2\dots\mu_{n-m}}$ also obviously have this property.

A further important property of the $\gamma$ tensors is
\beq
\label{eq.A57}
{(\gamma^{\mu_1\mu_2\dots\mu_{n+m}})_\alpha}^\rho{(\gamma^{\delta_1\delta_2\dots\delta_{n+m}})_\rho}^\beta 
k_{\mu_1} k_{\mu_2} \dots k_{\mu_{n+m}} k_{\delta_1} k_{\delta_2} \dots k_{\delta_{n+m}}
= (g^{\mu\nu}k_\mu k_\nu)^{n+m} {\delta_\alpha}^\beta
\eeq
for any vector $k_\mu$. The related identity 
\beq
\label{eq.A58}
{(\gamma^{\mu_1\mu_2\dots\mu_{n+m}})_\alpha}^\rho{(\gamma^{\delta_1\delta_2\dots\delta_{n+m}})_\rho}^\beta 
\partial_{\mu_1} \partial_{\mu_2} \dots \partial_{\mu_{n+m}} \partial_{\delta_1} \partial_{\delta_2} \dots \partial_{\delta_{n+m}}
= (g^{\mu\nu}\partial_\mu \partial_\nu)^{n+m} {\delta_\alpha}^\beta
\eeq
is of central importance in Section \ref{sec.hs}.

We define $\gamma^5$ on  $D^{(n)}$ and on $D^{(n,m)}$ as 
\beq
\gamma^5=\left( \begin{array}{cc}
\ii I & 0 \\
0 & -\ii I \\
\end{array}\right),
\eeq
where the matrix form is understood with respect to the decomposition $D^{(n)}= (n/2,0)\oplus (0,n/2)$ and 
$D^{(n,m)}= (n/2,m/2)\oplus (m/2,n/2)$.
$\gamma^5$ anticommutes with
$\gamma^{\mu_1\mu_2\dots\mu_n}$ and
$\gamma^{\mu_1\mu_2\dots\mu_n\mu'_1\mu'_2\dots\mu'_m}$, 
$\gamma_{(1)}^{\mu_1\mu_2\dots\mu_{n-1}\mu'_1\mu'_2\dots\mu'_{m-1}}$, $\dots$, 
$\gamma_{(m)}^{\mu_1\mu_2\dots\mu_{n-m}}$.
$\gamma^5$ is not defined for  $\tilde{D}^{(n)}$.

\subsection{Invariant complex conjugation, basis vectors and projectors in 
$D^{(n)}$, $D^{(n,m)}$ and $\tilde{D}^{(n)}$}
\label{app.a5}

\subsubsection{The representations $D^{(n)}$}
\label{app.a51}

In the following we focus on the representations $(n/2,0)$, $(0,n/2)$ and $D^{(n)}$, and we return to 
$D^{(n,m)}$ and $\tilde{D}^{(n)}$ subsequently in \ref{app.a52} and \ref{app.a53}.

The action of $\gamma^{00\dots 0}$ as a linear mapping on $D^{(n)}$ is given by
\begin{eqnarray}
\gamma^{00\dots 0}E_i & = & F_i\\
\gamma^{00\dots 0}F_i & = & E_i\ ,\qquad i=1,\dots, (1+n)\ . 
\end{eqnarray}
We also have 
$\gamma^5 E_l=\ii E_l$, $\gamma^5 F_l=-\ii F_l$, $l=1,\dots, (1+n)$.

A complex conjugation can be defined on $D^{(1)}$ in the following way: 
\beq
e_1^*= -e_4\ ,\qquad e_2^*= e_3\ ,
\eeq
and the complex conjugate of an arbitrary Dirac spinor $c_ie_i$, $c_i\in\CC$, is given by
$(c_ie_i)^*=c_i^*e_i^*$.
This is an invariant complex conjugation in the sense that it commutes with the action of $M_i$, $N_i$.
The complex conjugate of the dual basis vectors is defined in the same way:
\beq
\hat{e}_1^*= -\hat{e}_4\ , \qquad \hat{e}_2^*= \hat{e}_3\ .
\eeq

Real basis vectors (with respect to the complex conjugation above) can be defined as follows: 
\beq
\label{eq.realb}
v_1=\frac{e_2+e_3}{\sqrt{2}}\ ,\quad
v_2=\frac{\ii e_1+\ii e_4}{\sqrt{2}}\ ,\quad
v_3=\frac{-e_1+e_4}{\sqrt{2}}\ ,\quad
v_4=\frac{\ii e_2-\ii e_3}{\sqrt{2}}\ . 
\eeq
The representations of $M_i$, $N_i$ on $D^{(1)}$ are real matrices with respect to this basis, and the gamma matrices are imaginary: 
$$
\gamma^0=\left( \begin{array}{cccc}
0 & \ii & 0 & 0\\
-\ii & 0 & 0 & 0\\
0 & 0 & 0 & \ii\\
0 & 0 & -\ii & 0\\
\end{array}\right)\quad 
\gamma^1=\left( \begin{array}{cccc}
0 & 0 & 0 & -\ii\\
0 & 0 & -\ii & 0\\
0 & -\ii & 0 & 0\\
-\ii & 0 & 0 & 0\\
\end{array}\right)\quad
\gamma^2=\left( \begin{array}{cccc}
\ii & 0 & 0 & 0\\
0 & -\ii & 0 & 0\\
0 & 0 & \ii & 0\\
0 & 0 & 0 & -\ii\\
\end{array}\right) 
$$
$$
\gamma^3=\left( \begin{array}{cccc}
0 & -\ii & 0 & 0\\
-\ii & 0 & 0 & 0\\
0 & 0 & 0 & \ii\\
0 & 0 & \ii & 0\\
\end{array}\right). 
$$
$\gamma^5$ takes the form
$$
\gamma^5=
\left( \begin{array}{cccc}
0 & 0 & 0 & -1 \\
0 & 0 & -1 & 0 \\
0 & 1 & 0 & 0\\
1 & 0 & 0 & 0\\
\end{array}\right) .
$$ 
$\epsilon$ has the canonical form
$$
\epsilon=
\left( \begin{array}{cccc}
0 & \ii & 0 & 0 \\
-\ii & 0 & 0 & 0\\
0 & 0 & 0 & \ii\\
0 & 0 & -\ii & 0\\
\end{array}\right) .
$$

The complex conjugation defined on $D^{(1)}$ gives rise to the following $SL(2,\CC)$-invariant complex conjugation on $D^{(n)}$:
\begin{eqnarray}
&& E_l^*=\frac{1}{n!} {n\choose {l-1}}^{1/2}\, \mathrm{Sym}_{i_1,i_2,\dots, i_n} e_{i_1}^*\otimes e_{i_2}^*\otimes \dots \otimes e_{i_n}^*\ , 
\qquad l=1,\dots, (n+1) \nonumber \\
&& i_1=1,\ i_2=1,\ \dots,\ i_{n-l+1}=1,\quad
i_{n-l+2}=2,\ i_{n-l+3}=2,\ \dots,\  i_n=2
\end{eqnarray}
\begin{eqnarray}
&& F_l^*=\frac{1}{n!} {n\choose {l-1}}^{1/2}\, \mathrm{Sym}_{i_1,i_2,\dots, i_n} e_{i_1}^*\otimes e_{i_2}^*\otimes \dots \otimes e_{i_n}^*, 
\qquad l=1,\dots, (n+1) \nonumber \\
&& i_1=3,\ i_2=3,\ \dots,\ i_{n-l+1}=3\ ,\quad
i_{n-l+2}=4,\ i_{n-l+3}=4,\ \dots,\ i_n=4\ .
\end{eqnarray}
This can be further written as
\begin{eqnarray}
E_l^* & = & (-1)^{n+l+1} F_{n+2-l}\ ,\qquad l=1,\dots, (n+1)\\
F_l^* & = & (-1)^{l+1} E_{n+2-l}\ ,\qquad l=1,\dots, (n+1)\ .
\end{eqnarray}
The complex conjugates of the dual basis vectors $\hat{E}_l$, $\hat{F}_l$ are given by the same formulas, i.e.\
\begin{eqnarray}
\hat{E}_l^* & = & (-1)^{n+l+1} \hat{F}_{n+2-l}\ ,\qquad l=1,\dots, (n+1)\\
\hat{F}_l^* & = & (-1)^{l+1} \hat{E}_{n+2-l}\ ,\qquad l=1,\dots, (n+1)\ .
\end{eqnarray}

On real vectors, $\epsilon$ is purely imaginary if $n$ is odd and real if $n$ is even. 
As mentioned after (\ref{eq.A17}), $\epsilon$ is also symmetric if $n$ is even, 
therefore it has a well-defined signature, which is 
$(n+1,n+1)$. 
The tensors ${(\gamma^{\mu_1\mu_2\dots\mu_n})_\alpha}^\beta$ are also imaginary if $n$ is odd and real if $n$ is even. 
$\gamma^5$ is real for any value of $n$.

The following relations are also important:
\begin{eqnarray}
&& \epsilon(E_l^*,F_k)  =  \epsilon(F_l^*,E_k) = \delta_{lk}\\
&& \epsilon(E_l,F_k)  = \epsilon(E_l^*, F_k^*) = 0 \\
&& \epsilon(E_l^* + F_l^*, E_k+F_k)=2\delta_{lk} \\
&& \epsilon(E_l^* - F_l^*, E_k-F_k)=-2\delta_{lk}\\
&& \epsilon(E_l^* + F_l^*, E_k-F_k)= \epsilon(E_l^* - F_l^*, E_k+F_k) = 0\ .
\end{eqnarray} 
\begin{eqnarray}
&& \epsilon(\hat{E}_l^*,\hat{F}_k)  =  \epsilon(\hat{F}_l^*,\hat{E}_k) = (-1)^n\delta_{lk}\\
&& \epsilon(\hat{E}_l,\hat{F}_k)  = \epsilon(\hat{E}_l^*, \hat{F}_k^*) = 0 \\
&& \epsilon(\hat{E}_l^* + \hat{F}_l^*, \hat{E}_k+\hat{F}_k)=2\cdot (-1)^n\delta_{lk} \\
&& \epsilon(\hat{E}_l^* - \hat{F}_l^*, \hat{E}_k-\hat{F}_k)=-2\cdot (-1)^n\delta_{lk}\\
&& \epsilon(\hat{E}_l^* + \hat{F}_l^*, \hat{E}_k-\hat{F}_k)= \epsilon(\hat{E}_l^* - \hat{F}_l^*, \hat{E}_k+\hat{F}_k)=0\ .
\end{eqnarray}

We introduce the basis vectors
\begin{eqnarray}
u_i & = & \frac{1}{\sqrt{2}}(E_i+ (-1)^n F_i)\ , \qquad i=1,\dots, (n+1)\\
v_i & = & \frac{1}{\sqrt{2}}(E_i - (-1)^n  F_i)\ , \qquad i=1,\dots, (n+1)
\end{eqnarray}
and their duals
\begin{eqnarray}
\hat{u}_i & = & \frac{1}{\sqrt{2}}(\hat{E}_i+ (-1)^n \hat{F}_i)\ , \qquad i=1,\dots, (n+1)\\
\hat{v}_i & = & \frac{1}{\sqrt{2}}(\hat{E}_i- (-1)^n \hat{F}_i)\ , \qquad i=1,\dots, (n+1)\ .
\end{eqnarray}
We also define the vectors $u_i(k)$, $v_i(k)$ as
\begin{eqnarray}
\label{eq.uik}
u_i(k) & = &  \Lambda_{D^{(n)}}(k)u_i\ , \qquad i=1,\dots, (n+1)\\
\label{eq.vik}
v_i(k) & = & \Lambda_{D^{(n)}}(k)v_i\ , \qquad i=1,\dots, (n+1)\ ,
\end{eqnarray}
where $\Lambda_{D^{(n)}}(k)$ represents in $D^{(n)}$ 
the unique $SL(2,\CC)$ element $\Lambda(k)$ determined by the properties that 
$\Lambda(k)$ is a continuous function of $k$, $\Lambda(0)=I$, and the Lorentz transformation
corresponding to $\Lambda(k)$ is the Lorentz boost that
takes the dual four-vector $(\mass,0)$ to $(\omega(k),k)$. 
The vectors dual to $u_i(k)$, $v_i(k)$ are denoted by $\hat{u}_i(k)$, $\hat{v}_i(k)$.

The vectors $\hat{u}_i(k)$ and $\hat{v}_i(k)$ satisfy the orthogonality relations
\beq
\braketv{\hat{u}_i(k)}{\hat{u}_j(k)}=\delta_{ij}\qquad \braketv{\hat{v}_i(k)}{\hat{v}_j(k)}=-\delta_{ij}\qquad
\braketv{\hat{u}_i(k)}{\hat{v}_j(k)}=0\ ,
\eeq
where $\braketv{\ }{\ }$ denotes the scalar product introduced at the beginning of Section  \ref{sec.hilsp}.

The following obvious completeness relations are also important to note:
\begin{eqnarray}
\label{eq.c1}
{\delta_\alpha}^\beta & = &
\sum_{i=1}^{n+1} u_{i\alpha}(k)\hat{u}_i^\beta(k) + \sum_{i=1}^{n+1} v_{i\alpha}(k)\hat{v}_i^\beta(k)\\
\label{eq.c2}
{\delta_\alpha}^\beta & = &
\sum_{i=1}^{n+1} u_{i\alpha}(k)^*\hat{u}_i^\beta(k)^* + \sum_{i=1}^{n+1} v_{i\alpha}(k)^*\hat{v}_i^\beta(k)^* \ .
\end{eqnarray}
The second relation is obtained from the first one by complex conjugation. 
The $SL(2,\CC)$-invariant complex conjugation defined above is understood to be applied, 
which coincides with the componentwise complex conjugation 
if the indices $\alpha$ and $\beta$ correspond to a real basis.  
Not only here, but also throughout the appendix we use $SL(2,\CC)$-invariant complex conjugation.

The complex conjugates of $u_i(k)$ and $v_i(k)$ are
\begin{eqnarray}
u_i(k)^* & = & (-1)^{i+1}u_{n+2-i}(k) \\
v_i(k)^* & = & (-1)^{i} v_{n+2-i}(k), \qquad i=1,\dots, (n+1)
\end{eqnarray}
if $n$ is even, and 
\begin{eqnarray}
u_i(k)^* & = & (-1)^{i}v_{n+2-i}(k) \\
v_i(k)^* & = & (-1)^{i+1} u_{n+2-i}(k), \qquad i=1,\dots, (n+1)
\end{eqnarray}
if $n$ is odd. The same formulas apply to the dual vectors $\hat{u}_i(k)$ and $\hat{v}_i(k)$.

The matrices 
$$\frac{1}{2\mass^n}  {(\mass^n+(-1)^n k_{\mu_1} k_{\mu_2}\dots k_{\mu_n}\gamma^{\mu_1\mu_2\dots\mu_n} )_\alpha}^\beta $$
and 
$$\frac{1}{2\mass^n}  {(\mass^n - (-1)^n k_{\mu_1} k_{\mu_2}\dots k_{\mu_n}\gamma^{\mu_1\mu_2\dots\mu_n} )_\alpha}^\beta $$
are projectors if $k_\mu k^\mu=\mass^2$, specifically we have the identities
\begin{eqnarray}
\label{eq.pr1}
\frac{1}{2\mass^n}   {(\mass^n+(-1)^n k_{\mu_1} k_{\mu_2}\dots k_{\mu_n}\gamma^{\mu_1\mu_2\dots\mu_n} )_\alpha}^\beta 
u_{i\beta}(k)
 & = & u_{i\alpha}(k) \\
\label{eq.pr2}
\frac{1}{2\mass^n}  {(\mass^n+(-1)^n k_{\mu_1} k_{\mu_2}\dots k_{\mu_n}\gamma^{\mu_1\mu_2\dots\mu_n} )_\alpha}^\beta 
v_{i\beta}(k)
 & = & 0\\
\label{eq.pr3}
\frac{1}{2\mass^n}  {(\mass^n-(-1)^n k_{\mu_1} k_{\nu_2}\dots k_{\mu_n}\gamma^{\mu_1\mu_2\dots\mu_n} )_\alpha}^\beta 
u_{i\beta}(k)
 & = & 0 \\
\label{eq.pr4}
\frac{1}{2\mass^n}  {(\mass^n-(-1)^n k_{\mu_1} k_{\mu_2}\dots k_{\mu_n}\gamma^{\mu_1\mu_2\dots\mu_n} )_\alpha}^\beta 
v_{i\beta}(k)
 & = & v_{i\alpha}(k)
\end{eqnarray}
for all $i=1,\dots, (n+1)$, which imply
\begin{eqnarray}
\label{eq.pp1}
\sum_{i=1}^{n+1} u_{i\alpha}(k) \hat{u}_i^\beta (k) & = &  \frac{1}{2\mass^n}
{(\mass^n+(-1)^n  k_{\mu_1} k_{\mu_2}\dots k_{\mu_n}\gamma^{\mu_1\mu_2\dots\mu_n} )_\alpha}^\beta \\
\label{eq.pp2}
\sum_{i=1}^{n+1} v_{i\alpha}(k) \hat{v}_i^\beta (k) & = &  \frac{1}{2\mass^n}
{(\mass^n - (-1)^n  k_{\mu_1} k_{\mu_2}\dots k_{\mu_n}\gamma^{\mu_1\mu_2\dots\mu_n} )_\alpha}^\beta \ .
\end{eqnarray}
Complex conjugation of these formulas gives
\begin{eqnarray}
\label{eq.pp3}
\sum_{i=1}^{n+1} u_{i\alpha}(k)^* \hat{u}_i^\beta (k)^* & = &  \frac{1}{2\mass^n}
{(\mass^n+  k_{\mu_1} k_{\mu_2}\dots k_{\mu_n}\gamma^{\mu_1\mu_2\dots\mu_n} )_\alpha}^\beta \\
\label{eq.pp4}
\sum_{i=1}^{n+1} v_{i\alpha}(k)^* \hat{v}_i^\beta (k)^* & = &  \frac{1}{2\mass^n}
{(\mass^n -  k_{\mu_1} k_{\mu_2}\dots k_{\mu_n}\gamma^{\mu_1\mu_2\dots\mu_n} )_\alpha}^\beta \ .
\end{eqnarray}

\subsubsection{The representations $D^{(n,m)}$}
\label{app.a52}

We consider now the representations $D^{(n,m)}$ with $n,m\ne 0$, $n\ge m$. 
The vectors 
\beq
E_{ij}=E_i\otimes F_j\ ,\quad F_{ji}=E_j\otimes F_i\ ,\qquad i=1,\dots, (1+n)\ ,\ j=1,\dots, (1+m)
\eeq
are basis vectors in  $D^{(n,m)}$; $E_{ij}$ span the subspace $(n/2,m/2)$, whereas $F_{ji}$ span the subspace 
$(m/2,n/2)$.
The corresponding dual basis vectors are 
\beq
\hat{E}_{ij}=\hat{E}_i\otimes \hat{F}_j\ ,\quad \hat{F}_{ji}=\hat{E}_j\otimes \hat{F}_i\ ,\qquad i=1,\dots, (1+n)\ ,\ j=1,\dots, (1+m)\ .
\eeq
The action of $\gamma^{00\dots 0}$ on $D^{(n,m)}$ is 
\begin{eqnarray}
\gamma^{00\dots 0} E_{ij} & = & F_{ji} \\
\gamma^{00\dots 0} F_{ji} & = & E_{ij}\ , \qquad i=1,\dots, (1+n)\ ,\ j=1,\dots, (1+m)\ .
\end{eqnarray}
$\gamma^{00\dots 0}$ commutes with 
$\gamma_{(1)}^{00\dots 0}$, $\gamma_{(2)}^{00\dots 0}$, $\dots$, $\gamma_{(m)}^{00\dots 0}$.
The action of $\gamma^5$ on $E_{ij}$ and $F_{ji}$ is 
$\gamma^5 E_{ij}=\ii E_{ij}$, $\gamma^5 F_{ji}=-\ii F_{ji}$.

The complex conjugates of the basis vectors are
\begin{eqnarray}
E_{ij}^* & = &  F_j^* \otimes E_i^* \ =\  (-1)^{n+i+j} F_{m+2-j,n+2-i} \\
F_{ji}^* & = &  F_i^* \otimes E_j^* \ =\  (-1)^{m+i+j} E_{n+2-i,m+2-j}\ ,
\end{eqnarray}
and the same formulas apply to the dual basis vectors.
$\epsilon$ is purely imaginary on real vectors if $n+m$ is odd and real if $n+m$ is even. The signature of $\epsilon$  is
$((n+1)(m+1),(n+1)(m+1))$ when $n+m$ is even. 
The $\gamma$ tensors are also purely imaginary if $n+m$ is odd and real if $n+m$ is even.
$\gamma^5$ is real for all values of $n+m$.

The nonzero matrix elements of $\epsilon$ with respect to the basis vectors $E_{ij}$ are 
\begin{eqnarray}
&& \epsilon(E_{ij},E_{n-i+2,m-j+2}) = (-1)^{n+i+j}\\
&& \epsilon(F_{ji},F_{m-j+2,n-i+2}) = (-1)^{m+j+i}\ .
\end{eqnarray}
The nonzero matrix elements of the inverse of $\epsilon$ with respect to the dual basis vectors $\hat{E}_{ij}$ are 
\begin{eqnarray}
&& \epsilon(\hat{E}_{ij},\hat{E}_{n-i+2,m-j+2}) = (-1)^{m+i+j}\\
&& \epsilon(\hat{F}_{ji},\hat{F}_{m-j+2,n-i+2}) = (-1)^{n+j+i} \ .
\end{eqnarray}
The following relations are also important to note:
\begin{eqnarray}
&& \epsilon(E_{ij}^*,F_{lk})  =  \epsilon(F_{ji}^*,E_{kl}) = \delta_{ik}\delta_{jl}\\
&& \epsilon(E_{ij},F_{lk})  = \epsilon(E_{ij}^*, F_{lk}^*) = 0 \\
&& \epsilon(E_{ij}^* + F_{ji}^*, E_{kl}+F_{lk})=2\delta_{ik}\delta_{jl} \\
&& \epsilon(E_{ij}^* - F_{ji}^*, E_{kl}-F_{lk})=-2\delta_{ik}\delta_{jl}\\
&& \epsilon(E_{ij}^* + F_{ji}^*, E_{kl} - F_{lk})= \epsilon(E_{ij}^* - F_{ji}^*, E_{kl} + F_{lk})=  0\ ,
\end{eqnarray} 
\begin{eqnarray}
&& \epsilon(\hat{E}_{ij}^*,\hat{F}_{lk})  =  \epsilon(\hat{F}_{ji}^*,\hat{E}_{kl}) = (-1)^{n+m}\delta_{ik}\delta_{jl}\\
&& \epsilon(\hat{E}_{ij},\hat{F}_{lk})  = \epsilon(\hat{E}_{ij}^*, \hat{F}_{lk}^*) = 0 \\
&& \epsilon(\hat{E}_{ij}^* + \hat{F}_{ji}^*, \hat{E}_{kl}+\hat{F}_{lk})=2\cdot (-1)^{n+m}\delta_{ik}\delta_{jl} \\
&& \epsilon(\hat{E}_{ij}^* - \hat{F}_{ji}^*, \hat{E}_{kl}-\hat{F}_{lk})=-2\cdot (-1)^{n+m}\delta_{ik}\delta_{jl}\\ 
&& \epsilon(\hat{E}_{ij}^* + \hat{F}_{ji}^*, \hat{E}_{kl}-\hat{F}_{lk}) =
\epsilon(\hat{E}_{ij}^* - \hat{F}_{ji}^*, \hat{E}_{kl}+\hat{F}_{lk}) = 0\ .
\end{eqnarray} 

We introduce the basis vectors
\begin{eqnarray}
u_{ij} & = & \frac{1}{\sqrt{2}}(E_{ij}+ (-1)^{n+m} F_{ji})\ , \qquad i=1,\dots, (n+1),\ j=1,\dots, (m+1)\\
v_{ij} & = & \frac{1}{\sqrt{2}}(E_{ij} - (-1)^{n+m}  F_{ji})\ , \qquad i=1,\dots, (n+1),\ j=1,\dots, (m+1)
\end{eqnarray}
and their duals
\begin{eqnarray}
\hat{u}_{ij} & = & \frac{1}{\sqrt{2}}(\hat{E}_{ij}+ (-1)^{n+m} \hat{F}_{ji})\ , \qquad i=1,\dots, (n+1),\ j=1,\dots, (m+1)\\
\hat{v}_{ij} & = & \frac{1}{\sqrt{2}}(\hat{E}_{ij}- (-1)^{n+m} \hat{F}_{ji})\ , \qquad i=1,\dots, (n+1),\ j=1,\dots, (m+1)\ .
\end{eqnarray}
We also define the vectors $u_{ij}(k)$, $v_{ij}(k)$ in the same way as $u_i(k)$, $v_i(k)$ in (\ref{eq.uik}) 
and (\ref{eq.vik}).
The dual vectors
$\hat{u}_{ij}(k)$ and $\hat{v}_{ij}(k)$ satisfy the orthogonality relations
\beq
\braketv{\hat{u}_{ij}(k)}{\hat{u}_{kl}(k)}=\delta_{ik}\delta_{jl}\qquad 
\braketv{\hat{v}_{ij}(k)}{\hat{v}_{kl}(k)}=-\delta_{ik}\delta_{jl}\qquad
\braketv{\hat{u}_{ij}(k)}{\hat{v}_{kl}(k)}=0\ ,
\eeq
where $\braketv{\ }{\ }$ denotes the scalar product introduced at the beginning of Section  \ref{sec.hilsp}.

The  completeness relations analogous to (\ref{eq.c1}) and (\ref{eq.c2}) take the form
\begin{eqnarray}
\label{eq.ccc1}
{\delta_\alpha}^\beta & = &
\sum_{i=1}^{n+1}\sum_{j=1}^{m+1} u_{ij\alpha}(k)\hat{u}_{ij}^\beta(k) + 
\sum_{i=1}^{n+1}\sum_{j=1}^{m+1} v_{ij\alpha}(k)\hat{v}_{ij}^\beta(k)\\
\label{eq.ccc2}
{\delta_\alpha}^\beta & = & 
\sum_{i=1}^{n+1}\sum_{j=1}^{m+1} u_{ij\alpha}(k)^* \hat{u}_{ij}^\beta(k)^* + 
\sum_{i=1}^{n+1}\sum_{j=1}^{m+1} v_{ij\alpha}(k)^* \hat{v}_{ij}^\beta(k)^* \ .
\end{eqnarray}

The complex conjugates of $u_{ij}(k)$, $v_{ij}(k)$, $i=1,\dots, (n+1)$, $j=1,\dots, (m+1)$, are
\begin{eqnarray}
u_{ij}(k)^* & = & (-1)^{n+i+j}u_{n+2-i,m+2-j}(k) \\
v_{ij}(k)^* & = & (-1)^{n+i+j+1} v_{n+2-i,m+2-j}(k)
\end{eqnarray}
if $n+m$ is even, and 
\begin{eqnarray}
u_{ij}(k)^* & = & (-1)^{n+i+j}v_{n+2-i,m+2-j}(k) \\
v_{ij}(k)^* & = & (-1)^{n+i+j+1} u_{n+2-i,m+2-j}(k)
\end{eqnarray}
if $n+m$ is odd. The same formulas apply to the dual vectors $\hat{u}_{ij}(k)$ and $\hat{v}_{ij}(k)$.

The matrices 
$$\frac{1}{2\mass^{n+m}}  {(\mass^{n+m}+(-1)^{n+m} k_{\mu_1} k_{\mu_2}\dots k_{\mu_{n+m}}\gamma^{\mu_1\mu_2\dots\mu_{n+m}} )_\alpha}^\beta $$
and 
$$\frac{1}{2\mass^{n+m}}  {(\mass^{n+m} - (-1)^{n+m} k_{\mu_1} k_{\mu_2}\dots k_{\mu_{n+m}}\gamma^{\mu_1\mu_2\dots\mu_{n+m}} )_\alpha}^\beta $$
are projectors if $k_\mu k^\mu=\mass^2$, and identities analogous to (\ref{eq.pr1})-(\ref{eq.pp4}) hold.

The space spanned by $u_{ij}\equiv u_{ij}(k=0)$ can be decomposed into irreducible representations with respect to 
the $SU(2)$ (rotation) little group generated by $M_1$, $M_2$, $M_3$. The decomposition is 
$((n+m)/2)\oplus ((n+m)/2-1) \oplus \dots \oplus ((n-m)/2)$. These invariant subspaces are orthogonal 
with respect to the scalar product $\braketv{\ }{\ }$ introduced in Section  \ref{sec.hilsp}.
One can also introduce orthonormal (with respect to $\braketv{\ }{\ }$) 
basis vectors in these subspaces, dual basis vectors, and then the boosted versions of these.
We denote these vectors by $u_{(s),i}(k)$ and  $\hat{u}_{(s),i}(k)$, where $s$ denotes the spin and $i$ is an index 
labeling the basis vectors within the subspace of spin $s$.
Projection operators on the invariant subspaces and their boosted versions can also be formed using these basis vectors
and the dual basis vectors. These projection operators 
are $\sum_{i=1}^{2s+1} u_{(s),i\alpha}(k)\hat{u}_{(s),i}^\beta (k)$.
Similar statements can be made also for the space spanned by $v_{ij}$.

\subsubsection{The representations $\tilde{D}^{(n)}$}
\label{app.a53}

We consider finally the representations  $\tilde{D}^{(n)}$. 
The tensors $\tau^{\mu_1\mu_2\dots\mu_n\mu_1'\mu_2'\dots\mu_n'}$ have the properties 
\begin{eqnarray}
{(\tau^{\mu_1\mu_2\dots\mu_n\mu_1'\mu_2'\dots\mu_n'})_\alpha}^\rho \epsilon_{\rho\beta} & = & 
{(\tau^{\mu_1'\mu_2'\dots\mu_n'\mu_1\mu_2\dots\mu_n})_\beta}^\rho \epsilon_{\rho\alpha}\\
{(\tau^{\mu_1\mu_2\dots\mu_n\mu_1'\mu_2'\dots\mu_n'})_\rho}^\alpha \epsilon^{\rho\beta} & = & 
{(\tau^{\mu_1'\mu_2'\dots\mu_n'\mu_1\mu_2\dots\mu_n})_\rho}^\beta \epsilon^{\rho\alpha}\ ,
\end{eqnarray}
which are analogous to 
(\ref{eq.s1}) and (\ref{eq.s2}).
The formulas
\begin{eqnarray}
{(\tau^{\mu_1\mu_2\dots\mu_{2n}})_\alpha}^\rho  {(\tau^{\delta_1\delta_2\dots\delta_{2n}})_\rho}^\beta
k_{\mu_1}k_{\mu_2}\dots k_{\mu_{2n}}k_{\delta_1}k_{\delta_2}\dots k_{\delta_{2n}} & = & (g^{\mu\nu}k_\mu k_\nu)^{2n} {\delta_\alpha}^\beta \\
{(\tau^{\mu_1\mu_2\dots\mu_{2n}})_\alpha}^\rho  {(\tau^{\delta_1\delta_2\dots\delta_{2n}})_\rho}^\beta
\partial_{\mu_1}\partial_{\mu_2}\dots \partial_{\mu_{2n}}\partial_{\delta_1}\partial_{\delta_2}\dots \partial_{\delta_{2n}} & = & (g^{\mu\nu}\partial_\mu \partial_\nu)^{2n} {\delta_\alpha}^\beta\ , 
\end{eqnarray}
which are analogous to (\ref{eq.A57}) and (\ref{eq.A58}), also hold. 
The vectors 
\beq
E_{ij}=E_i\otimes F_j\ ,\qquad i=1,\dots, (1+n)\ ,\quad j=1,\dots, (1+n)\ ,
\eeq
are basis vectors in  $\tilde{D}^{(n)}$.
The corresponding dual basis vectors are 
\beq
\hat{E}_{ij}=\hat{E}_i\otimes \hat{F}_j\ ,\qquad i=1,\dots, (1+n)\ ,\quad j=1,\dots, (1+n)\ .
\eeq
The action of $\tau^{00\dots 0}$ on $\tilde{D}^{(n)}$ is given by 
\beq
\tau^{00\dots 0} E_{ij}  =  E_{ji}\ .
\eeq

The complex conjugates of the basis vectors are defined as
\begin{eqnarray}
E_{ij}^* & = &  F_j^* \otimes E_i^* \ =\  (-1)^{n+i+j} E_{n+2-j,n+2-i} \ ,
\end{eqnarray}
and the same formulas apply to the dual basis vectors.
$\epsilon$ is real on real vectors and is symmetric.
The signature of $\epsilon$  is
$\left(\frac{(n+2)(n+1)}{2},\frac{n(n+1)}{2}\right)$.
The $\tau$ tensor is  real.

The nonzero matrix elements of $\epsilon$ with respect to the basis vectors $E_{ij}$ are 
\beq
\epsilon(E_{ij},E_{n-i+2,n-j+2}) = (-1)^{n+i+j}\ .
\eeq
The nonzero matrix elements of the inverse of $\epsilon$ with respect to the dual basis vectors $\hat{E}_{ij}$ are 
given by the same formula,
\beq
\epsilon(\hat{E}_{ij},\hat{E}_{n-i+2,n-j+2}) = (-1)^{n+i+j} \ .
\eeq
The following relations are also worth noting:
\begin{eqnarray}
&& \epsilon(E_{ij}^*,E_{lk})  = \delta_{ik}\delta_{jl}\\
&& \epsilon(E_{ij}^*+E_{ji}^*, E_{kl}+E_{lk}) = 2\delta_{ik}\delta_{jl}\ , \qquad i>j,\ k\ge l\\
&& \epsilon(E_{ii}^*+E_{ii}^*, E_{kl}+E_{lk}) = 4\delta_{ik}\delta_{il}\\
&& \epsilon(E_{ij}^*-E_{ji}^*, E_{kl}-E_{lk}) = -2\delta_{ik}\delta_{jl}\ , \qquad i>j,\ k> l\\
&& \epsilon(E_{ij}^*+E_{ji}^*, E_{kl}-E_{lk}) = \epsilon(E_{ij}^*-E_{ji}^*, E_{kl}+E_{lk}) = 0 \ . 
\end{eqnarray} 
The same formulas apply to the dual vectors and the inverse of $\epsilon$.

We introduce the basis vectors
\begin{eqnarray}
u_{ij} & = & \frac{1}{\sqrt{2}}(E_{ij}+  E_{ji})\ , \qquad i>j\\
u_{ii} & = & E_{ii}\\
v_{ij} & = & \frac{1}{\sqrt{2}}(E_{ij}-  E_{ji})\ , \qquad i>j
\end{eqnarray}
and their duals
\begin{eqnarray}
\hat{u}_{ij} & = & \frac{1}{\sqrt{2}}(\hat{E}_{ij}+ \hat{E}_{ji})\ , \qquad  i>j\\
\hat{u}_{ii} & = & \hat{E}_{ii}\\
\hat{v}_{ij} & = & \frac{1}{\sqrt{2}}(\hat{E}_{ij}- \hat{E}_{ji})\ , \qquad i>j\ .
\end{eqnarray}
We also define the vectors $u_{ij}(k)$, $v_{ij}(k)$ in the same way as $u_i(k)$, $v_i(k)$ 
in (\ref{eq.uik}) and (\ref{eq.vik}).

The complex conjugates of $u_{ij}(k)$, $v_{ij}(k)$ are
\begin{eqnarray}
u_{ij}(k)^* & = & (-1)^{n+i+j}u_{n+2-j,n+2-i}(k) \\
v_{ij}(k)^* & = & (-1)^{n+i+j} v_{n+2-j,n+2-i}(k)\ .
\end{eqnarray}
The same formulas apply to the dual vectors $\hat{u}_{ij}(k)$ and $\hat{v}_{ij}(k)$.

The dual vectors
$\hat{u}_{ij}(k)$ and $\hat{v}_{ij}(k)$ satisfy the orthogonality relations
\beq
\braketv{\hat{u}_{ij}(k)}{\hat{u}_{kl}(k)}=\delta_{ik}\delta_{jl}\qquad 
\braketv{\hat{v}_{ij}(k)}{\hat{v}_{kl}(k)}=-\delta_{ik}\delta_{jl}\qquad
\braketv{\hat{u}_{ij}(k)}{\hat{v}_{kl}(k)}=0\ ,
\eeq
where $\braketv{\ }{\ }$ denotes the scalar product introduced at the beginning of Section  \ref{sec.hilsp}.

The space spanned by $u_{ij}\equiv u_{ij}(k=0)$ can be decomposed into irreducible representations with respect to 
the $SU(2)$ little group generated by $M_1$, $M_2$, $M_3$. The decomposition is 
$(n)\oplus (n-2) \oplus (n-4) \oplus \dots$. These invariant subspaces are orthogonal 
with respect to the scalar product $\braketv{\ }{\ }$.
One can also introduce orthonormal (with respect to $\braketv{\ }{\ }$)
basis vectors in these subspaces, dual basis vectors, and the boosted versions of these.
Projection operators on the invariant subspaces and their boosted versions can also be formed using these basis vectors
and the dual basis vectors. 
The decomposition of the space spanned by $v_{ij}$ is 
$(n-1)\oplus (n-3) \oplus (n-5) \oplus \dots$; otherwise it can be treated in the same way as 
the space spanned by $u_{ij}$.

\section{Equal-time anticommutator of the spin-$3/2$ field transforming according to $(3/2,0)\oplus (0,3/2)$}
\label{app.b}

\renewcommand{\theequation}{B.\arabic{equation}} 
\setcounter{equation}{0}

In this appendix the equal-time anticommutator $[\psi_\alpha(x,t),\psi_\beta^\dagger(y,t)]_+$ is calculated 
in the case when $\psi$ transforms according to the representation $(3/2,0)\oplus (0,3/2)$, with the aim 
of illustrating 
the general rule described in Section \ref{sec.dn}. The anticommutator of the Dirac field is 
calculated in the same way in Section \ref{sec.qed}. 

According to (\ref{eq.174}) we have
\beq
[\psi_\alpha(x,t),\psi_\beta^\dagger(y,t)]_+ = 
\frac{1}{2\mass^3}[{(\mass^3 + \ii^3\gamma^{\mu_1\mu_2\mu_3}\partial_{\mu_1}\partial_{\mu_2}\partial_{\mu_3})_\alpha}^\delta
\Psi_\delta(x,t),\Psi^\dagger_\beta(y,t)]_+ \ .
\eeq
By applying (\ref{eq.elim}) and considering the equal-time anticommutation relations of $\Psi$, 
only those terms on the right hand side give nonzero
contribution which contain an odd number of time derivatives, thus   
\begin{eqnarray}
[\psi_\alpha(x,t),\psi_\beta^\dagger(y,t)]_+ & = & \nonumber \\ 
&&\hspace{-4cm} \frac{\ii^3}{2\mass^3}\left( 3{(\gamma^{0ij})_\alpha}^\delta  
[\partial_i\partial_j \Pi_\delta(x,t),\Psi_\beta^\dagger(y,t)]_+ 
+
{(\gamma^{000})_\alpha}^\delta [(\partial_i^2-\mass^2)\Pi_\delta(x,t),\Psi^\dagger_\beta(y,t)]_+ \right)\ .
\end{eqnarray}
Taking into account the anticommutation relations (\ref{eq.ac5ps}) and (\ref{eq.ac6ps})  of $\Psi$, we get the final result
\beq
[\psi_\alpha(x,t),\psi_\beta^\dagger(y,t)]_+ = 
\frac{1}{2\mass^3}\left(
3{(\gamma^{0ij})_\alpha}^\delta \partial_i\partial_j
+ {(\gamma^{000})_\alpha}^\delta (\partial_i^2-\mass^2) \right) \epsilon_{\beta\delta} \delta^3(x-y)\ .
\eeq
In this formula the derivations on the right hand side are understood to be derivations with respect to the components of $x$.

\end{document}